\documentclass[longauth]{aa} % for the long lists of affiliations 
\usepackage{graphicx}
\usepackage{lipsum} 
\usepackage{txfonts}
\usepackage{amssymb}
\usepackage{amsmath}
\usepackage{sidecap}
\usepackage{multirow}
\usepackage{xcolor}
\usepackage{natbib}
\usepackage{upgreek}
\usepackage{soul}
\usepackage{float}
\bibpunct{(}{)}{;}{a}{}{,}
\usepackage[hidelinks]{hyperref}

\begin{document} 

   \title{The high-albedo, low polarization disk around HD 114082 harbouring a Jupiter-sized transiting planet 
   \thanks{Based on data collected at the European Southern Observatory (ESO), Chile under programs 096.C-0388, 098.C-0505 and 198.C-0209.} }
\subtitle{Constraints from VLT/SPHERE completed by TESS, GAIA and radial velocities}
  
%   \thanks{The reduced images as FITS files are available in electronic form at the CDS via anonymous ftp to cdsarc.u-strasbg.fr (130.79.128.5) or via http://cdsweb.u-strasbg.fr/cgi-bin/qcat?J/A+A/} }
%   \author{Author list\inst{\ref{instch1}}
   \author{N.~Engler\inst{\ref{instch1}} 
   \and J.~Milli\inst{\ref{instf1}} 
   \and R.~Gratton\inst{\ref{insti1}}
   \and S.~Ulmer-Moll\inst{\ref{instch2}} 
   \and A.~Vigan\inst{\ref{instf3}} 
   \and A.-M.~Lagrange\inst{\ref{instf1}, \ref{instf4}} 
   \and F.~Kiefer\inst{\ref{instf4}} 
   \and P.~Rubini\inst{\ref{compf1}} 
   \and A.~Grandjean\inst{\ref{instf1}} 
   \and H.M.~Schmid\inst{\ref{instch1}} 
   \and S.~Messina\inst{\ref{insteso2}} 
   \and V.~Squicciarini\inst{\ref{insti1}} 
   \and J.~Olofsson\inst{\ref{instd1},\ref{instcl1},\ref{instcl2}} \and P.~Th\'{e}bault\inst{\ref{instf4}}
   \and R.G.~van Holstein\inst{\ref{instcl5}}   
   \and M.~Janson\inst{\ref{insts1}}
   \and F.~M\'{e}nard\inst{\ref{instf1}} 
   \and J.~P.~Marshall\inst{\ref{instt1},\ref{instau1}} 
   \and G.~Chauvin\inst{\ref{instf1},\ref{instcl6}} 
   \and M.~Lendl\inst{\ref{instch2}}   
   \and T.~Bhowmik\inst{\ref{instcl3}}
   \and A.~Boccaletti\inst{\ref{instf4}} 
   \and M.~Bonnefoy\inst{\ref{instf1}} 
   \and C.~del Burgo\inst{\ref{instm1}} 
   \and E.~Choquet\inst{\ref{instf3}} 
   \and S.~Desidera\inst{\ref{insti1}}   
   \and M.~Feldt\inst{\ref{instd1}}   
   \and T.~Fusco\inst{\ref{instf8},\ref{instf3}} 
   \and J.~Girard\inst{\ref{instam2}}     
   \and D.~Gisler\inst{\ref{instch1}}  
   \and J.~Hagelberg\inst{\ref{instch2}} 
   \and M.~Langlois\inst{\ref{instf3},\ref{instf7}}  
   \and A.-L.~Maire\inst{\ref{instf1}} 
   \and D.~Mesa\inst{\ref{insti1}} 
   \and M.R.~Meyer\inst{\ref{instam1}}
   \and P.~Rabou\inst{\ref{instf1}}
   \and L.~Rodet\inst{\ref{instam3}}  
   \and T.~Schmidt\inst{\ref{instf4}} 
   \and A.~Zurlo\inst{\ref{instcl3},\ref{instcl4},\ref{instf3}}   
%   \and N.~Pawellek\inst{\ref{insten1},\ref{instun1}} 
%   \and S.~Brown\inst{\ref{instf1}} 
%   \and T. Buey\inst{\ref{instf4}}
%   \and F.~Cantalloube\inst{\ref{instf1}}
%   \and M.~Carle\inst{\ref{instf3}}
%   \and A.~Cheetham\inst{\ref{instd1}}
%   \and C.~Ginski\inst{\ref{instnl2}} 
%   \and Th.~Henning\inst{\ref{instd1}}  
 %  \and S.~Hunziker\inst{\ref{instch1}}    
%   \and O.~Moeller-Nilsson\inst{\ref{instd1}}
%   \and C.~Petit\inst{\ref{instf6}}
%   \and S.~Petrus\inst{\ref{instf3}}
%   \and S.P.~Quanz\inst{\ref{instch1}}  
%   \and E.~Rickman\inst{\ref{instch2}}
%   \and E.~Stadler\inst{\ref{instf1}}
%   \and T.~Stolker\inst{\ref{instch1}}
%   \and F.~Wildi\inst{\ref{instch2}}  
        }    

\institute{
ETH Zurich, Institute for Particle Physics and Astrophysics, 
Wolfgang-Pauli-Strasse 27, 
CH-8093 Zurich, Switzerland \\ \email{englern@phys.ethz.ch}\label{instch1}
\and
Universit\'{e} Grenoble Alpes, CNRS, IPAG, 38000 Grenoble, France\label{instf1}
\and
INAF – Osservatorio Astronomico di Padova, Vicolo
dell’Osservatorio 5, 35122 Padova, Italy\label{insti1}
\and
Observatoire de Gen{\`e}ve, Universit{\'e} de
Gen{\`e}ve, Chemin Pegasi 51, 1290 Versoix, Switzerland\label{instch2}
\and
Aix Marseille Universit\'e, CNRS, LAM - Laboratoire d'Astrophysique de Marseille, 13388, Marseille, France\label{instf3}
\and
LESIA, Observatoire de Paris, Universit\'{e} PSL, CNRS, Sorbonne
Universit{\'e}, Universit{\'e} de Paris, 5 place Jules Janssen, 92195 Meudon, France\label{instf4}
\and
Pixyl, 5 Avenue du Grand Sablon, 38700 La Tronche, France\label{compf1}
\and
European Southern Observatory, Alonso de Cordova 3107, Casilla
19001 Vitacura, Santiago 19, Chile\label{insteso2}
\and
Max-Planck-Institut f\"{u}r Astronomie, K\"{o}nigstuhl 17, 69117
Heidelberg, Germany\label{instd1}
\and
Instituto de F\'isica y Astronom\'ia, Facultad de Ciencias, Universidad de Valpara\'iso, Av. Gran Breta\~na 1111, Playa Ancha, Valpara\'iso, Chile\label{instcl1}
\and
N\'ucleo Milenio Formaci\'on Planetaria - NPF, Universidad de Valpara\'iso, Av. Gran Breta\~na 1111, Valpara\'iso, Chile\label{instcl2}
\and
European Southern Observatory, Alonso de C\'{o}rdova 3107, Casilla 19001, Vitacura, Santiago, Chile\label{instcl5}
\and
Department of Astronomy, Stockholm University, AlbaNova University Center, SE-10691 Stockholm, Sweden\label{insts1}
\and
Academia Sinica Institute of Astronomy and Astrophysics, 11F of AS/NTU Astronomy-Mathematics Building, No.1, Sect. 4, Roosevelt Rd, Taipei 10617, Taiwan\label{instt1}
\and
Centre for Astrophysics, University of Southern Queensland, Toowoomba, QLD 4350, Australia\label{instau1}
\and
Unidad Mixta Internacional Franco-Chilena de Astronomía,
CNRS/INSU UMI 3386 and Departamento de Astronomía, Universidad de Chile, Casilla 36-D, Santiago, Chile\label{instcl6}
\and
N\'ucleo de Astronom\'ia, Facultad de Ingenier\'ia y Ciencias, Universidad Diego Portales, Av. Ejercito 441, Santiago, Chile\label{instcl3}
\and
Instituto Nacional de Astrof\'{\i}sica, \'Optica y Electr\'onica, Luis Enrique Erro 1, Sta. Ma. Tonantzintla, Puebla, Mexico\label{instm1}
\and
DOTA, ONERA, Universit\'{e} Paris Saclay, F-91123, Palaiseau, France\label{instf8}
\and
Space Telescope Science Institute, Baltimore, MD 21218, USA\label{instam2}
\and
Centre de Recherche Astrophysique de Lyon, CNRS/ENSL
Universit\'{e} Lyon 1, 9 av. Ch. Andr\'{e}, 69561 Saint-Genis-Laval, France\label{instf7}
\and
Department of Astronomy, University of Michigan, 311 West Hall, 1085 S. University Avenue, Ann Arbor, MI 48109, USA\label{instam1}
\and
Center for Astrophysics and Planetary Science, Department of Astronomy, Cornell University, Ithaca, NY 14853, USA\label{instam3}
\and
Escuela de Ingenier\'ia Industrial, Facultad de Ingenier\'ia y Ciencias, Universidad Diego Portales, Av. Ejercito 441, Santiago, Chile\label{instcl4}
%\and
%Institute of Astronomy, University of Cambridge, Madingley Road, Cambridge CB3 0HA, UK\label{insten1}
%\and
%IKonkoly Observatory, Research Centre for Astronomy and Earth Sciences, Konkoly-Thege Miklós út 15-17, H-1121 Budapest, Hungary\label{instun1}
%\and
%Anton Pannekoek Astronomical Institute, University of Amsterdam,
%PO Box 94249, 1090 GE Amsterdam, The Netherlands\label{instnl2}
%\and
%ONERA, The French Aerospace Lab BP72, 29 avenue de la
%Division Leclerc, 92322 Ch\^{a}tillon Cedex, France\label{instf6}
            }

   \date{Received ...; accepted ...}
\abstract
%context
{} 
% aim
{We present new optical and near-infrared images of the debris disk around the F-type star \object{HD 114082} in the Scorpius-Centaurus (Sco-Cen) OB Association. We obtained direct imaging observations and analysed the TESS photometric time series data of this target with a goal to search for planetary companions to \object{HD 114082} and characterise the morphology of the debris disk and the scattering properties of dust particles.}
% method
{HD 114082 was observed with the VLT/SPHERE instrument in different modes: the IRDIS camera in the K band ($2.0-2.3\, \upmu$m) together with the IFS in the Y, J and H band ($0.95-1.66\, \upmu$m) using the angular differential imaging (ADI) technique as well as IRDIS in the H band ($1.5-1.8\, \upmu$m) and ZIMPOL in the I\_PRIME band ($0.71-0.87\, \upmu$m) using the polarimetric differential imaging (PDI) technique. To constrain the basic geometrical parameters of the disk and scattering properties of dust grains, scattered light images were fitted with a 3D model for single scattering in an optically thin dust disk using a Markov Chain Monte Carlo (MCMC) approach. We performed aperture photometry in order to derive the scattering and polarized phase functions, polarization fraction and spectral scattering albedo for the dust particles in the disk. This method was also used to obtain the reflectance spectrum of the disk to retrieve the disk color and study the dust reflectivity in comparison to the debris disk \object{HD\,117214}. We also performed the modeling of the \object{HD\,114082} light curve measured by TESS using models for planet transit and stellar activity to put constraints on the radius of the detected planet and its orbit. Last we searched for additional planets in the system by combining archival radial velocity data, astrometry and direct imaging.  
}
% results
{The debris disk \object{HD\,114082} appears as an axisymmetric debris belt with a radius of $\sim$0.37$''$ (35 au), inclination of $\sim$83$^\circ$ and a wide inner cavity. Dust particles in \object{HD\,114082} have a maximum polarization fraction of $\sim$17\% and a higher reflectivity when compared to the debris disk \object{HD\,117214}. This high reflectivity results in a spectral scattering albedo of $\sim$0.65 for the \object{HD\,114082} disk at near-infrared wavelengths. The disk reflectance spectrum exhibits a red color at the position of the planetesimal belt and shows no obvious features, whereas that of \object{HD\,117214} might indicate the presence of CO$_2$ ice. The analysis of TESS photometric data reveals a transiting planetary companion to \object{HD\,114082} with a radius of $\sim$1~$R_{\rm Jup}$ on an orbit with a semi-major axis of $0.7 \pm 0.4$~au. No additional planet is detected in the system by combining the SPHERE images with constraints from astrometry and radial velocity. We reach deep sensitivity limits down to $\sim$$5\,M_{\rm Jup}$ at 50\,au, and $\sim$$10\,M_{\rm Jup}$ at 30\,au from the central star.  }  
% Conclusions
{}
%  \abstract


\authorrunning{Engler et al.}

\titlerunning{HD 114082 debris disk}

   \maketitle
%
%________________________________________________________________

\section{Introduction}
Numerous mid- and far-infrared surveys of the last two decades discovered debris disks around hundreds of young stars \citep[age < 100 Myr,][and references therein]{Hughes2018}. These discoveries have opened unprecedented possibilities for us to study the native environments of stars with newly born and still rapidly evolving planetary systems. 

Based on these studies we expect that nearly each star is surrounded by some amount of dust and gas, depending on its mass and evolutionary stage. The formation pathway of a star from a molecular cloud to a main-sequence star is accompanied by a transition of primordial gas and dust, building a protoplanetary disk in the first five million years, to a processed (secondary) disk referred to as a debris disk. Such a disk contains dust particles and sometimes gas generated by collisions of planetesimals or evaporated from the surfaces of large icy bodies. This secondary material mostly consists of silicates, carbonaceous minerals, H$_2$O, CO, and CO$_2$ ices and gases and provides building blocks for the growing cores of young terrestrial planets \citep{Wyatt2018}. 

While at a stellar age of ten million years the formation of gas giants shall be almost completed, this process takes longer for terrestrial planets \citep[e.g.,][]{Rice2008, Morbidelli2007}. They are expected to grow during the next few hundred million years collecting material on their orbits around a star and can experience violent collisions with the other planetary bodies from time to time.

The SpHere INfrared survey for Exoplanets \citep[SHINE,][]{Vigan2021, Desidera2021, Langlois2021} using the Spectro-Polarimetric High-contrast Exoplanet REsearch \citep[SPHERE;][]{Beuzit2019} imager was designed to search for and study young planets at early stages of their evolution. Many debris disks have been observed in the course of the SHINE program (Engler et al. in prep.), since they are the integral components of planetary systems. One of them is \object{HD\,114082}, a debris disk resolved for the first time in the course of the SPHERE High Angular Resolution Debris Disk Survey \cite[SHARDDS,][PI: J. Milli]{Wahhaj2016,Milli2017_SHARDDS,Choquet2018}, a survey targeting cold nearby debris disks at less than 100\,pc and with high infrared (IR) excess ($L_\mathrm{IR}/L_\ast>10^{-4}$).

\object{HD\,114082} is an F3V star \citep{Houk1975} located at a distance of $95.1 \pm 0.2$ parsecs \citep{GaiaCollaboration2021} in the Sco-Cen association and a member of the Lower Centaurus Crux (LCC) subgroup. The age of young stars in the LCC subgroup is estimated to be $\sim$17 Myr \citep{Mamajek2002}.

\begin{table*} 
      \caption[]{Log of IRDIS, IFS and ZIMPOL observations with atmospheric conditions.}
      \centering
         \label{t_114082_Settings}
                \renewcommand{\arraystretch}{1.3}
         \begin{tabular}{cccccccccc}
            \hline 
            \hline
            \multirow{ 3}{*}{Date} & {Instrument} & \multirow{ 3}{*}{Filter}& Field & \multicolumn{2}{c}{Integration Time} && \multicolumn{3}{c}{Observing conditions$^1$} \\ \cline{5-6}\cline{8-10} 
            &{mode} & & rotation & DIT & Total && Airmass & Seeing & Coherence time \\
            & & & ($^\circ$)& (sec) & (min) && & ($''$) & (ms) \\
            \hline
            \hline
            \noalign{\smallskip}
 17-05-2017 & IRDIS DBI & K1K2 & 38.9 & 64 & 102.4 && 1.23 -- 1.29 &$0.92\pm 0.09$&$2.4\pm 0.4$\\
 17-05-2017 & IFS Y-H &  Y--H & 33.2 & 64 & 85.3 && 1.23 -- 1.27 &$0.78\pm 0.10$ & $2.4\pm 0.3$\\
 15-03-2017 & ZIMPOL P2 & I\_PRIME & / & 10 & 37.3 && 1.24 -- 1.27 &$0.42\pm 0.05$&$7.3\pm 1.0$\\
 14-03-2017 & ZIMPOL P2 & I\_PRIME & / & 10 & 37.3 && 1.32 -- 1.43& $0.43\pm 0.05$& $9.0\pm 1.6$\\
 06-03-2017 & IRDIS DPI & BB\_H & / & 32 & 65.1 && 1.23 -- 1.24 &$0.49\pm 0.06$&$7.8\pm 1.6$\\
 17-02-2017 & ZIMPOL P2 & I\_PRIME & / & 10 & 93.3 && 1.23 -- 1.27 &$1.21\pm 0.66$&$3.2\pm 0.9$\\
 14-02-2016 & IRDIS CI & BB\_H & 16.7 & 16 & 53.6 && 1.23 -- 1.26 &$1.37\pm 0.15$&$2.6\pm 0.4$\\
            \hline
            \hline
            \noalign{\smallskip}
            
\end{tabular}
\tablefoot{
\tablefoottext{1}{For seeing condition and coherence time the mean value with standard deviation of the distribution are given.}
}
\end{table*}   

The circumstellar dust around \object{HD\,114082} was first detected at far-IR wavelengths \citep{Chen2011}, from observations with the Multiband Imaging Photometer for \textit{Spitzer} \citep[MIPS, ][]{Rieke2004}. % chen paper masse planetesimals. 
In these observations, the star showed a very large excess flux over the predicted stellar flux at 24 and 70\,$\upmu$m yielding a fractional luminosity of $L_{\mathrm{IR}}/L_{\ast}$ of $3.01\times 10^{-3}$ \citep{Jang-Condell2015}.
%a spectrally integrated IR excess $L_{\mathrm{IR}}/L_{\ast}$ of $3.01\times 10^{-3}$ compared to the bolometric luminosity of the star $L_{\ast}$ \citep{Jang-Condell2015}. 
%The direct imaging of HD\,114082 at optical and near-IR wavelengths with the instrument and the Gemini Planet Imager (GPI) in 2016 and 2018 revealed a bright exo-Kuiper belt in this system \citep[][]{Wahhaj2016, Esposito2020}.

\object{HD\,114082} was also targeted by the Gemini Planet Imager \citep[GPI,][]{Esposito2020} and several times by the Atacama Large Millimeter/submillimeter Array \cite[ALMA, e.g.][]{Lieman-Sifry2016}. In the GPI images of total intensity, the disk west extension appears to be by 0.2$''$ longer and up to 80\% brighter than the east extension. In the observations at mm wavelengths, the debris disk was marginally resolved \citep{Kral2017}, suggesting a spatial extent of the disk less than 1.69$''$ and a disk flux of $0.72 \pm 0.05$ mJy at 1.27 mm \citep{Kral2020}. CO gas was not detected. 
 
This work presents new optical and near-IR observations of the \object{HD\,114082} debris disk performed with the SPHERE Infra-Red Dual-beam Imager and Spectrograph \citep[IRDIS,][]{Dohlen2008}, the Integral Field Spectrograph \citep[IFS,][]{Claudi2008} and the Zurich Imaging POLarimeter \citep[ZIMPOL,][]{Schmid2018}, and an analysis of disk structure and scattering properties of its dust particles. We also report a detection of transiting planet in the cavity of this disk in the time series data taken with the Transiting Exoplanet Survey Satellite \citep[TESS,][]{Ricker2014}.

In Sects.~\ref{s_Observations} and \ref{s_data} we describe the acquisition of the SPHERE data and their reduction process. Section~\ref{s_Results} discusses disk morphology observed in the images of total and polarized intensities and presents the results of the disk modeling. Section~\ref{s_SPF_PF} addresses the scattering phase functions and the polarization fraction of scattered light. 
In Sect.~\ref{s_SED} we discuss scattering properties of dust particles by analysing the disk reflectance spectrum and comparing it to the spectrum of the HD\,117214 debris disk. HD\,117214 is another young F star in the LCC subgroup harbouring a bright debris disk \citep{Engler2020} with a fractional luminosity of $2.53\times 10^{-3}$ \citep{Jang-Condell2015} and a physical disk size which are close to those of the HD\,114082 disk. The optical properties of dust particles in both disks could therefore be similar. So far, HD\,117214 is the only F star from the same sky region which was observed under excellent observing conditions, making a comparison of disk reflectance spectra possible.

Finally in Sect.~\ref{s_planets}, we report about the discovery of transiting planet in the disk and companion candidates detected in the \object{HD\,114082} images, and present the sensitivity limits on bound companions for a combination of direct imaging and radial velocity data with Gaia astrometry. Our results are summarized in Sect.~\ref{s_Summary}.

%__________________________________________________________________
   
\section{Observations} \label{s_Observations}
We present in this paper new SHINE and SHARDDS observations of \object{HD\,114082} with SPHERE in three different observing modes:
\begin{itemize}
\item IRDIFS observations
\item IRDIS polarimetric observations
\item ZIMPOL polarimetric observations
\end{itemize}
These observations complement archival data of the disk presented in \citet{Wahhaj2016}. The archival observations used the classical imaging mode of IRDIS, with the broad-band H filter and the apodized Lyot coronagraph N\_ALC\_YJH\_S \citep[diameter of 185~mas,][]{Carbillet2011, Guerri2011}.

The new IRDIS and ZIMPOL polarimetric observations were part of the SHARDDS survey (Dahlqvist et al. submitted) as a follow-up and characterization strategy of the disk. 

The log and details of all observations are presented in Table~\ref{t_114082_Settings}.

\subsection{IRDIFS observations}
The IRDIFS observations of \object{HD\,114082} were performed on 2017 May 17 in the IRDIFS-EXT mode which provides a simultaneous data acquisition with IRDIS and the IFS. We used the pupil-stabilized mode to record a sequence of images consisting of $16 \times 6 \times 64$s IRDIS exposures and $16 \times 5 \times 64$s IFS exposures while the field of view was rotating. The observation started shortly before the meridian passage and covered a total range of parallactic angle variation of 39$^{\circ}$ in the IRDIS dataset (33$^{\circ}$ in the IFS data because of shorter duration of the IFS observation). The detector integration time (DIT) per frame, the total exposure time and the observing conditions can be found in Table~\ref{t_114082_Settings}.  

The IRDIS data were taken in the dual-band imaging mode \citep[DBI,][]{Vigan2010} using the filter pair K1K2 ($\lambda_{\rm K1}=2.110\,\upmu$m, $\Delta\lambda_{\rm K1} = 0.102\,\upmu$m; $\lambda_{\rm K2}=2.251\,\upmu$m, $\Delta\lambda_{\rm K2} = 0.109\,\upmu$m). The IFS was operated in the \textit{Y-H} mode ($0.95-1.66\, \upmu$m, with a spectral resolution $R_\lambda=35$). The field of view (FOV) of the IRDIS detector is approximately $11'' \times 12.5''$, and that of the IFS is $1.73'' \times 1.73''$. The apodized Lyot coronagraph N\_ALC\_Ks \citep[diameter of 240~mas,][]{Carbillet2011, Guerri2011} was used to block the stellar light. 

To measure the stellar flux, several short exposures, where the star was offset from the coronagraphic mask, were taken at the beginning of the observation using a neutral density (ND) filter ND1.0 with a transmission of about 15\% (see SPHERE User Manual\footnote{https://www.eso.org/sci/facilities/paranal/instruments/sphere/doc.html}). The DIT of these flux calibration frames is 2\,s (DIT = 4\,s for the IFS). 

Additionally, the ``center frames'' were taken before and after the science sequence using the deformable mirror waffle mode to provide a measurement of the star position behind the coronagraph. 

\subsection{IRDIS polarimetric observations}
To extract the near-IR linearly polarized intensity of the disk scattered light, we observed the disk with the dual-beam polarimetric imaging \citep[DPI,][]{Boer2020, Holstein2020} mode of IRDIS in the broadband H filter ($\lambda=1.625\,\upmu$m, $\Delta \lambda=0.290\,\upmu$m) using the apodized Lyot coronagraph N\_ALC\_YJH\_S. These polarimetric observations were field-stabilised, because pupil-stabilised observations were not yet offered in DPI at the epoch of our observations.
%\citep[diameter of 185~mas,][]{Carbillet2011, Guerri2011} was used to block the stellar light. 

A raw IRDIS frame has a format of $2048 \times 1024$ pixels and contains two side by side star images provided by a beam splitter in front of the instrument detector. In DPI mode, two polarizers in the optical paths downstream from the beam splitter ensure that the images of two opposite polarization states are simultaneously recorded on the left and right halves of detector. The Stokes parameters $+Q, -Q, +U$ and $-U$ are  measured using a half-wave plate (HWP) inserted upstream from the beam splitter and rotated by 0$^{\circ}$, 45$^\circ$, 22.5$^\circ$ and 67.5$^\circ$, respectively.

We performed seven polarimetric cycles of consecutive measurements of the Stokes parameters $+Q, -Q, +U$ and $-U$ consisted of five exposures with a DIT of 32\,s for each parameter. The stellar flux was obtained with an off-axis image of the star using a neutral density filter ND2.0 with a 1.5\% transmissivity in the broadband H filter (see SPHERE User Manual) and a DIT of 4\,s. 

\subsection{ZIMPOL polarimetric observations}
The SPHERE instrument also has an optical arm, equipped with the ZIMPOL polarimeter. We present here polarimetric observations obtained in the I\_PRIME filter ($\lambda=0.790\,\upmu$m, $\Delta\lambda = 0.153\,\upmu$m) in the field-stabilized polarimetric mode (P2 mode) of the instrument using the slow polarimetry mode of detector. The FOV of ZIMPOL is $3.6'' \times 3.6''$.

The principle of polarimetric data recording with ZIMPOL is different from that using IRDIS in DPI mode. In ZIMPOL, the incoming signal is modulated with a ferro-electric liquid crystal retarder and a polarization beam splitter downstream from the HWP, and sent to two demodulating CCD detectors/cameras of ZIMPOL. The demodulation technique provides quasi-simultaneous recording of intensities of opposite polarization states with the same detector pixels \citep{Thalmann2008}. The modulation/demodulation cycle frequency can be set to 967.5 Hz (fast polarimetry mode) or 26.97 Hz (slow polarimetry mode). A high detector gain of 10.5 e$^-$/ADU in the fast polarimetry mode allows for the measurements of bright sources with short integration times without detector saturation. The slow polarimetry mode has a lower detector gain of 1.5 e$^-$/ADU and a much lower read-out noise level providing a higher sensitivity for longer integration times compared to the fast polarimetry mode.

During three observing nights (see Table~\ref{t_114082_Settings}) we have recorded 20 polarimetric cycles arranged in six blocks. For each cycle, four measurements of the Stokes parameters $+Q, -Q, +U$ and $-U$ were performed by rotation of the HWP by 0$^{\circ}$, 45$^\circ$, 22.5$^\circ$ and 67.5$^\circ$, respectively. All frames were taken without a coronagraph and with a DIT of 10~s. Such a long exposure saturated the science frames inside a circular region with a radius of $\sim$10 pixels ($\sim$36 mas). 

In order to determine the position of the star in the saturated frames and correct for the beam shift in frames of opposite polarization states \citep{Schmid2018}, two additional unsaturated polarimetric cycles were recorded at the beginning and end of each block (on 2017 February 17 only at the block beginning) using a neutral density filter ND2.0 with a transmission of 0.95\% (see SPHERE User Manual). These frames were also used for the measurement of the stellar flux to calibrate the photometry.

\begin{figure*}   
\includegraphics[width=16.0 cm]{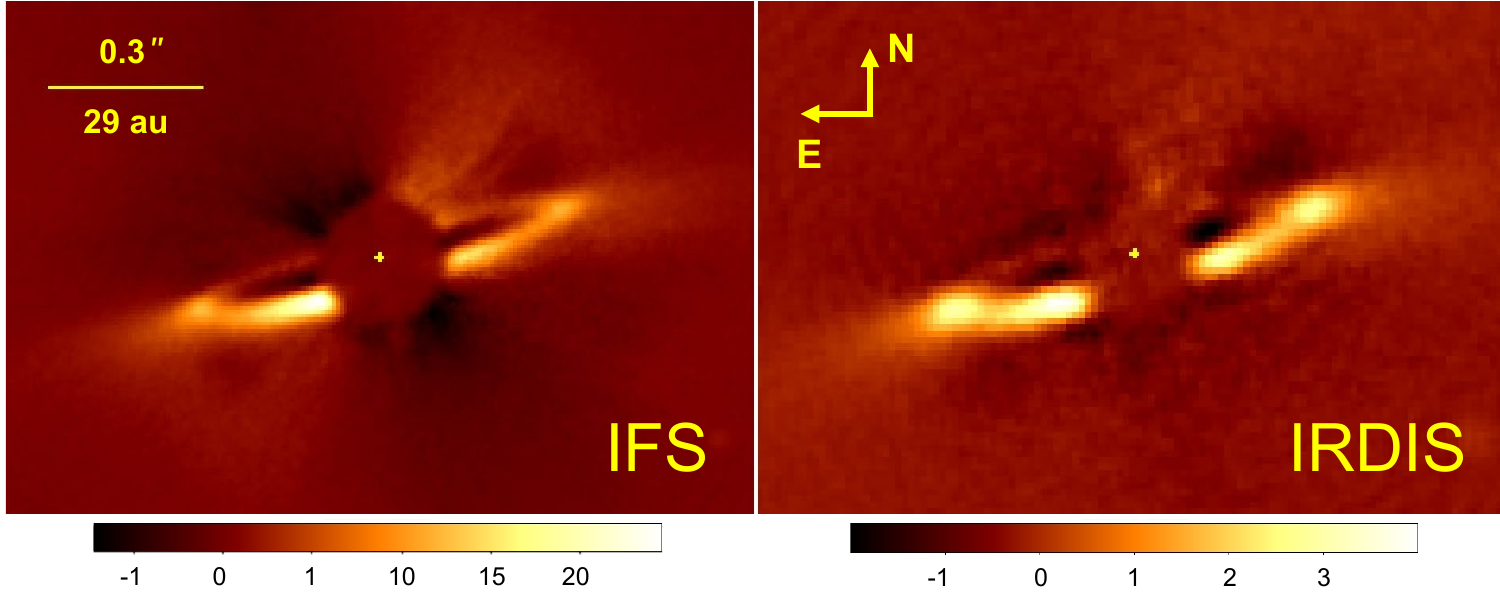} 
    \centering
    \caption{Total intensity images of the \object{HD\,114082} debris disk obtained with the cADI data reduction of the spectrally combined IFS data (\textit{left panel}) and IRDIS data (\textit{right panel}). The position of the star is marked by a yellow cross. Color bars show the flux values in counts per second (cts/s). \label{f_imaging}}
\end{figure*} 
\begin{figure*}   
\includegraphics[width=16.0cm]{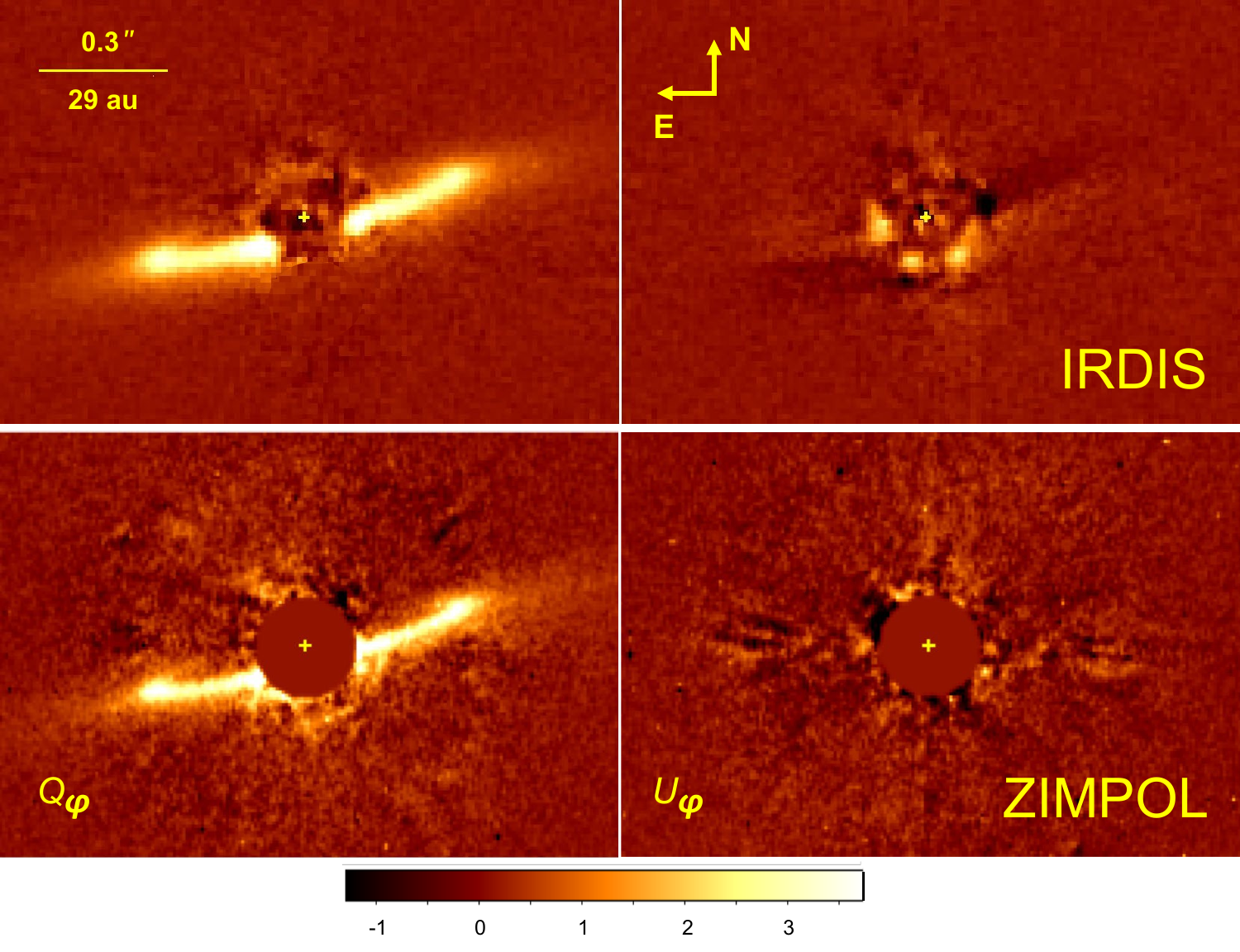} 
    \centering
    \caption{Polarized intensity images of the \object{HD\,114082} debris disk. The $Q_\varphi$ images are shown on the left panels and $U_\varphi$ images on the right panels. IRDIS data taken in the H band are displayed in the top row. ZIMPOL data recorded with the I\_PRIME filter are shown in the bottom row. In the IRDIS $Q_\varphi$ image, a strong residual signal at the coronagraph edge is masked out for clarity of the picture. The positions of the star are marked by yellow crosses. All images are smoothed via a Gaussian kernel with $\sigma_{\rm kernel}=1$ pixel and shown in the same spatial scale indicated in the top left image. The color bar shows the flux values in cts/s. \label{f_QphiUphi_IRDIS}} 
\end{figure*} 

\section{Data reduction}\label{s_data}
\subsection{IRDIFS observations}
The IRDIS and IFS data were calibrated with the SPHERE Data Reduction and Handling (DRH) 
pipeline {\it esorex} \citep{Pavlov2008} and processed at the SPHERE Data Center \citep{Delorme2017}. The calibration of raw data consisted of background subtraction, bad pixel correction, flat fielding, correction of the pixel distortion \citep{Maire2016}, and extraction of the IFS spectral data cube. 

The final calibrated datasets include two IRDIS temporal data cubes (K1 and K2 filters) and 39 IFS temporal data cubes (39 wavelength channels) with 48 frames each. The pixel scale of the IRDIS detector is 12.27 mas (K band), and a science frame is $1024 \times 1024$ pixels. In the IFS data the pixel scale is 7.46 mas and a science frame has a format of $290 \times 290$ pixels.

To subtract the stellar light, we used the SpeCal pipeline \citep{Galicher2018}, which provides several algorithms based on angular differential imaging \citep[ADI;][]{Marois2006} such as classical ADI (cADI) and template locally optimized combination of images \citep[TLOCI;][]{Marois2014}, which were used to process both the IRDIS and IFS datasets.
Figure \ref{f_imaging} shows the cADI images of the spectrally combined IFS (left panel) and IRDIS (right panel) data. 

\subsection{IRDIS polarimetric observations}
For the data reduction of IRDIS polarimetric dataset we used the IRDIS Data reduction for Accurate Polarimetry (IRDAP) pipeline \citep{Holstein2020}. This pipeline provides the basic data calibration steps such as sky-subtraction, flat-fielding and bad pixel masking. The processed frames are centered using ``center'' calibration frame with the satellite spots. 

IRDAP then corrects for the instrumental polarization and polarization crosstalk using a Mueller matrix model describing the telescope and instrument. The pipeline also offers an additional correction for the linear polarization of stellar light if needed. We measured no significant polarization from the halo of starlight in the images of \object{HD\,114082}.

The calibrated images of the Stokes parameters $Q$ and $U$ are finally obtained with a double difference method. They are used to calculate images of the azimuthal Stokes parameters $Q_\varphi$ and $U_\varphi$ presented in Fig.~\ref{f_QphiUphi_IRDIS}: 
\begin{equation}
Q_\varphi = - Q\cos 2\varphi-U\sin 2\varphi
\end{equation}
\begin{equation}
U_\varphi = Q\sin 2\varphi - U\cos 2\varphi,
\end{equation}
where $\varphi$ is the polar angle measured East of North (EoN) in the coordinate system centered on the star and the sign convention for the $Q_\varphi = - Q_{\rm r}$ and $U_\varphi = - U_{\rm r}$ parameters defined in \citet{Schmid2006} is adopted. 

The azimuthal Stokes parameter $Q_\varphi$ image contains the linearly polarized intensity of the disk scattered light as expected for an optically thin debris disk \citep[e.g.,][]{Engler2017}. In this work, we refer to the Stokes parameter $Q_\varphi$ as simply polarized intensity because the component of circular polarization is relatively very small in our observation and can be neglected. 

\subsection{ZIMPOL polarimetric observations}
The ZIMPOL polarimetric dataset was reduced with the ZIMPOL data reduction pipeline developed at ETH Zurich. The pipeline includes preprocessing and calibration of the raw data: subtraction of the bias and dark frames, flat-fielding and correction for the modulation/demodulation efficiency. The instrumental polarization is corrected through the forced equalization of fluxes measured in the frames of perpendicular polarization states within an annulus of inner radius 110 pixels (400 mas) and outer radius 200 pixels (730 mas). 

To determine the shift between positions of the star in frames of different polarization states \citep{Schmid2018} and position of the star in the combined intensity image, we fitted a 2D Gaussian function to the unsaturated PSF profiles in the frames taken with the ND filter. The position of the star on detector and the beam offset in the saturated science frames were interpolated as a function of the local siderial time using the measurements with the neutral density filter recorded immediately before and after the respective block of science frames.

The format of the reduced images is $1024 \times 1024$ pixels with a pixel size of $3.6 \times 3.6$ mas. The spatial resolution of our data is about 25 mas.

For the analysis of ZIMPOL data we selected only the best quality frames where the star position could be well determined from the unsaturated frames: all cycles from 2017 March 14 and~15 and the first block of science data taken on 2017 February~17. 

\section{Disk morphology}\label{s_Results}
\subsection{Total intensity images} 
The IFS and IRDIS images of \object{HD\,114082} (Fig.~\ref{f_imaging}) show a nearly symmetric debris ring with a radius larger than 0.3$''$ inclined at $\sim$80$^{\circ}$ ($\sim$10$^{\circ}$ away from an edge-on configuration). The south-east side of the disk, close to the edge of the coronagraph, appears to be brighter than the south-west side. This could be a result of a higher particle number density on the south-east side or an offset of the ring with respect to the star towards the north-west.  
However our data do not show any significant or measurable ring offset.

A similar surface brightness (SB) asymmetry is observed in the image of the HD~117214 debris disk \citep{Engler2020}, which has no noticeable ring offset either. 

A possible reason for such an asymmetric brightness seen in images of both targets after data post-processing, could be instrumental effects, for instance the wind-driven halo, the low-wind effect (LWE) or stellar leakages outside the coronagraph \cite[see][for a review of these effects]{Cantalloube2019}. This explanation is also supported by the images of polarized intensity (Fig.~\ref{f_QphiUphi_IRDIS}, left panels) which show a rather symmetric east-west SB distribution.

On the contrary, the north-south SB asymmetry of the debris ring clearly results from the asymmetric scattering of stellar light by dust particles which are commonly assumed to scatter more light in the forward direction.  

Both images of total intensity (left and right panels in Fig.~\ref{f_imaging}) show a cavity inside the ring from $\sim$0.3$''$ down to at least the inner working angle of coronagraph at 0.11$''$. 

\subsection{Polarized intensity images}
Figure \ref{f_QphiUphi_IRDIS} displays the polarized intensity data taken in the H band with IRDIS (top row) and in the I\_PRIME filter with ZIMPOL (bottom row). Both $Q_\varphi$ images (left panels) reveal the same disk morphology as seen in the images of total intensity (Fig.~\ref{f_imaging}). The back side of the disk is particularly visible in the ZIMPOL image. 

The $U_\varphi$ images (right panels in Fig.~\ref{f_QphiUphi_IRDIS}) contain only stellar residuals and, in the IRDIS image, a faint residual disk signal that comes from the effect of convolution of the point spread function (PSF) with the disk image \cite[see Appendix in ][for description of the effect]{Engler2018}. This spurious $U_\varphi$ signal is not seen in the ZIMPOL image (right lower panel in Fig.~\ref{f_QphiUphi_IRDIS} and Fig.~\ref{f_Uphi_colored}) because the spatial resolution of ZIMPOL is $\sim$3 times higher than the resolution provided by IRDIS. 

To support this statement, we modeled the images of the Stokes parameters $Q$ and $U$ to calculate the parameter $U_\varphi$ for the \object{HD\,114082} disk. Figure~\ref{f_Uphi_model} shows the modeled $U_\varphi$ image resulting from the convolution of parameters $Q$ and $U$ with the IRDIS PSF. This image reproduces the pattern of the IRDIS $U_\varphi$ image (right upper panel in Fig.~\ref{f_QphiUphi_IRDIS}) well.
%\textbf{Additionally, we show the ZIMPOL $U_\varphi$ image in Fig.~\ref{f_Uphi_colored} with a different scale range and colormap to underline the absence of polarized signal from the disk in this image. }

\begin{figure*}
\centering
\includegraphics[width=16.5cm]{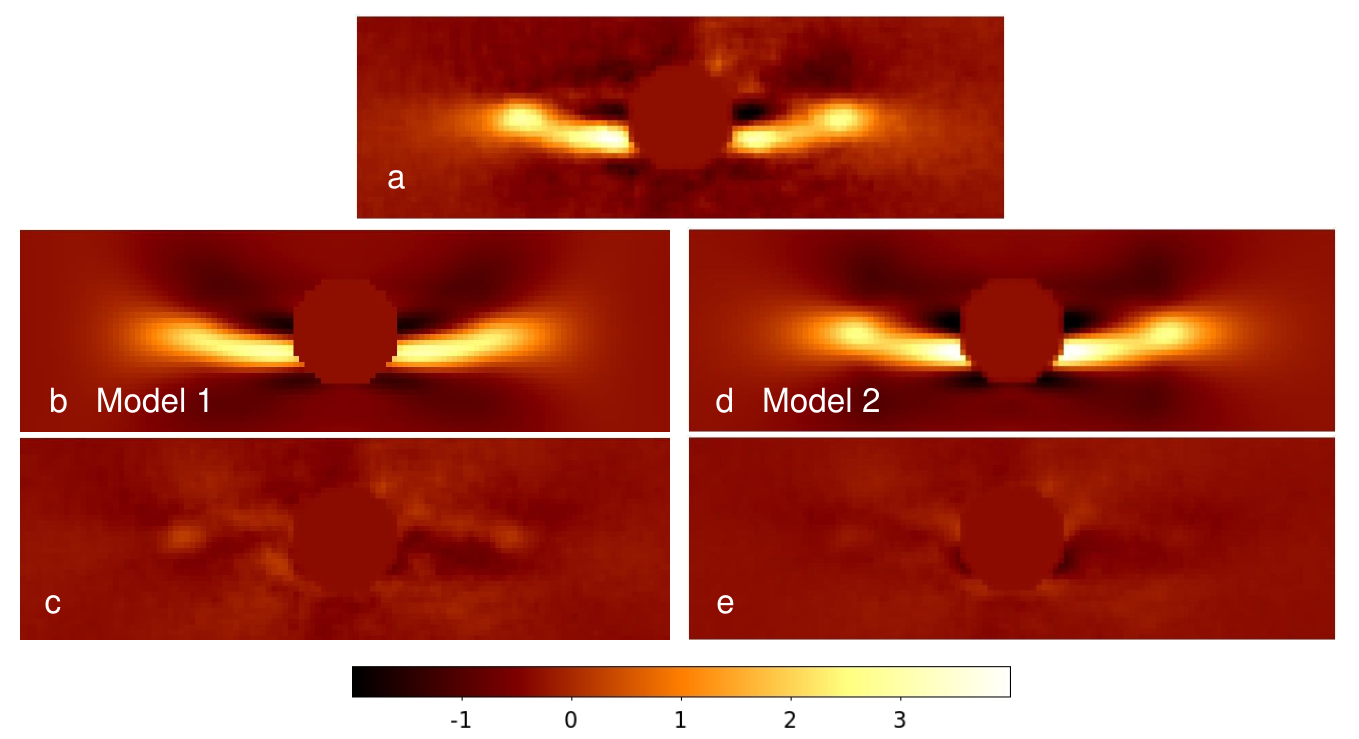} 
\caption{Comparison of the total intensity image ({\it panel a}) with the forward models of scattered light with a single HG function ({\it panel b}) and a two-component HG function ({\it panel d}). {\it Panels c} and {\it e}: Residual images obtained after subtracting the forward model images from the total intensity image. The color bar shows the flux values in cts/s.  \label{f_model}}
\end{figure*}  

\begin{table*}  
  \caption[]{Disk model parameters. \label{t_results}}
             \centering
         \begin{tabular}{lccc}
           \hline 
            \hline
            \noalign{\smallskip}
           \multirow{ 2}{*}{Optimized parameter} &  \multirow{ 2}{2 cm}{Priors} & Model 1 ($\chi^2_\nu = 1.79$ ) & Model 2 ($\chi^2_\nu = 1.67$ )  \\
            & & 1 HG parameter & 2 HG parameters \\
            \hline
            \hline
            \noalign{\smallskip}
       Reference radius $r_0$ ($''$ (au)) &[0.25, 0.45] & 0.32$^{+0.04}_{-0.03}$ (30.6$^{+3.5}_{-2.6})$ & 0.34$^{+0.03}_{-0.03}$ (32.5$^{+2.6}_{-3.5}$) \\[5pt]
       
       Scale height $H_0$ ($''$ (au)) &[0.0, 0.1] & 0.004$^{+0.003}_{-0.001}$ (0.4$^{+0.3}_{-0.1}$) & 0.007$^{+0.003}_{-0.003}$ (0.7$^{+0.3}_{-0.3}$) \\[5pt]
       
           Inner radial index $\alpha_{\rm in}$ & [0, 80] & 20$^{+7}_{-5}$ & 24$^{+18}_{-10}$ \\[5pt]
           
           Outer radial index $\alpha_{\rm out}$ & [-15, 0] & -4.5$^{+1.2}_{-1.5}$ & -5.0$^{+1.0}_{-1.5}$ \\[5pt]
           
           Flare index $\beta$ &[0, 10] & 6.6$^{+0.3}_{-0.3}$  & 4.0$^{+0.4}_{-0.3}$  \\[5pt]
           
           Inclination $i$ ($^\circ$) &[75, 90] & 82.3$^{+1.6}_{-1.4}$ & 83.0$^{+1.5}_{-1.6}$  \\[5pt]
           
           Position angle ($^\circ$) &[100, 110] & 105.8$^{+0.9}_{-0.9}$ & 105.9$^{+0.9}_{-0.9}$ \\[5pt]
           
           HG parameter $g_{1}$ &[0, 1] & 0.56$^{+0.11}_{-0.11}$ & 0.83$^{+0.13}_{-0.13}$  \\[5pt]
           
           HG parameter $g_{2}$ &[-0.4, 0.5] & (...) & 0.06$^{+0.05}_{-0.06}$  \\[5pt]
           
           Scaling parameter $w$ &[0, 1] & (...) & 0.73$^{+0.09}_{-0.11}$  \\[5pt]
           
           Scaling factor $A$ & [0, 10] & 1.17$^{+0.70}_{-0.70}$ & 1.65$^{+0.95}_{-0.95}$  \\[5pt]
           \noalign{\smallskip}
           \hline
           \hline
            \noalign{\smallskip}
\end{tabular}
\end{table*}

\subsection{Model for scattered light} \label{s_Modeling}
To estimate the geometrical parameters of the disk we used a 3D model for single-scattering of stellar photons by dust particles. The detailed description of models for scattered and polarized intensity images can be found in Appendix~\ref{s_modeling_app}. All model parameters are listed in the first column of Table~\ref{t_results}.  

In order to find a set of parameters that best fits the total intensity image in the K1 band (Fig.~\ref{f_model}a), we ran a custom MCMC code. %using the Python package {\it emcee} by \cite{Foreman-Mackey2013} 
At each iteration in this code, a new model with a parameter set drawn from the uniform prior distributions (Col.~2 in Table~\ref{t_results}) is created, convolved with the instrument PSF and inserted into an empty data cube at different position angles to mimic the rotation of the sky field during the observation. A cADI forward-modeling is then performed and the final model image is compared to the imaging data.

We tested two families of models with two different scattering phase functions (SPFs): the first SPF with a single Henyey-Greenstein (HG) function (hereafter we refer to this model as model 1), and the second SPF with a two-component HG function referred to as model 2. The detailed description of the HG functions and the applied MCMC fitting method is given in Appendix~\ref{s_modeling_app}.

The posterior distributions of fitted parameters for both families of models are shown in Figs.~\ref{f_mcmc_1g} and \ref{f_mcmc_2g_1}. We consider the median values of these distributions as the best-fit parameters and list them in Cols.~3 and 4 of Table~\ref{t_results}. Both models yield similar best-fit parameter sets with values for the radius of the planetesimal belt $R_{\rm belt}$ of $0.33 \pm 0.03$ (model 1) and $0.35 \pm 0.03$~arcsec (model 2), and disk inclinations of $82-83^\circ$. The index of the inner radial power law for the particle number density $\alpha_{\rm in}$ is the only parameter which remains unconstrained. The MCMC sample for this parameter shows a clear tendency to values larger than 40 indicating a sharp inner edge of the disk.

\cite{Wahhaj2016} fitted a different model for the spatial distribution of dust particles to the scattered light images of the \object{HD\,114082} disk obtained in the H band with SPHERE. The authors modeled the disk surface density within the range of radial distances $r_{\rm in}$ and $r_{\rm in} + \Delta r$ with $r_{\rm in}$ being the inner radius  and $\Delta r$ the width of the disk using three images from various data reductions (PCA, matched LOCI \citep[MLOCI,][]{Wahhaj2015}+ADI, MLOCI+RDI). Their results related to the radial location of the belt are comparable with the radius for the planetesimal belt $R_{\rm belt}$ we derived for our models 1 and 2: $r_{\rm in} + \Delta r = 0.31'' + 0.05''$ for the PCA image. However, the disk inclination of $84.9^\circ$ in the PCA model is slightly outside of the 68\% credible intervals for the inclination provided by our models. In contrast, the other fitting procedures by \cite{Wahhaj2016} (MLOCI+ADI, MLOCI+RDI) preferred also a lower inclination of the disk down to $81^\circ$. As a SPF, \cite{Wahhaj2016} used a single HG function with one asymmetry parameter~$g$. Their best-fit values for $g$ are between 0.07 and 0.23 which are much lower than our result ($g=0.56$ in model 1). These differences can be explained by a lower quality of the images due to unfavourable observing conditions\footnote{The SHARDDS data taken on 2016 February 14 are impacted by the LWE \cite[][]{Milli2018} which distorts the shape of the instrument PSF.} and more aggressive data reduction techniques. 

With the best-fit parameters (Cols.~3 and 4 of Table~\ref{t_results}) we created two images for a visual and statistical comparison of model 1 and model 2 to the data. Figure~\ref{f_model} displays forward-model images of both models %after convolution with the instrument PSF and post-processing 
(Figs.~\ref{f_model}b and \ref{f_model}d). It shows also the residual images (Figs.~\ref{f_model}c and \ref{f_model}e) obtained after subtraction of forward-model images from the data (Fig.~\ref{f_model}a). The visual comparison of the residual images suggests that model 2 fits the data slightly better.

The $\chi^2$ analysis, which we performed in order to assess the goodness-of-fit of both models, also supports this conclusion. We calculated the reduced $\chi_\nu^2$ metric according to:
\begin{equation} \label{eq_chi2}
\chi^2_{\nu} = \frac{1}{\nu}\sum_{i=1}^{N_{\rm data}} \frac{\left[ F_{i,\, {\rm data}}- F_{i,\, {\rm model}}(\vec{p})\right]^2 }{ \sigma_{i,\, {\rm data}}^2}
\end{equation}
where $N_{\rm data}$ is a number of modeled resolution elements, $F_{i,\, {\rm data}}$ is the flux measured in element $i$ with an uncertainty $\sigma_{i,\, {\rm data}}$ and $F_{i,\, {\rm model}}(\vec{p})$ (Eq.~\ref{eq_F}) is the modeled flux of the $i$ element. The degree of freedom of the fit is denoted by $\nu = N_{\rm data}-N_{\rm par}(\mathcal{M})$, where $N_{\rm par}(\mathcal{M})$ represents the number of variable parameters $\vec{p}=(p_{1}, p_{2}, ..., p_{N_{\rm par}})$ of model $\mathcal{M}$. In total, the model 1 has nine parameters, and the model~2 eleven (Col.~1 of Table~\ref{t_results}). The FWHM of the stellar profile in the IRDIS flux images is $\sim$3 pixels. Therefore we consider an image area of $3\times3$ binned pixels as one resolution element. The uncertainties $\sigma_{i,\, {\rm data}}$ are computed as the standard deviation of the flux distribution in concentric annuli of the total intensity image excluding regions with the disk flux.

We obtained a reduced $\chi^2_\nu = 1.79$ for the model 1 ($\nu = 458$) and $\chi^2_\nu = 1.67$ for the model 2 ($\nu = 456$).
Although both $\chi^2_\nu$ values are relatively close to each other, the Bayesian information criterion (BIC) indicates that the model~2 provides a better fitting result. We calculate the BIC according to \citet{Schwarz1978}:
\begin{equation} \label{eq_BIC}
BIC(\mathcal{M}) = \chi^2_{\nu} \nu + N_{\rm par}(\mathcal{M}) \, \ln(N_{\rm data}) + const,
\end{equation}
For models 1 and 2 the difference $BIC(model\, 1) - BIC(model\, 2)$ is positive and equal to 46. The lower BIC value of model 2 implies that this model should be preferred over model 1 even if the latter has less fitting parameters. We have to note, however, that a correlation between the adjacent resolution elements is not taken into account in Eqs.~\ref{eq_chi2} and \ref{eq_BIC}, and, thus, the actual BIC difference value could deviate from the adopted one.

\begin{figure}
\centering
\includegraphics[width=8.5cm]{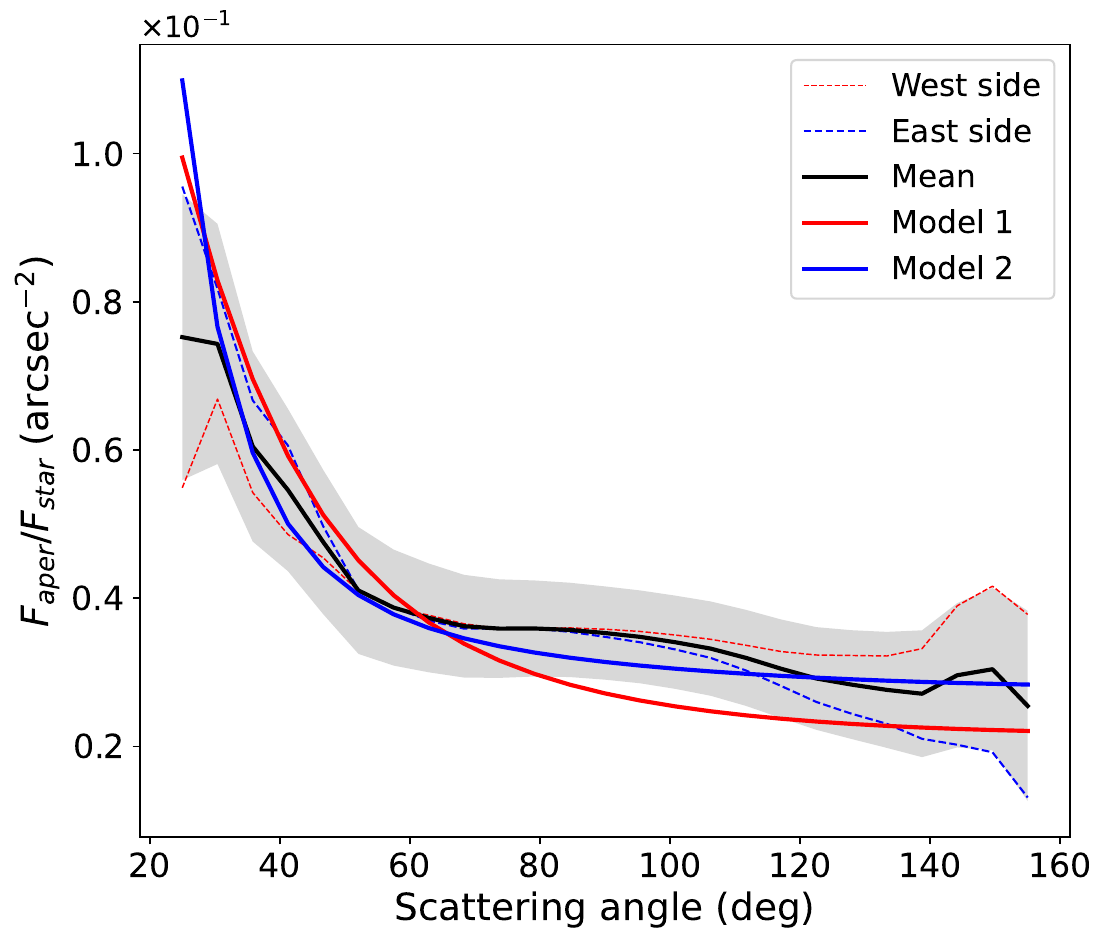} 
   \caption{Scattering phase function of the \object{HD\,114082} debris disk normalized by the stellar flux. The grey-shaded area indicates the uncertainty on the mean flux (black line) of the east and west disk sides. \label{f_SPF}}
\end{figure}  

\section{Phase functions and polarization fraction} \label{s_SPF_PF}
The SPF (respectively the polarized SPF, pSPF) of a debris disk is defined as a function that describes how the total (respectively the polarized) flux scattered by a population of dust particles is spatially distributed as a function of scattering angle $\theta$ ($0 < \theta < 180^\circ$). The polarized flux represents only a fraction of the total scattered flux. This fraction depends on the scattering angle as well and can be defined as a polarization fraction (PF) phase function $p(\theta)$:
\begin{equation} \label{eq_p}
p(\theta) = \frac{pSPF(\theta)}{SPF(\theta)}
\end{equation}
In this work, when we refer to the PF, we always mean a degree of linear polarization of scattered light because the circular polarization is negligibly small in the observations of debris disks and, therefore, not taken into account.

All three phase functions together with the value of maximum PF of scattered light $p_{\rm max}$ are unique for each debris disk and, as a tool, offer an opportunity to study the reflectivity of small dust particles in the disks surrounding distant stars. 
Light scattering theories, like the Mie theory, predict, and laboratory measurements \citep[e.g.,][]{Munoz2017, Frattin2019} show that the phase functions contain an information about dust optical properties. Therefore, many previous debris disk studies at optical and near-IR wavelengths used the phase function analysis to infer the characteristics of observed dust populations, for instance particle effective size or composition \citep[e.g.,][]{Graham2007, Milli2017,  Milli2019, Esposito2018, Ren2019, Engler2020}. 

\begin{figure*}
\centering
\includegraphics[width=18cm]{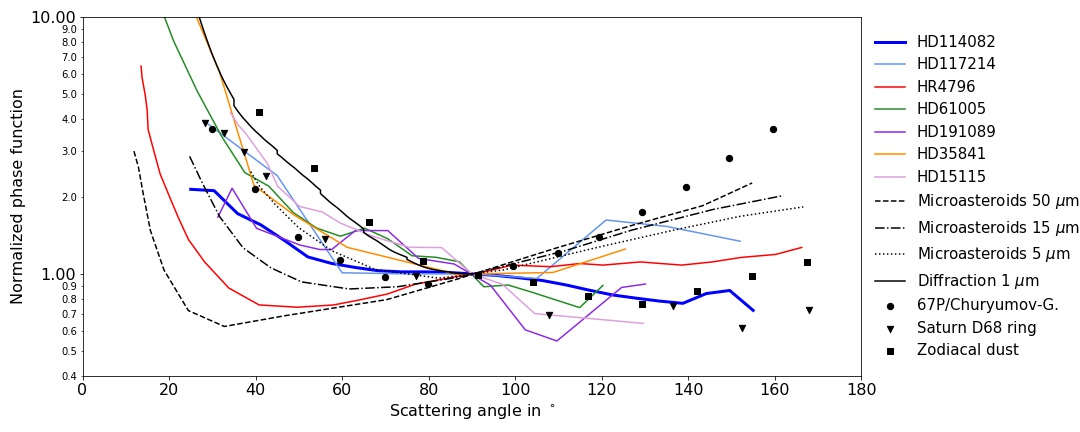}
 \caption{Scattering phase functions measured for different debris disks and dust populations in the Solar System in comparison with the theoretical micro-asteroid phase functions and diffraction model for micron-sized particles.  \label{f_all_SPFs}}
\end{figure*}

For such a kind of analysis the phase functions can be measured from the images of debris disks in scattered light. However, this is a challenging task that requires data with a high signal-to-noise ratio (S/N), additional data calibrations and corrections for different effects such as disk flux losses due to the data post-processing technique or convolution with the instrument PSF. The correction factors for these effects can be estimated only by modeling  the disk geometry and forward-modeling of the scattered (polarized) light image. 

Furthermore, the calculation of PF requires two images in the same wavelength range: a total intensity image and a polarized intensity image of the disk. 

\object{HD\,114082} was observed in field-stabilized polarimetric mode with both IRDIS and ZIMPOL. Therefore two images showing the disk polarized intensity are available (in the H band and in the I\_PRIME filter, Fig.~\ref{f_QphiUphi_IRDIS}) but there are no suitable images of the disk in total intensity from the same observations. The IRDIS broadband H image of the disk from the previous SHARDDS observation suffers a lot from the LWE in the regions of interest, as mentioned in Sect.~\ref{s_Modeling}. Therefore, in order to measure the SPF and PF of the \object{HD\,114082} disk at near-IR wavelengths, we created a combined total intensity image corresponding to the H band by summing up the total intensity images in the IFS channels from 1.45 to 1.64~$\upmu$m. This wavelength range covers approximately the range of the broadband H IRDIS filter used in the DPI observations of \object{HD\,114082}. Thus, the obtained IFS image for the total intensity is directly comparable to the polarized intensity image in the broadband H filter shown in Fig.~\ref{f_QphiUphi_IRDIS} (left panel).

\subsection{Scattering phase function}\label{s_SPF}
To derive the SPF of the \object{HD\,114082} disk shown in Fig.~\ref{f_SPF}, we measured the total intensity of scattered light within circular apertures with a radius of 19~mas ($0.5\, \lambda/D$, that is the half width at half maximum of the PSF) or an area of $1.13 \times 10^{-3}$ arcsec$^2$ placed along the disk rim in the combined IFS image. Before applying aperture photometry, we corrected the IFS ADI data for the disk flux losses resulting from the ADI post-processing. For this correction, we multiplied the IFS image by a scaling factor map which was computed as a flux ratio in each pixel between the image of the best-fit model convolved with the instrument PSF and the image obtained after the ADI post-processing of the same model. The post-processing was applied to a data cube consisting of 80 frames, where the model image (convolved with the instrument PSF) was injected at different PAs corresponding to the parallactic angle change during observation. 

The flux values shown in Fig.~\ref{f_SPF} are normalized to a sum of stellar fluxes in the IFS channels corresponding to the H band (wavelengths 1.45 - 1.64~$\upmu$m). Because of the high inclination and small angular size of the disk, we were able to perform aperture photometry only within the range of scattering angles $25^\circ < \theta < 155^\circ$. 

By using this measurement method, we assume that the disk is thin in the vertical direction so that the derived aperture flux can be associated with a single scattering angle. The vertical thickness of the disk, however, affects the shape of the measured SPF at scattering angles close to $90^\circ$ for highly inclined disks \citep{Olofsson2020}. This effect, sometimes referred to as limb brightening, due to the increased column density of dust particles around the disk major axis ($\theta = 90^\circ$) is also observed in our measurements: the SPF curve (black solid line in Fig.~\ref{f_SPF}) shows a slight bump between $\theta = 60^\circ$ and $\theta = 120^\circ$ compared to the smoothly decreasing model SPFs which are shown by the red and blue solid lines for the best-fitting model 1 and model 2, respectively. As seen in Fig.~\ref{f_SPF}, the SPF of model 2 seems to match the measurement better. 

We compare the measured SPF of \object{HD\,114082} disk in Fig.~\ref{f_all_SPFs} with the empirically derived phase functions of other debris dust systems as compiled in \citet{Engler2020}. This figure includes the SPFs of several debris disks: HD 117214 \citep{Engler2020}, \object{HR 4796 A} \citep{Milli2017}, \object{HD 61005} \citep{Olofsson2016}, \object{HD 191089} \citep{Ren2019}, \object{HD 35841} \citep{Esposito2018} and \object{HD~15115} \citep{Engler2019}. The phase functions obtained for the zodiacal dust in the solar system \citep{Leinert1976}, ring D68 of Saturn \citep{Hedman2015} and comet 67P/Churyumov–Gerasimenko \citep[dataset MTP020 taken on 2015 August 28,][]{Bertini2017} are also plotted in this figure, as well as the theoretical SPFs derived using the Fraunhofer diffraction model for the particles with radii of 1, 5, 15, and 50 $\upmu$m \citep{Min2010, Engler2020} and Hapke reflectance theory \citep{Hapke1981} for the dust grains covered by small regolith particles \citep[microasteroids SPFs in Fig.~\ref{f_all_SPFs},][]{Min2010}. For easier comparison, all displayed SPFs are normalized to their values at scattering angle of 90$^\circ$.

Between $\theta= 40^\circ$ and $\theta= 90^\circ$ the SPF of the HD\,114082 disk has a negative slope like the SPFs of other debris disks except for \object{HR 4796 A}. For scattering angles larger than $90^\circ$, the \object{HD\,114082} SPF seems to decrease in a similar way as the SPFs of the solar zodiacal dust and Saturn's ring. In the whole range of measured scattering angles ($25^\circ < \theta < 155^\circ$), the trend of the \object{HD\,114082} phase function is most similar to that of the debris disk surrounding \object{HD 191089}. The latter is a young F star (F5V) in the $\beta$ Pictoris moving group which harbours an exo-Kuiper belt with a radius of $\sim$45 au \citep{Ren2019}.

As seen in Fig.~\ref{f_all_SPFs}, the SPF curve of the \object{HD\,114082} disk rises for scattering angles smaller than $90^\circ$, and is contained between the microasteroids curves of 5 and 15~$\upmu$m.
If the forward-scattering peak of phase function is governed by Fraunhofer diffraction \cite[see Sect.~6 in][]{Engler2020}, the average effective size of particles probed with this near-IR observation (H band) is expected to be between 5 and 15 $\upmu$m. 
The size of $5\,\upmu$m would be consistent with the blow-out size of 1 - 3 $\upmu$m for the dust particles which are mainly removed from the system by the radiation pressure of the host star. Adopting stellar luminosity $L_{\ast} = 3.86 L_{\odot}$ and mass $M_{\ast} = 1.4 M_{\odot}$ \citep{Pecaut2012}, and the average bulk density of dust particles $\rho$ of 2 - 3 g cm$^{-3}$, we derive this blow-out size from relation between radiation pressure force and gravitational force acting on dust particles \citep{Burns1979}:
\begin{equation}
a_{\rm min} = \frac{3 L_{\ast} Q_{\rm pr}}{16 \pi G M_{\ast} c \rho },
\end{equation}
where $Q_{\rm pr}$ is the radiation pressure coupling coefficient, $G$ is the gravitational constant and $c$ is the speed of light. For a F~star, most of the stellar energy is concentrated at wavelengths near 0.5 $\upmu$m. Therefore for dust particles with sizes larger than 1 $\upmu$m, the size parameter $x = 2 \pi a / \lambda \gg 1$ and the geometrical optics approximation holds. This means that the effective cross-section of particles can be approximated by their geometric cross-section resulting in $Q_{\rm pr} \approx 1$, and thus the particle minimum size $a_{\rm min}$ is around 3 $\upmu$m for HD\,114082.

If we assume that dust is produced through a collisional cascade, then its size distribution should follow a power-law of index $q=-3.5$, for which most of the geometrical cross section should be in the smallest (bound) particles with sizes $a> 3$~$\upmu$m, so that the scattered flux should be dominated by these particles, at least at wavelengths smaller than $2 \pi \, a_{\rm min}$ \citep[e.g.,][]{Thebault2007}. Thus, we can expect to trace dust particle population with the effective particle size of $\sim$5~$\upmu$m in this observation.

\begin{figure}
\centering
\includegraphics[width=8cm]{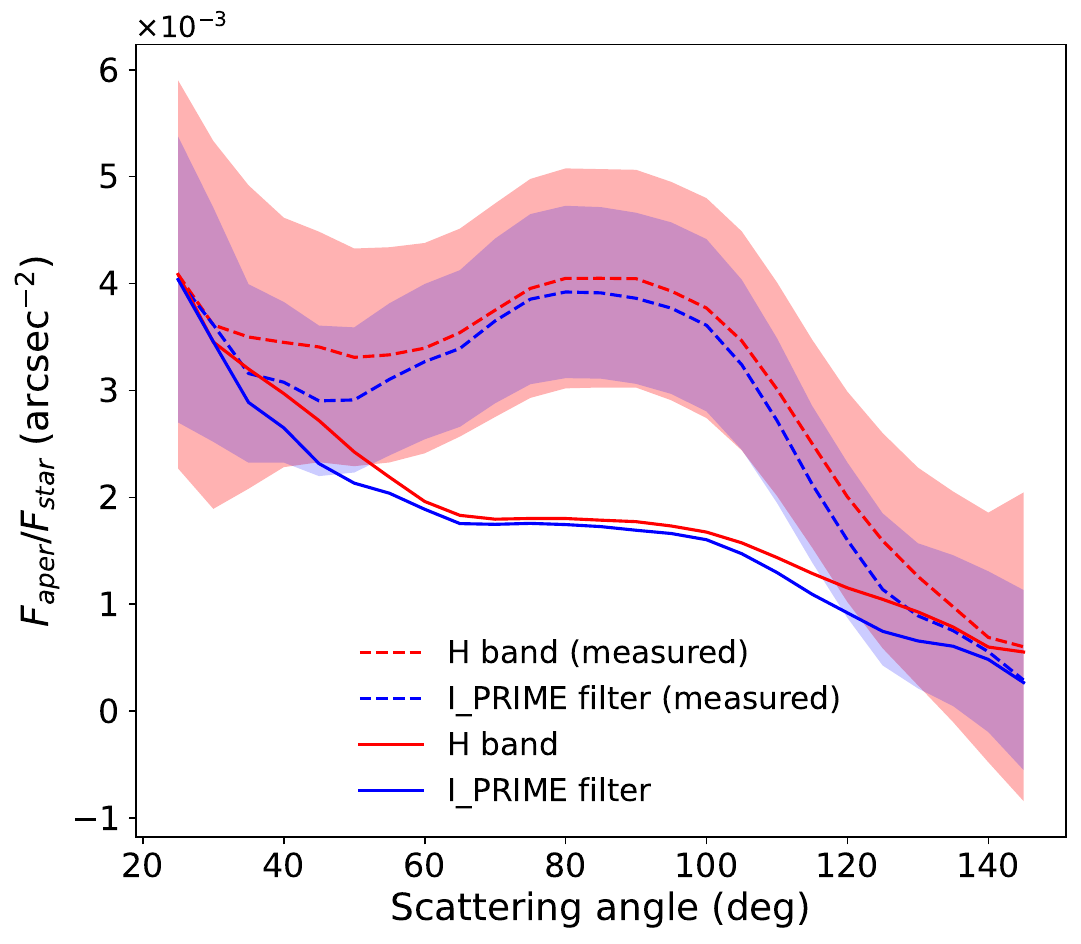}
   \caption{Polarized scattering phase functions of the \object{HD\,114082} debris disk in H band and I\_PRIME filter normalized by the stellar fluxes. The pSPFs corrected for the limb brightening effect are shown by the solid lines. The light red-shaded area indicates the uncertainty on the measured disk contrast in H band. The light blue-shaded area indicates the uncertainty on the measured disk contrast in I\_PRIME filter. \label{f_PPF}}
\end{figure}

\subsection{Polarized scattering phase function}\label{s_PPF}
To retrieve the pSPFs (dashed lines in Fig.~\ref{f_PPF}), we applied the same aperture photometry to the $Q_\varphi$ images in the H band and I\_PRIME filter (Fig.~\ref{f_QphiUphi_IRDIS}, left panels): the circular apertures with a radius of 19\,mas were placed at the same positions in the disk as in the IFS H band image of the total intensity. 
The measured aperture fluxes $F_{\rm aper}$ plotted in Fig.~\ref{f_PPF} are normalized by the stellar flux $F_{\rm star}$ in the corresponding filter and therefore represent the disk contrast in polarized light at measured position.

Both retrieved pSPFs have a similar shape and exhibit a bow between $\theta = 80^\circ$ and $\theta = 110^\circ$. The highest disk contrast is measured at the disk ansae (around $\theta = 90^\circ$) and at the smallest scattering angle $\theta = 25^\circ$ we are able to measure. This contrast reaches $(4.0 \pm 1.0)\times 10^{-3}$ arcsec$^{-2}$ in the H band and $(3.9 \pm 0.8)\times 10^{-3}$ arcsec$^{-2}$ in the I\_PRIME filter. A higher polarized contrast at the disk ansae can be explained by a higher column density of dust particles along the LOS at the position of the planetesimal belt (limb brightening effect) and by the highest PF of scattered light at angles close to $90^\circ$. At smaller scattering angles, the increasing polarized flux results from the rapidly growing amount of scattered light due to the forward-scattering behavior of dust particles. 
For $\theta > 100^\circ$ the polarized flux is continuously decreasing and apparently diminishing towards zero at $\theta > 140^\circ$. The shapes of both measured pSPFs thus are quite similar to that of HR 4796~A \citep{Milli2019}. 

The impact of limb brightening on the shape of measured pSPFs can be estimated with the model of a disk that scatters light isotropically. The result of such an estimation is model-dependent but, if corrected, the measured phase function should reproduce the shape of the actual phase function better. Therefore, we applied the aperture photometry to the image of model 1 generated with the HG asymmetry parameter $g=0$ (isotropic scattering) to calculate the correction factor for the pSPFs in dependence on scattering angle $\theta$. The correction factor was derived as a ratio between the flux value measured within a circular aperture at a given position and the smallest flux value we obtained for $\theta =25^\circ$. Multiplying the measured pSPFs by this factor, we derived the corrected phase functions which are flat between $\theta = 60^\circ$ and $\theta = 110^\circ$ (solid lines in Fig.~\ref{f_PPF}).

\subsection{Degree of linear polarization of scattered light}\label{s_pol_fraction}
\subsubsection{Polarization fraction in H band} \label{s_pol_frac_Hband}
To measure the PF of disk scattered flux in the H band we used the aperture photometry results presented in Sects.~\ref{s_SPF} and \ref{s_PPF}. 
We preferred to use the IFS H band image for this analysis instead of the archival IRDIS H broadband image of total intensity as mentioned above. Nevertheless, we utilized the latter to put constraints on the PF at the disk ansae where the disk flux is least disturbed by the LWE.

\begin{figure}
\centering
\includegraphics[width=8cm]{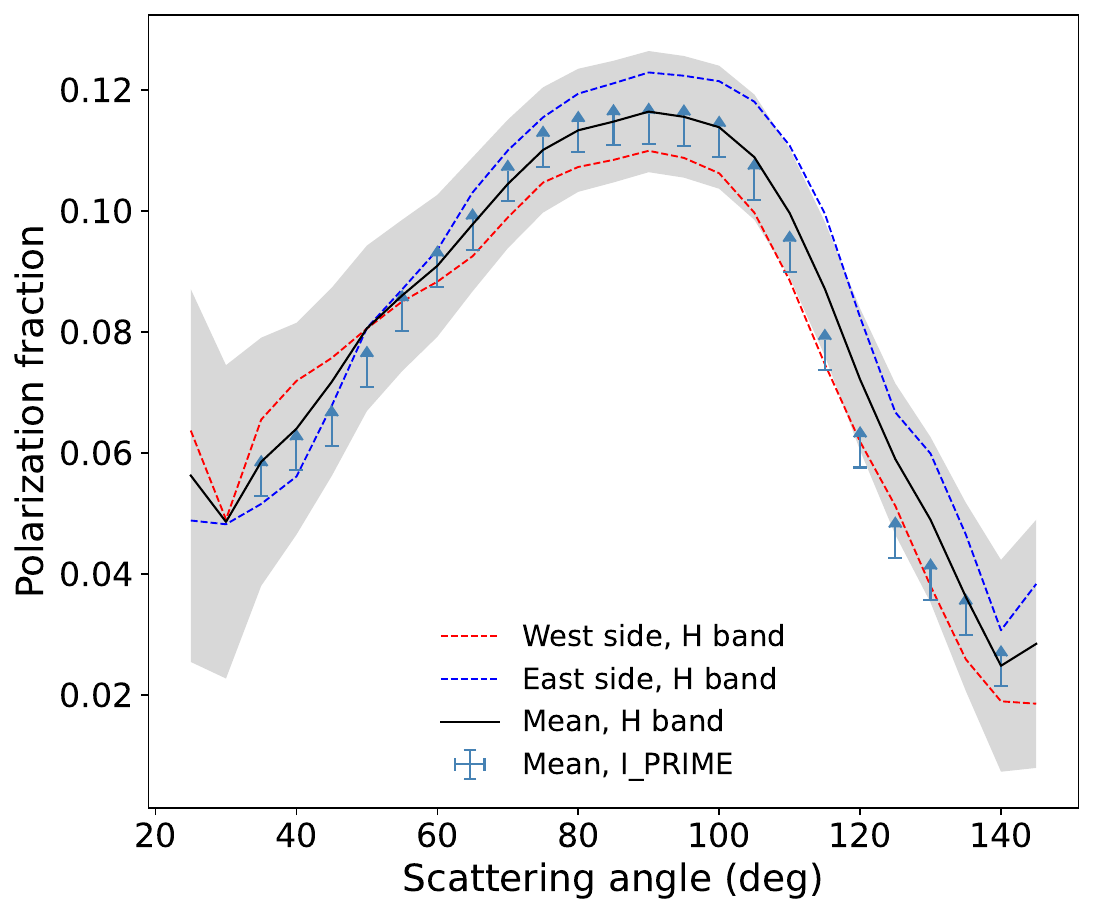}
 \caption{Polarization fraction of the debris disk \object{HD\,114082} derived as a mean value of the measurements on the west and east disk sides in the H band (black line). The gray-shaded area indicates the uncertainty on this mean polarization fraction from the aperture photometry. The blue arrows show a lower limit for the PF in the I\_PRIME filter. \label{f_pol_frac_zimpol}}
\end{figure}

The PF $p(\theta)$ in Fig.~\ref{f_pol_frac_zimpol} is derived by dividing polarized intensity by the total intensity, corrected for the disk flux losses caused by the data post-processing (see Sect.~\ref{s_SPF}), which are both normalized to the median value of the stellar flux measured in the IRDIS broadband H ($F_\star(IRDIS)$) and in the IFS H band image ($F_\star(IFS$)):
\begin{equation} \label{e_pol_frac}
p(\theta)= \frac{F_{\rm aper, Q_\phi}(\theta)}{F_{\rm aper, IFS}(\theta)} \cdot \frac{F_\star(IFS)}{F_\star(IRDIS)}
\end{equation}

The purpose of this flux normalization is to take into account the disk flux variations due to different observing techniques with two different instruments as well as observing conditions and slightly different wavelength ranges which we compare with each other. All these factors contribute to the error of the PF measurement but their effects are essentially cancelled out when applying the flux normalization. %by the stellar flux which  but not completely eliminated. 

Figure \ref{f_pol_frac_zimpol} shows the PF derived from the aperture photometry for both disk sides and their mean value (black line) that represents the measured PF  before the corrections for the dilution of the polarized light. Within the range of investigated scattering angles ($25^\circ < \theta < 155^\circ$), the PF has a bell shape with a maximum value of $p_{\rm max}\approx 11.5$\% between $\theta = 70^\circ$ and $\theta = 115^\circ$ and a minimum value of $p_{\rm min} < 2.5$\% beyond $\theta = 140^\circ$. 

The PF derived from archival H broadband image of total intensity reaches similar values at disk ansae ($\theta = 90^\circ$) although with a larger uncertainty. For the west ansa of the disk we obtain $p_{\rm max} = (11.2 \pm 2.5)\%$ and for the east ansa $p_{\rm max} = (10.8 \pm 2.5)\%$ which are in good agreement with the PF estimated using the IFS data.

The grey-shaded area in Fig.~\ref{f_pol_frac_zimpol} shows the measurement uncertainty that takes into account only statistical noise of the flux measurement when applying aperture photometry to the $Q_\varphi$ and IFS H band images. This uncertainty does not include the methodological errors mentioned above which are difficult to quantify. In particular, the observing conditions, e.g. seeing, coherence time or atmospheric transparency, can strongly decrease disk flux and have a significant impact on observational results. For instance, if the coherence times are short ($\tau < 2-3$ ms), the position of the star behind the coronagraph is unstable. In this case, a very accurate registering of frames is not possible, since the position of the star is not precisely known and deviates from the stellar position in the ``center'' frames. For a disk with a small angular size, such as the \object{HD\,114082} disk, the location of the maximum value of the PF (scattering angle at which the PF maximum occurs) depends on position of the star in the image. Thus, the frame centering should be performed as accurately as possible.   

The uncertainty indicated in Fig.~\ref{f_pol_frac_zimpol} neither includes the uncertainty resulting from the model-dependent correction of the disk flux losses caused by the ADI data post-processing. 

By this point, we have considered the disk flux losses only in the IFS total intensity data. However, the disk flux is reduced not only in the total intensity image but also in the polarized intensity image. The polarized flux we measure in the $Q_\varphi$ image, and hence the estimated PF shown by black line in Fig.~\ref{f_pol_frac_zimpol}, is lower compared to the actual disk polarized flux. The reason for this flux reduction is the extended instrument PSF which leads to the polarization cancellation effect in the Stokes $Q$ and $U$ signals \citep{Schmid2006}. %The magnitudes of this effect can be estimated only with model calculations \citep[e.g.,][]{Engler2018}. 

There is another important effect which is responsible for the underestimation of the polarized flux amount in the range of observed scattering angles. This phenomenon is not related to the instrument but to the disk morphology itself. When performing the aperture photometry,
we often measure a smaller PF than the intrinsic PF of scattered light in the disk midplane as a result of polarized flux dilution along the LOS. We call such an attenuation of polarized flux the LOS effect in this work and discuss it in more details in Sect.~\ref{s_LOS_effect}.
%we measure the polarized flux within a circular aperture from locations in the disk with different scattering angles because we integrate the flux along the LOS. At most of these locations less polarized flux is produced by scattering compared to the location in the disk midplane. Therefore the measured PF attributed to the latter is lower compared to the intrinsic PF. 
%the intrinsic PF of disk scattered light is certainly higher because the disk morphology causes a phenomenon of polarized flux dilution along the LOS  We call this attenuation of polarized flux the LOS effect in this work and discuss it in Sect.~\ref{s_LOS_effect}. 

In order to derive the ``true'' PF of scattered light, the measurement needs to be corrected for both effects by estimating their magnitudes with forward-modeling \citep[e.g.,][]{Engler2018}. For this purpose, we created several images of disk polarized light using the best-fit parameters of both models and iterating the shape of PF phase function each time.

\begin{figure}
\centering
\includegraphics[width=8cm]{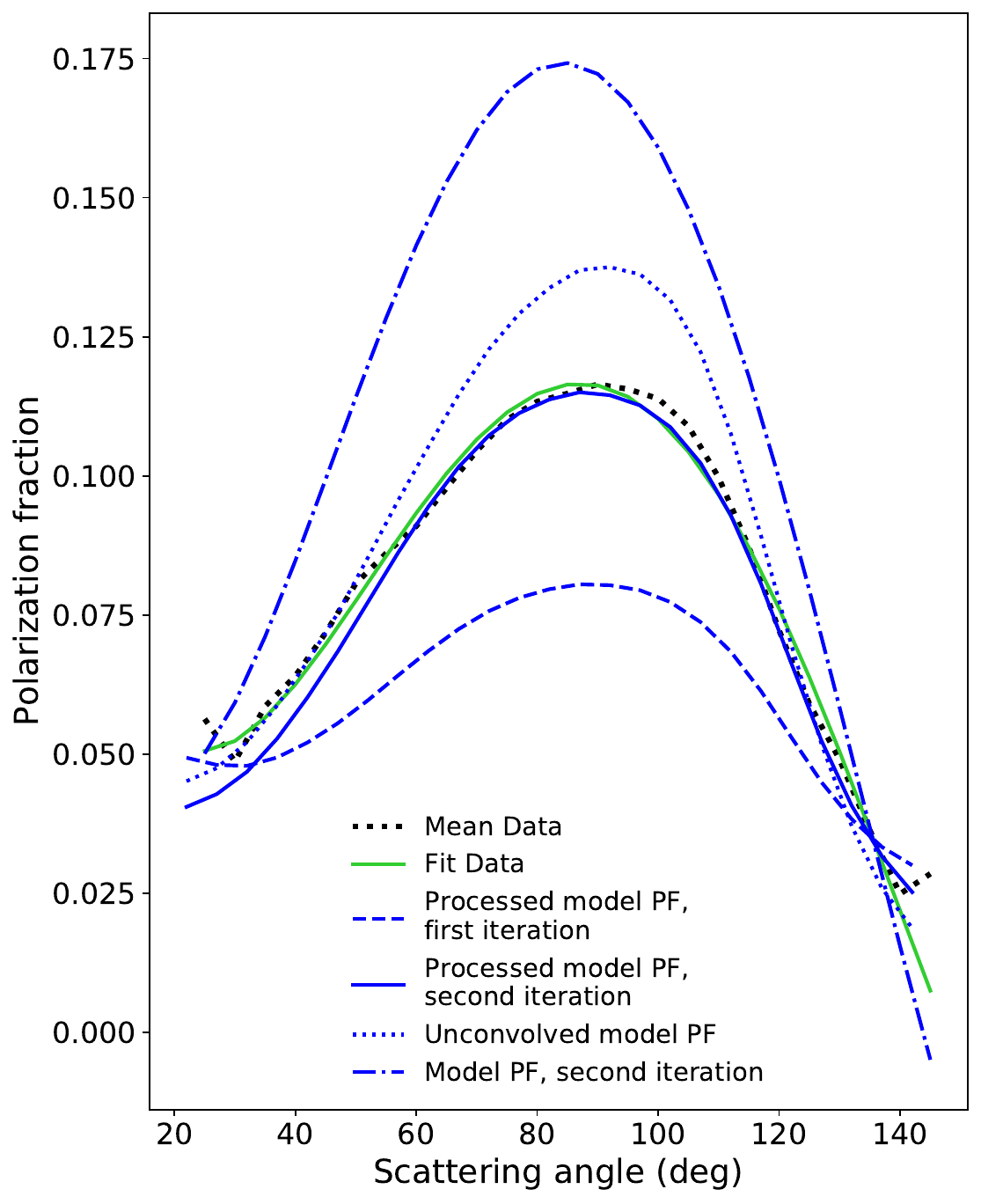} 
   \caption{Polarization fraction of scattered light in debris disk HD\,114082. The black dotted line shows a mean value of PF derived for the west and east disk sides. The green line shows a polynomial fit to it.  \label{f_pol_frac}}
\end{figure}

For the first iteration, the measurement from the data was used as a first guess for the PF. %The measured mean PF of the disk (black dotted line in Fig.~\ref{f_pol_frac}) was fitted with a }.   
To model the image we adopted a pSPF given by a product of the model SPF and a polynomial fit (green solid line in Fig.~\ref{f_pol_frac}) to the measured mean PF of the disk (black dotted line in Fig.~\ref{f_pol_frac}). %The polynomial fit represents the initial model PF phase function.

%\textcolor{red}{After convolution of the Stokes $Q$ and $U$ model images with the instrument PSF and re-calculation of the final $Q_\varphi$ image, the latter contain less polarized flux compared to the $Q_\varphi$ image modeled initially.} 
The forward-model $Q_\varphi$ image contains less polarized flux compared to the initial guess because the LOS and PSF convolution effects reduce the polarized flux as discussed above.
Therefore the PF that we measure from the forward-model $Q_\varphi$ image is lower than that given by the initial PF. As an example, the blue dashed line in Fig.~\ref{f_pol_frac} shows the PF measured from the forward-model images of total and polarized intensities of model 2\footnote{The PF measured from the forward-model images of model 1 is shown in Fig.~\ref{f_pol_frac_2models}.}. As expected, it is lower than the polynomial fit used as the initial guess for the PF (green solid line). It is also flatter.

Therefore, for the second iteration, we compute a new PF for our model multiplying the measured PF (black dotted line) by a ratio between the first polynomial fit (green solid line) and the PF measured from the forward model after first iteration (blue dashed line). The blue dashdotted line in Fig.~\ref{f_pol_frac} shows a polynomial fit to the data corrected by this ratio. This fit is used as the PF in the second iteration. 

The PF measured from the forward-model images of second iteration (blue solid line in Fig.~\ref{f_pol_frac}) matches the data measurement very well. At smaller ($\theta = 30^\circ$) and larger ($\theta = 130^\circ$) scattering angles there is no large discrepancy in values between the PF used to create model images (blue dashdotted line) and the PF measured from the processed model. However, the actual PF maximum value $p_{\rm max}$ used to model the data is much higher than 11.5\% measured from the data and reaches 17.4\% in model 2 (16.6\% in model 1). We conclude that the actual maximum PF should be about 17\% for the dust particles in \object{HD\,114082} disk observed at near-IR wavelengths, and estimate the modeling uncertainty on this value to be 3-4\% including modeling uncertainty from the ADI post-processing effect.  

\subsubsection{Model validation} \label{s_Modeling_polarised}

\begin{figure}
\centering
\includegraphics[width=8.5cm]{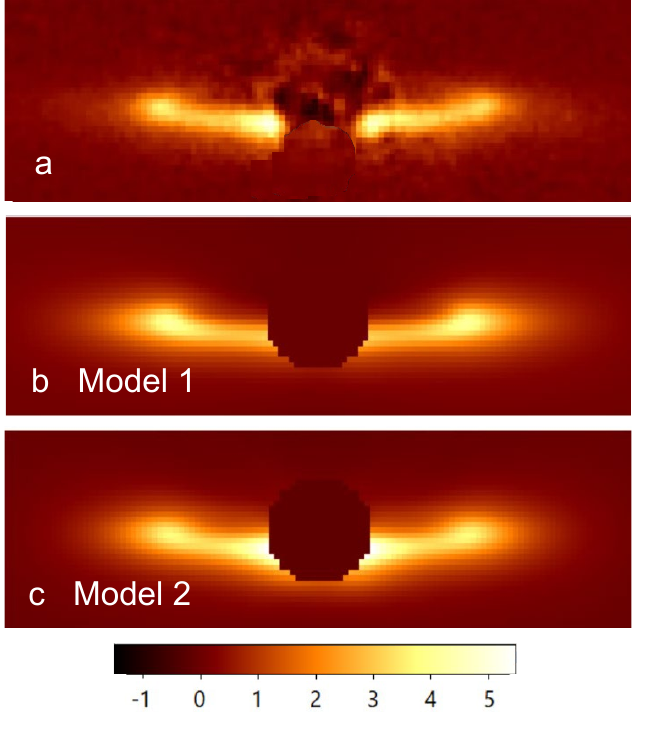} 
   \caption{Comparison of the $Q_\varphi$ image ({\it panel a}) to the model images of polarized light obtained with parameters of model 1 ({\it panel b}) and model\,2 ({\it panel c}). The color bar shows flux in cts/s.  \label{f_pol_model}}
\end{figure}

For a visual examination of the goodness of fit, we compare the images of models 1 and 2 showing the disk in scattered polarized light with the $Q_\varphi$ image in the H band in Fig.~\ref{f_pol_model}. The model images shown in Fig.~\ref{f_pol_model}b and Fig.~\ref{f_pol_model}c are created using the pSPF given by the product of the SPFs of both models and the intrinsic PF ((blue dashdotted line in Fig.~\ref{f_pol_frac})) found with the forward-modeling approach in Sect.~\ref{s_pol_frac_Hband}.

Both models provide similar images which fit the $Q_\varphi$ image well, except for the regions very close to the coronagraph edge and where the coronagraph throughput is higher than 50\%. At these radial separations ($0.10'' < r < 0.15''$) model 2 seems to match the SB distribution of the debris ring better because the SPF of model~2 has a much higher forward-scattering peak ($g_1=0.83$) compared to the model 1 ($g=0.56$). The higher scattered flux means that the polarized scattered flux is also higher in model 2 at smaller scattering angles because the same fraction of light polarization was adopted in both models. The stronger forward-scattering becomes apparent through the higher polarized SB close to the coronagraph edge in the image of model\,2 (Fig.~\ref{f_pol_model}c). Similar SB peaks are observable in the $Q_\varphi$ image (Fig.~\ref{f_pol_model}a) which would support the preference for model 2 over model 1. However, we should note that this region in the $Q_\varphi$ image is relatively noisy because of the coronagraph and telescope spider effects. 

\subsubsection{Line-of-sight effect} \label{s_LOS_effect}

Our modeling results confirm that not only does the effect of the convolution with the instrument PSF significantly influence the outcome of the PF measurement, but the disk morphology also has a large impact. 

As mentioned in Sect.~\ref{s_SPF}, we attributed the disk flux measured within the circular apertures to a single scattering angle corresponding to the dust location in the midplane of a narrow planetesimal belt. By doing so, we neglected the effect of disk thickness on the amount of polarized flux measured along the LOS direction. However, the non-zero disk extension in both radial and vertical directions means that scattering from many locations with different scattering angles contributes to the disk flux which we measure within the circular apertures in the images of both total and polarized intensities. Through the integration along the LOS, in particular the polarized light can become diluted because of contributions from the scattering angles with lower degree of polarization compared to the disk midplane. This leads to the underestimation of the disk PF, for instance the maximum PF value at $\theta = 90^\circ$. 

We can probe this LOS effect %, hereafter LOS effect, 
by measuring the PF from the unconvolved model images (total and polarized intensities) using the same aperture photometry method. This measurement is indicated by a blue dotted line in Fig.~\ref{f_pol_frac} (for model 2). It shows the impact of disk morphology on the measured PF.
Even without taking into account the effect of the instrument PSF, we derive a lower degree of disk polarization and a lower PF maximum value of 13.9\% compared to that used to create the model images.

Not only is the magnitude of the measured maximum PF $p_{\rm max}$ influenced by the LOS effect but its position is also impacted. As seen in Fig.~\ref{f_pol_frac}, the intrinsic PF of the model has a maximum at $\theta = 85^\circ$ (blue dashdotted line) whereas the measured PF maximum is shifted to $\theta = 95^\circ$ for both convolved and non-convolved models (blue solid and dotted line, respectively). 

In Appendix~\ref{s_modeling_app}, we discuss how the inclination and scale height of the disk affect the PF measurement. Depending on these two parameters the measured maximum PF $p_{\rm max}$ can be reduced by 30\%  compared to the intrinsic maximum PF of the disk. Since the inclination of the HD\,114082 disk is constrained, we estimated the impact of the scale height on the measured $p_{\rm max}$. We used model 2 to create images of the disk with an inclination of 83$^\circ$ and different scale heights: $H_0 = 0.004''$ (best-fit parameter of model 1), 0.007$''$ (best-fit parameter of model 2), 0.01$''$ and 0.1$''$. Our results show that the scale height has a rather small effect in this particular case. The measured maximum PF $p_{\rm max}$ is 13.9\% for models with the scale heights of $0.007'', 0.01''$ and 0.1$''$. It is 14.2\% in case the scale height equals to $0.004''$. 

\subsubsection{Polarization fraction along the disk major axis}
\begin{figure}
\centering
\includegraphics[width=8.4cm]{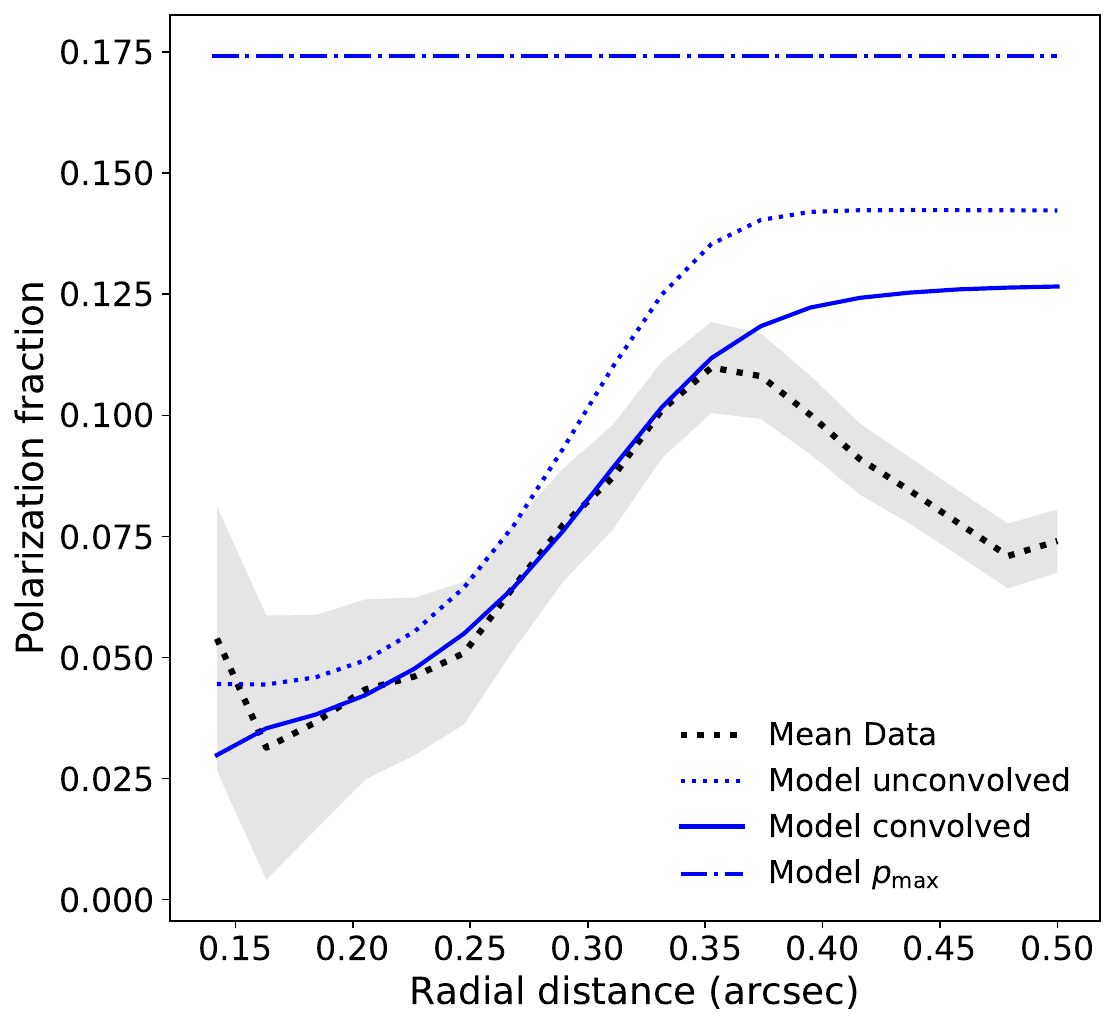} 
   \caption{Degree of linear polarization of scattered light for the debris disk \object{HD\,114082} calculated along the disk major axis. The black line shows a mean value of the west and east disk sides. The radial dependence of PF is significantly moderated by the difference in scattering angles such as in Fig.~\ref{f_pol_frac} due to the LOS and aperture size effects.  \label{f_PF_maxis}}
\end{figure}

Using the same aperture photometry, we measured the PF along the disk major axis. The derived PF mean value between the east and west disk sides is shown by the black dotted line in Fig.~\ref{f_PF_maxis}. It continuously increases when approaching the radial position of the planetesimal belt, where the peak SB is located ($r \approx 0.36''$). The highest mean PF value of $\sim$11.2\% is measured between 0.35$''$ and 0.37$''$, thus at the location of the belt. 

In our ideal model, shown by the dashdotted line in Fig.~\ref{f_PF_maxis}, the PF is constant and equal to $p_{\rm max}$ independently of the separation, because all dust particles located in the disk along its major axis scatter stellar radiation at a scattering angle of $90^\circ$ towards the observer. In our best-fit model this scattering angle corresponds to a maximum PF of 17.4\% (see the blue dashdotted line in Fig.~\ref{f_pol_frac}). This value does not depend on the distance from the star because we assume an uniform population of dust particles throughout the disk in the model. This assumption is certainly valid for the range of particle sizes which we probe in the observations at near-IR wavelengths. %This ideal case is shown by the blue dashdotted line in Fig.~\ref{f_PF_maxis}.   

However, when applying this measurement to the image of the non-convolved best-fit model, we obtain the maximum PF value $p_{\rm max}$ of 14.2\% for $r > 0.37''$ (blue dotted line in Fig.~\ref{f_PF_maxis}). As expected, the PF measured from the model image is lower than the actual model value of 17.4\% due to the LOS effect discussed in Sect.~\ref{s_LOS_effect}. The PF also varies with the radial distance ($0.14'' < r < 0.37''$) because the aperture encloses dust grains emitting at different scattering angles, in particular decreasing the PF at smaller separations.

The effect of convolution with the instrument PSF additionally decreases the observed PF. To demonstrate this, we perform the measurement with the convolved best-fit model and show the result with the blue solid line in Fig.~\ref{f_PF_maxis}. The PF measured from the convolved model image matches the data very well reaching $\sim$11.5\% at a radial separation of 0.35$''$. Beyond $r = 0.35''$, it slowly increases up to 12.6\% and remains at this level following the PF trend derived from the non-convolved model image. This is in contrast to decreasing PF measured from the data. The apparent discrepancy between the model and the data could be explained by the low polarized flux measured in the $Q_\varphi$ image at larger distances. 

In general, we expect that the PF of the scattered light measured along the disk major axis is slowly growing beyond the planetesimal belt because the effects of LOS and PSF convolution are getting smaller with the distance to the star. 

In our $Q_\varphi$ image, the polarized flux at larger separations might be diminished through the applied correction for the instrumental polarization. Though, the other reasons, for instance different optical properties of dust particles at larger distances because of different shape or composition, cannot be completely ruled out.

\subsubsection{Constraints on the polarization fraction from ZIMPOL data}

As mentioned above, only the polarized intensity image of the disk is available in the ZIMPOL I\_PRIME filter. The ZIMPOL data were taken in the field-stabilized mode that does not allow to detect the disk in total intensity because the region of interest is contaminated with the stellar PSF. For this reason, it is not possible to derive the PF of scattered light  using the ZIMPOL dataset alone. 

However, we can put constraints on PF in the I\_PRIME filter comparing the magnitude of the polarized flux measured in the ZIMPOL $Q_\varphi$ image (Fig.~\ref{f_QphiUphi_IRDIS}, bottom left panel) with the total intensity of the disk measured at shortest IFS wavelengths. 
For this purpose, we applied the aperture photometry to the $Q_\varphi$ image in the I\_PRIME filter ($\lambda=0.790\,\upmu$m, $\Delta\lambda = 0.153\,\upmu$m) and the total intensity IFS image combining wavelength channels from 0.95 to 0.99 $\upmu$m as described in Sect.~\ref{s_pol_frac_Hband}.
The measured fluxes were also normalized according to Eq.~\ref{e_pol_frac}.

The PF estimated in this way is indicated by the blue arrows in Fig.~\ref{f_pol_frac_zimpol}. It is slightly lower than the PF obtained for the H band and represents a lower limit on the PF in the I\_PRIME filter because the corrections for the attenuation of polarized flux (LOS effect and convolution with the instrument PSF effect) are not applied to this measurement.
%Debris disks often exhibit a red color at the position of planetesimal belt \citep{Debes2008, Engler2019}. This seems to be valid for the disks HD\,114082 and HD 117214 (see Sect.~\ref{s_SED}). The red color of a disk means, 
 %In addition, the HD\,114082 disk reflectance spectrum seems to show a red color (Fig.~\ref{f_SED}). This means }that the ratio between scattered flux and incoming stellar flux could be higher at longer wavelengths. Therefore in the PF estimation, we possibly used a higher \textcolor{red}{value for this relation than the actual ratio valid for the I\_PRIME filter.}
 
\subsubsection{Comparison with other debris disks and dust populations}
Based on modeling results (Sect.~\ref{s_pol_frac_Hband}), we conclude that the PF maximum value in the H band is $17 \pm 4$\% at the radial position of the \object{HD\,114082} planetesimal belt. 

Similar PFs were measured for a number of other exo-Kuiper belts, although the results of previous studies are most probably underestimated because they do not take into account the convolution of polarized signals with the instrument PSF and LOS effects decreasing the measured PF. Using the coronagraphic imager with adaptive optics HiCIAO at Subaru Telescope, \cite{Tamura2006} detected $\sim$10\% polarization in the K band (2.2 $\mu$m) in the $\beta$ Pictoris debris disk at radial separations $50 < r < 120$~au: at distances, where the edges of one or two planetesimal belts are expected to be \citep{Heap2000, Golimowski2006}. The optical (BVRI) imaging polarimetry of the $\beta$ Pictoris disk showed a higher level of polarization of $15-20$\% but at larger radial distances ($r > 150$ au) \citep{Gledhill1991, Wolstencroft1995}. \cite{Asensio-Torres2016} also used the HiCIAO instrument to measure $\sim$15\% of polarization for the outer belt in the HD 32297 disk.
\cite{Bhowmik2019} obtained a polarized flux fraction of $\sim$15\% in the J band (1.2 $\mu$m, SPHERE/IRDIS) for the HD 32297 disk as well.  A recent study of another double-belt system HD 15115 reported a maximum polarization fraction of $10-15$\% at radial location of the outer belt measured in the J band with SPHERE/IRDIS  \citep{Engler2019}. 

The shape of the measured \object{HD\,114082} PF is also very similar to that derived for the dust in debris belt HR 4796 A \citep{Milli2019}. It is interesting to note, that the \object{HD\,114082} PF seems to approach a zero value at scattering angle close to 160$^\circ$ like the HR 4796 A phase function does. 
Many studied samples of cometary dust in the solar system \citep{Hadamcik2016, Shestopalov2017} and their analogs \citep{Frattin2019} behave in the same way: they exhibit a negative polarization branch for scattering angles larger than 160$^\circ$. 

\section{Disk near-IR spectrum} \label{s_SED}
The IFS and IRDIS data allow us to investigate the scattering efficiency of dust particles in the \object{HD\,114082} disk at near-IR wavelengths. We perform this analysis comparing the dust reflectivity in two debris disks: \object{HD\,114082} and HD\,117214. This comparison was motivated by the idea to show that the \object{HD\,114082} disk is much brighter than a debris disk with a similar infrared excess, temperature of dust grains and host star characteristics. 

The HD\,117214 is a young F6V star \citep{Houk1975} located at the distance of $107.4 \pm 0.3$ parsecs \citep{GaiaCollaboration2021} in the LCC subgroup as well. It possesses a bright debris disk \citep{Engler2020} with a radius of $\sim$0.45$''$ (48 au) and fractional luminosity of $2.53\times 10^{-3}$ \citep{Jang-Condell2015} which are similar to the parameters of the \object{HD\,114082} disk. The HD\,117214 is the only F star thus far which was observed under very good observing conditions with SPHERE using the same instrument configuration \citep{Engler2020} making such a comparison possible. Since both targets are young F stars in the LCC subgroup, their circumstellar dust might potentially exhibit similar scattering properties. 

\subsection{Spectral analysis} \label{s_SED_1}
\begin{figure}
\centering
\includegraphics[width=8.5cm]{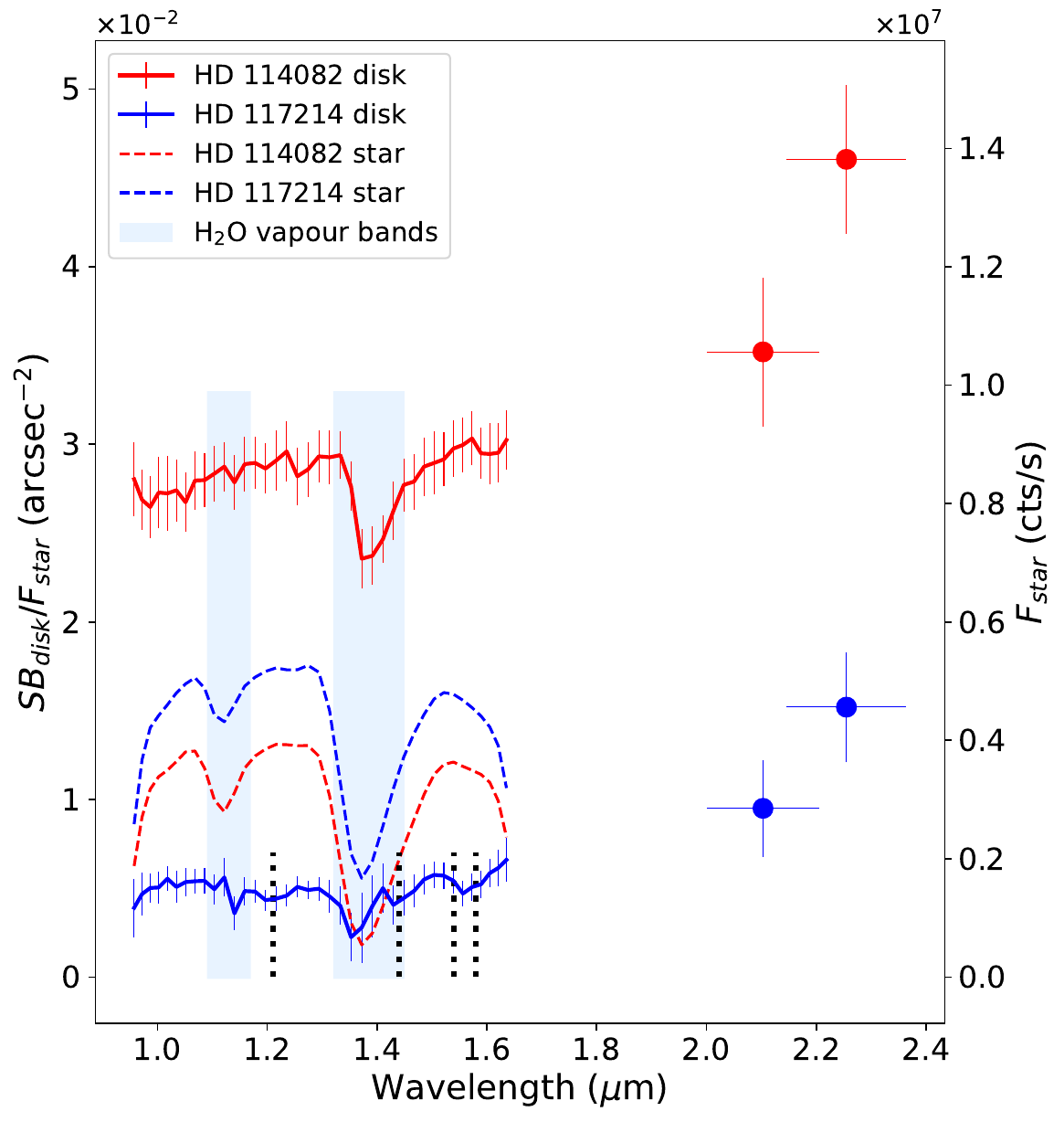} 
 \caption{Near-IR reflectance spectra of debris disks \object{HD\,114082} and HD\,117214 measured with the IFS (solid lines) and IRDIS K1 and K2 bands (dots). The HD 117214 spectrum has been additionally scaled by a factor of 1.55 for comparison purposes.
 The spectra are normalized by the stellar fluxes shown by the dashed lines for the IFS data. Stellar flux values are given on the right y-axis of the plot. Black dotted lines show the location of the CO$_2$ ice absorption bands at 1.21, 1.44 and $1.54 - 1.58$ $\upmu$m. The blue-shaded regions schematically indicate the absorption bands of the atmospheric water vapor. \label{f_SED}}
\end{figure}

In order to derive the spectra of \object{HD\,114082} and HD\,117214 debris disks, we applied the aperture photometry to the IFS and IRDIS images as described in Sect.~\ref{s_SPF}. The disk contrast presented in Fig.~\ref{f_SED} was measured along the disk major axes at positions of the peak SB in the $r^2$-scaled disk images which indicate the location of the planetesimal belt. These positions were selected because the disk flux is less affected (reduced) by the ADI data processing at larger radial distances. We placed the apertures at radial separations of 0.37$''$ (35 au) from the star in the image of \object{HD\,114082} disk, and at radial separations of 0.46$''$ (49 au) in the HD 117214 disk image. 

The displayed spectra (Fig.~\ref{f_SED}) are corrected for the disk flux losses (see Sect.~\ref{s_SPF}) and normalized by the value of stellar flux obtained from the flux measurement frames in the corresponding IFS wavelength channel or IRDIS band. The spectra show thus the disk SB in terms of contrast to the host star. This contrast was measured in the circular apertures with an area of $4.25 \times 10^{-3}$ arcsec$^2$. The normalization by the stellar flux allows us to remove disk flux fluctuations which reflect the variation of the stellar spectrum, among other reasons due to spectrally dependent transmission of both instruments and the Earth's atmosphere. As practice shows, especially the correction for the telluric water absorption bands close to 1.4 $\upmu$m is challenging, because the absorption of water molecules is very strong in this wavelength range. The error bars in Fig.~\ref{f_SED} indicate the uncertainty of the aperture photometry and do not include uncertainties from the modeling of flux losses or any calibration issues.

Except for the specific regions of telluric water vapor absorption (see Fig.~\ref{f_SED}), the derived reflectance spectra should represent the underlying disk spectra well, in particular because the spectrum of HD 117214 disk shows some dips at locations where the stellar spectrum is featureless, for instance at 1.2 $\upmu$m and in the region between 1.5 and 1.6 $\upmu$m. In these spectral ranges, the CO$_2$ ice exhibits relatively strong absorption bands: at 1.21, 1.44 and $1.54 - 1.58$ $\upmu$m \citep{Hansen2005}. Therefore the measured spectrum of HD 117214 disk maybe indicates the presence of CO$_2$ ice as a constituent of debris material at the investigated radial distance from the star. 

In fact, it would not be surprising to identify this ice in a young debris system. The CO$_2$ is one of the most abundant ices after H$_2$O ice. It was found in many lines of sight toward molecular clouds and young stellar objects \citep{Oeberg2007}. The CO$_2$ gas is also present in the Earth atmosphere but in a very small amount. The absorption features, which we possibly observe in the HD 117214 spectrum, should have another origin. Otherwise we would clearly see them in the stellar spectrum as well. 

The CO$_2$ gas freezes out over a narrow range of dust temperatures between 40-80 K \citep{Harsono2015}. Under the assumption of local temperature equilibrium, the dust temperature at location of planetesimal belt can be estimated using the canonical expression by \cite{Backman1993}:
\begin{equation} \label{eq_TdustBB}
T\mathrm{_{dust}}=(278\, \mathrm{K}) \left( \frac{L_{\ast}}{L_{\odot}}\right)^{0.25} \left( \frac{\rm{1}}{R_{\rm belt}}\right)^{0.5},
\end{equation}
where the $L_{\ast}$ is the stellar luminosity, and $R_{\rm belt}$ is the radius of the planetesimal belt in au. The luminosity of F6V star HD 117214 is found to be 5.64 $L_\odot$ \citep{Jang-Condell2015} whereas the luminosity of F3V star \object{HD\,114082} is 3.86 $L_\odot$ \citep{Pecaut2012}. The estimated dust temperature in the HD 117214 disk at the distance of 49~au from the star (disk ansae) is 61 K. For the \object{HD\,114082} disk we obtain a dust temperature of 66 K at the distance of planetesimal belt at 35 au. Thus both debris belts could be located at the CO$_2$ snowline.

The spectrum of \object{HD\,114082} disk shows no obvious absorption features except at 1.4 $\upmu$m, which most likely comes from the insufficient telluric correction of spectrum. Nevertheless, the presence of CO$_2$ cannot be ruled out for \object{HD\,114082}. 

In the wavelength range between 0.96 and 1.66 $\upmu$m both disk reflectance spectra are rather flat or have a grey color at measured radial position in the disk. However, they seem to show a red color at longer wavelengths than 1.66 $\upmu$m because the contrasts in K1 and K2 bands are higher than the contrasts measured in the IFS channels with the K2 band contrast being the highest. In this respect, the \object{HD\,114082} and 117214 disks are similar to the debris disks HR 4796A \citep{Milli2017} and HD 115600 \citep{Gibbs2019} which also show a red color in their reflectance spectrum. 

As seen in Fig.~\ref{f_SED}, the measured scattered flux is significantly higher in the \object{HD\,114082} disk compared to the HD 117214 disk. When comparing these two disks, we have to keep in mind that planetesimal belts are located at different radial distances from their host stars and have different inclinations. For a comparison of measured fluxes, these geometric parameters have to be taken into account. 
Therefore, the HD 117214 disk fluxes plotted in Fig.~\ref{f_SED} are multiplied by a factor of 1.55. This factor is given by the ratio of squared radial distances of the belts used in our measurement: $R^2_{\rm HD 117214} / R^2_{\rm HD 11482} = (0.46 / 0.37)^2 = 1.55$. The correction factor compensates for the different illumination of disks because of dilution of stellar light with a distance which scales like $1/r^2$.

Likewise, the amount of light scattered by dust particles into the LOS of an observer strongly depends on the disk morphology, in particular the disk inclination. With a larger inclination, more flux should be measured along the disk rim because more dust particles will be in the LOS of the observer due to the projection effect. Using the best-fit model for the HD 117214 disk \citep{Engler2020}, we estimated this effect by varying the disk inclination between 73$^\circ$ (inclination of HD 117214 disk) and 83$^\circ$ (inclination of HD\,114082 disk). %keeping all other parameters constant. 
In the disk model with a higher inclination, the scattered flux measured within the same circular apertures placed at the disk ansae is 1.75 times higher than in the disk with a lower inclination. Multiplying the HD 117214 data (Fig.~\ref{f_SED}) by this correction factor for the disk inclination, in addition to the previous correction factor of 1.55 for the flux dilution, we still obtain a $\sim$2.5 lower flux for HD 117214 compared to \object{HD\,114082}.  

\subsection{Constraints on the disk albedo}
We define a spectral albedo of a debris disk $\omega_{\rm \lambda \, disk}$ as a relation between the amount of attenuation of the stellar flux $F_{\rm \ast \, \lambda}$ by its scattering off dust particles in the whole debris disk (including reflection, refraction and diffraction) and the total amount of stellar flux attenuation by dust scattering and absorption. The disk spectral albedo can be expressed in terms of the averaged contrast of the disk representing the disk scattering cross section $\sigma_{\rm \lambda \, disk}$ at a wavelength $\lambda$ and the fractional IR luminosity of the disk $L_{\mathrm{IR \, disk}}/L_{\ast}$ which can be used as a proxy for the averaged absorption cross section of the disk $\kappa_{\rm disk}$:
\begin{equation} \label{e_albedo}
\omega_{\rm \lambda \,disk} = \frac{\sigma_{\rm sca \,\lambda \, disk}} {\sigma_{\rm sca \,\lambda \, disk} + \kappa_{\rm disk}} = \frac{F_{\rm sca \,\lambda}/F_{\rm \ast \, \lambda}}{F_{\rm sca \,\lambda}/F_{\rm \ast \, \lambda}  + L_\mathrm{IR \, disk}/L_{\ast} }
\end{equation}

The scattering albedo $\omega_{\rm disk}$ for the HD\,114082 disk at near-IR wavelengths can be evaluated based on the best-fitting model~1 and model~2. With their help we are able to estimate how much flux is scattered into the LOS of the observer and the entire space. 

For this purpose, we measured the scattered flux in the K1 band image of total intensity (Fig.~\ref{f_model}a) within two rectangular boxes with a height of 0.16$''$ centered on the disk major axis and covering the image area between radial distances $r= 0.12''$ and $r= 0.54''$ from the star. For the east side of the disk we obtained a total flux of $410\pm 20 $ cts/s, for the west side $340\pm 18 $ cts/s. This flux is significantly reduced compared to the actual disk flux scattered into the LOS because of the ADI-processing of the data. Using the models~1 and 2 we derive the total intrinsic flux for the left and right boxes together of $4400\pm 500 $~cts/s. 

The stellar flux in the K1 band sums up to $(1.13 \pm 0.00)\times 10^6$~cts/s. This gives a contrast for the measured total intrinsic flux of $(3.9 \pm 0.5)\times 10^{-3}.$ With this contrast we can estimate the effective spectral disk albedo, that means the albedo derived with the contrast of the probed regions only. Since one part of the disk is behind the coronagraph, the fractional IR luminosity in Eq.~\ref{e_albedo} needs to be reduced by a factor of 0.93 taking into account only visible part of the disk. With this roughly estimated factor we obtain the effective spectral albedo for the HD 114082 disk of 0.58:
\begin{equation} \label{e_eff_albedo1}
\begin{split}
&\omega_{\rm eff\, \lambda \,HD\,114082} = \frac{(3.90 \pm 0.50)\times 10^{-3}}{(3.90 \pm 0.50)\times 10^{-3} +  0.93 \cdot 3.01\times 10^{-3}} =\\
&= 0.58 \pm 0.08.    
\end{split}
\end{equation}

We note that this value does not include scattered flux from the region behind the coronagraphic mask where the peak of forward-scattering occurs for the highly inclined disks such as \object{HD\,114082} disk. Moreover, the effective albedo derived with the flux scattered into LOS depends on disk inclination. This means that if the \object{HD\,114082} disk would have a different inclination, then the effective albedo would be also different. Therefore in the following paragraphs, we estimate the average spectral albedo for \object{HD\,114082} disk based on the modeled disk luminosity. 

The total scattered flux in the images of non-convolved and unprocessed models for the K1 band data is $(2.0 \pm 0.4)\times 10^4$ cts/s. The scattered flux averaged over full sphere solid angle, which we define as scattering luminosity of the disk divided through 4$\pi$, is $F_{\rm sca \,\lambda} = (6.32 \pm 1.25)\times 10^3$ cts/s. The uncertainty on these values mostly results from the unknown amount of scattered flux from the disk part behind the coronagraph. Models with a large HG asymmetry parameter $g$ describing the diffraction peak of forward-scattering, like for instance model 2 ($g_1=0.73$), predict much higher unseen flux compared to the models with a smaller parameter $g$, like model 1 ($g=0.55$). 

 This yields an averaged contrast of the disk $F_{\rm sca \,\lambda}/F_{\rm \ast \, \lambda} = (5.61 \pm 1.10)\times 10^{-3}$ in scattered light at $\lambda=2.110\,\upmu$m. 
Based on the estimated averaged flux value, we derive a relatively high mean albedo of $0.65 \pm 0.15$ for the HD\,114082 disk:
\begin{equation} \label{e_albedo1}
\omega_{\rm \lambda \,HD\,114082} = \frac{(5.61 \pm 1.10)\times 10^{-3}}{(5.61 \pm 1.10)\times 10^{-3} +  3.01\times 10^{-3}} = 0.65 \pm 0.15.
\end{equation}

This value is however close to the derived effective spectral albedo of 0.58.

Our analysis of the SPF curve in Sect.~\ref{s_SPF} constrains the effective particle size in the disk observed at near-IR wavelengths to be within a range of 5 to 15 $\upmu$m. This relatively large effective particle size and the low maximum PF of 17\% (Sect.~\ref{s_pol_fraction}) agree well with the predictions of light scattering theory for high-albedo particles. Aggregates of debris particles with relatively large sizes %(larger than $3-5$ $\mu$m)
tend to produce multiple scattering of stellar photons and, thus, generate a higher scattered flux similar to  high-albedo surfaces. Due to multiple scattering stellar light becomes less polarized compared to the single scattering event. This de-polarization effect, the so-called Umov effect \citep{Umov1905}, could explain the low PF of scattered light in the HD\,114082 disk. 

The scattering albedo of the HD\,117214 disk might be lower than the value we estimated for the \object{HD\,114082} disk. According to the results of the spectral analysis (Sect.~\ref{s_SED_1}), the micron-sized particles from  HD\,117214 scatter $\sim$2.5 times less stellar flux than the particles from the \object{HD\,114082} disk at scattering angle $\theta=90^\circ$.
This means that, if the scattering albedo of the dust particles in both disks were similar, the mass of  micron-sized particles we trace in both observations should be $\sim$2.5 times lower in HD 117214 compared to \object{HD\,114082}. 

Assuming the same particle size distribution in both systems, we can roughly estimate the actual relation between these masses by comparing the disk IR luminosities $L_{\rm IR \, disk}$. Since the particle temperatures we derived for both disks (see Eq.~\ref{eq_TdustBB}) are approximately the same, the IR luminosities of disks should scale with the dust masses of the micron-sized particles $M\mathrm{_{dust}}$ like
\begin{equation} \label{eq_Mdust1}
M\mathrm{_{dust}}\propto L_{\rm IR \, disk} = f_{\rm IR \, disk} \cdot L_{\ast},
\end{equation}
where $f_{\rm IR \, disk}$ is the fractional IR luminosity of the disk which is equal to $3.01\times 10^{-3}$ for the HD\,114082 disk and $2.53\times 10^{-3}$ for the HD 117214 disk \citep{Jang-Condell2015}. The ratio of the disk masses should scale like 
\begin{equation} \label{eq_Mdust2}
\frac{M\mathrm{_{disk}}\mathrm{(HD\,117214)}} {M\mathrm{_{disk}}\mathrm{(HD\,114082)}} \sim \frac{2.53 \cdot 10^{-3} \cdot 5.64 L_\odot }{3.01 \cdot 10^{-3} \cdot 3.86 L_\odot} = 1.23
\end{equation}

If our assumptions are valid, the HD 117214 disk should possess a higher mass than the \object{HD\,114082} disk. This conclusion is consistent with the estimated stellar mass is $1.14$ times higher for HD 117214 compared to \object{HD\,114082} \citep{Jang-Condell2015, Pecaut2012}. This indicates that the mass of the original molecular cloud of the former was higher, and thus the mass of debris disk surrounding it is most likely also higher. Higher dust mass and lower scattered flux mean that the albedo of micron-sized particles in HD 117214 disk maybe $2 - 2.5$ times lower at near-IR wavelengths compared to the \object{HD\,114082} disk.  

Different albedo of the disks can be explained by different optical properties of dust: slightly different effective sizes, shapes or composition of dust particles. 

If the particle size is the most important parameter affecting the amount of scattered flux, then, from comparison of the particle reflectivity between \object{HD\,114082} and HD 117214 disks, we shall conclude that the effective particle size in the HD 117214 disk is smaller than 5 $\upmu$m. 

If the particle composition is responsible for the lower reflectivity of dust in HD 117214 disk, we can think of particles composed of ``dirty'' CO$_2$ and H$_2$O ice particles with inclusions of dark material, e.g., complex organic material \citep{Levasseur2018}. The reflectance of the CO$_2$ ice particles could be lower than 50\% in the near-IR, as the spectroscopic studies of the Martian polar regions covered by a layer of the CO$_2$ ice report \citep{Appere2011}. 

Different composition of dust particles, in particular the ice volume fraction, in \object{HD\,114082} and HD 117214 disks could be a direct consequence of different evolutionary stages at which both stars might be. In Appendix~\ref{s_HD117214}, we discuss in more details the possibility for HD 117214 to be much younger than the median age of 17 Myr assumed for the F stars in the LCC. At younger ages, the circumstellar disks possess more gas and ices. Their amounts decrease over time due to gas release, when icy particles collide with other debris, evaporate or undergo photodesorption with subsequent gas photodissociation caused by the stellar and interstellar radiation \citep[e.g.,][]{Grigorieva2007} or due to accretion onto growing giant planets.

The upcoming studies with the James Webb Space Telescope (JWST) can shed light on the content of various ice species such as H$_2$O, CO, CO$_2$, CH$_3$OH, NH$_3$, CH$_4$, in the dust particles of debris disks surrounding stars at different ages. In particular, the JWST Near Infrared Camera (NIRCam) will provide sensitive imaging and spectroscopy at near-IR wavelengths from 0.6 to 5.0 $\upmu$m which includes H$_2$O and CO$_2$ frost features. The future JWST observations can reveal the locations of ices, and thus locations of various snowlines in the disks and possibly clarify how much primordial gases remain bound in the debris around young stars entering the main sequence.

\section{Planetary system of HD\,114082 } \label{s_planets}

The planetary system of \object{HD\,114082} consists of at least two spatially separated planetesimal belts: a bright exo-Kuiper belt with a radius $R_{\rm belt}$ of $0.37''$ (35 au) discussed in this work and a warm dust belt. The latter was identified from the disk SED which is better matched with a thermal emission from two separate dust populations. The radial location of the warm dust belt is however unclear: different studies  placed the second dust component at different radial positions between 3 and 18 au from the star \citep[e.g.,][]{Mittal2015, Jang-Condell2015}, and even a dust population with a temperature of 500 K at 0.8~au provided a good fit to the \textit{Spitzer} IRS spectrum \citep{Chen2014}. It might be therefore that \object{HD\,114082} disk harbours multiple  components with inner debris rings which cannot be resolved with current instruments because of small angular sizes (the radius of the coronagraphic mask is 0.12$"$ or 11~au in our observations) and high inclination of the bright outer belt.

No planetary companions have been found in the cavity of the outer belt with direct imaging so far. The analysis of recent TESS data shows that a Jupiter-sized planet resides within the cavity and is hidden by the coronagraph in our SPHERE observations.

\subsection{Discovery of transiting planet} \label{s_exoJupiter}
\begin{figure}
\centering 
\includegraphics[width=0.5\textwidth]{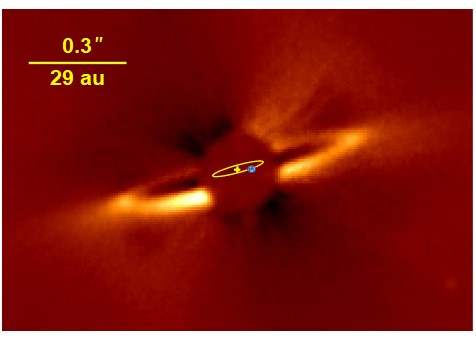}
\caption{Illustration of the transiting planet orbit. The orbit and planet sizes are shown schematically and not in the scale indicated in the upper left corner: the radius of the disk is 35~au whereas the radius of the planet orbit is 0.7~au.  } \label{f_planet}
\end{figure}

\begin{figure}
\centering 
\includegraphics[width=0.5\textwidth]{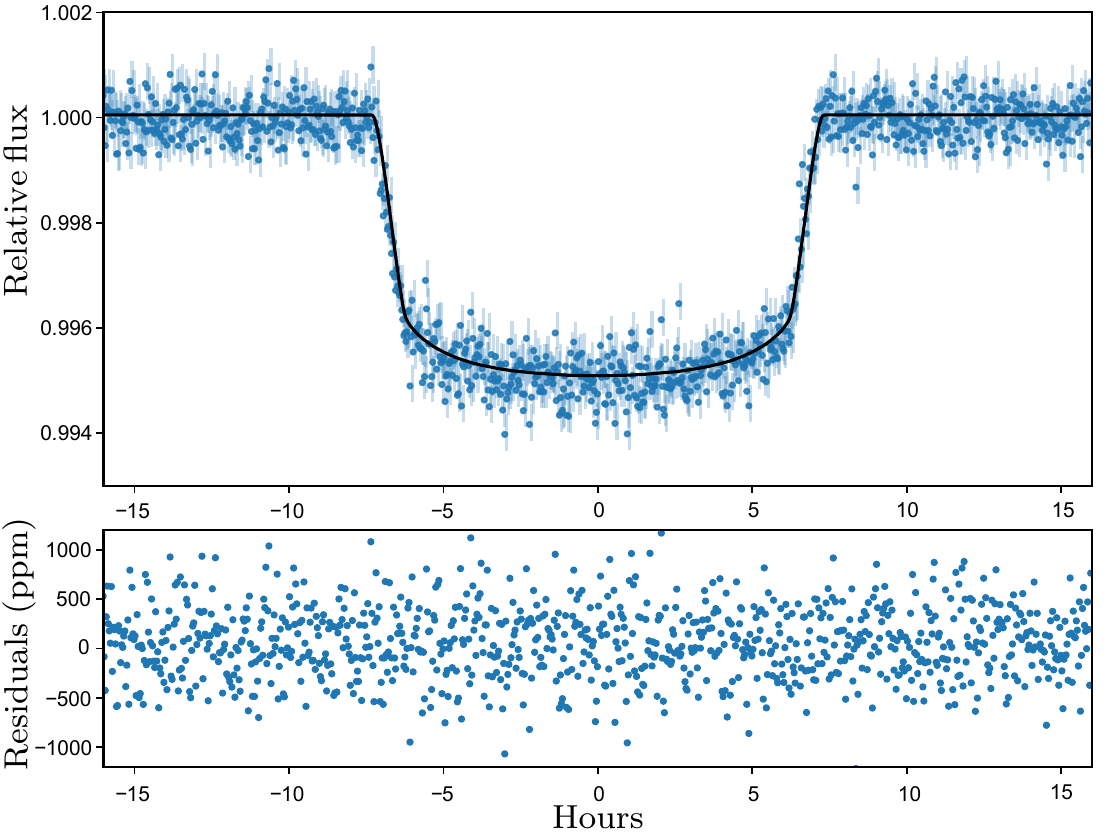}
\caption{Detrended TESS light curve with the associated residuals. The
standard deviation of the TESS residuals is 380 ppm.}  \label{f_light_curve}
\end{figure}
\object{HD\,114082} was observed by TESS in sector 38 (2021-Apr-28 to 2021-May-26) at a 2-min cadence and additional observations are planned in sectors 64 and 65 (2023-Apr-06 to 2023-Jun-02). A single transit event is detected and announced as a Community TESS Object of Interest (CTOI) in July 2021. This single transit event is not concurrent with any asteroid crossing or centroid shifts, which could indicate that the signal is on a background star. The transiting candidate is on target and passed the first photometric vetting stage. 

%\textbf{Looking at radial velocity data, HD\,114082 has been monitored with the Fibre-fed Extended Range Échelle Spectrograp (FEROS) over four years within the framework of a survey aimed at detecting planets around young stars with debris disks. The results of this survey are presented in Zakhozhay et al. (2022) where the authors report a radial velocity scatter of $\rm 75\,m s^{-1}$ for HD\,114082. They attribute this large scatter to stellar activity due to a strong correlation of the radial velocity data with a stellar activity indicator (the bisector of the cross-correlation function). While no planet detection or significant periodicity is highlighted in their paper, these radial velocity measurements allow one to reasonably discard scenarios where the single transit event is due to a spectroscopic binary. Thus we propose an estimation of the planetary parameters of the transiting planetary candidate based on the modeling of the TESS light curve. }

We downloaded the TESS light curve available on the Mikulski Archive for Space Telescopes\footnote{\href{https://mast.stsci.edu/portal/Mashup/Clients/Mast/Portal.html}{mast.stsci.edu/}} and we selected the Pre-search Data Conditioning Simple Aperture Flux (PDCSAP) and associated errors to use in our modeling. The PDCSAP flux is a product of the data reduction done at the Science Processing Operation Center \citep[SPOC,][]{Jenkins2016}. The PDCSAP flux is corrected for systematic trends and the photometric dilution due to contamination by other stars in the aperture. 

We perform the modeling of the photometric data with the software package \texttt{juliet} \citep{Espinoza2019}. The transit model is provided by \texttt{batman} \citep{Kreidberg2015} and we model the stellar activity with a Gaussian process. The fit is done using the nested sampler \texttt{dynesty} implemented in \texttt{juliet}. We set a log-uniform prior for the orbital period between 21.5 and 1000 days, the lower period bound being the minimum period allowed by the TESS data. We choose to fit for the stellar density with a normal period based on the values from the TICv8 catalog \citep{Stassun2019}. The stellar density information combined with the transit shape (e.g. transit duration, impact parameter) constrain the speed at which the planet is crossing the stellar disk and thus give us an estimate of its orbital period. 

We find that the planetary candidate corresponds to a $\rm 0.977 ^{+0.026} _{-0.025}\,R_{J}$ with an orbital period of $197 ^{+171} _{-109}$ days, which corresponds to a semi-major axis of $0.7 ^{+0.4} _{-0.3}$ au. The total transit duration equals to $14.58 ^{+0.06} _{-0.06}$ hours, and the impact parameter is $0.42 ^{+0.23} _{-0.18}$ that translates into inclination of the orbit of $89.78 ^{+0.10} _{-0.25}$ degree. 

In parallel with our work another study describing the discovery of this transiting planetary candidate appeared on the arxiv \citep{Zakhozhay2022b}. This study combines the constraints on the orbital period and planet mass from the TESS transit event in 2021 and radial velocity (RV) data obtained from the four-years monitoring campaign of \object{HD\,114082} with the Fibre-fed Extended Range Échelle Spectrograph (FEROS) and publicly available spectra from the the High Accuracy Radial velocity Planet Searcher \citep[HARPS,][]{Mayor2003} archive. The authors confirm the presence of a close-in giant planet in the system which was assigned an identification name \object{HD\,114082 b}. According to the analysis by \citet{Zakhozhay2022b}, the planet has a mass of $8.0 \pm 1.0$ $M_{\rm Jup}$ and moves around the star on an eccentric orbit ($e = 0.40 \pm 0.04$) with a semi-major axis of $a = 0.51 \pm 0.01$~au and orbital period of $109.8 \pm 0.4$~d. The last two parameters are within 1$\sigma$-interval of our parameter values from the fitting of TESS light curve.

Since the orbital plane of \object{HD\,114082 b} is almost edge-on, it is non-coplanar with the disk midplane and has an offset of more than six degrees from it. The possible ellipticity of the planet orbit is not apparent in the disk images presented in this work: the disk morphology is consistent with the rotationally symmetric model for the dust distribution.

\begin{figure*}
    \centering 
\includegraphics[width=16cm]{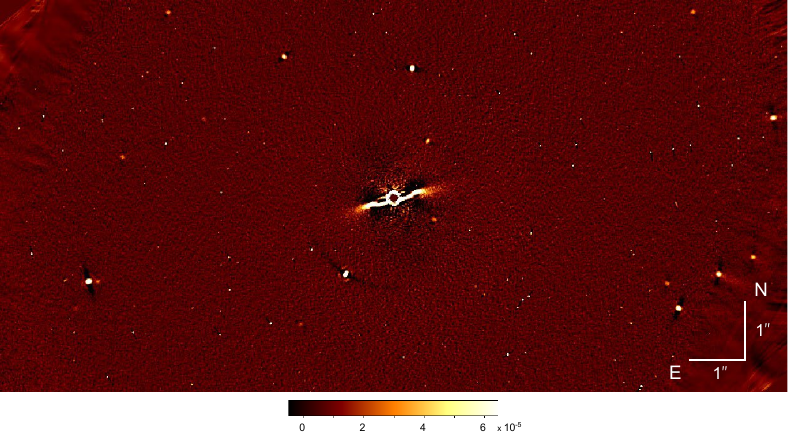}
\caption{Companion candidates for \object{HD\,114082} detected with SPHERE/IRDIS. The color bar shows flux units normalized by the stellar flux.}     \label{f_candidates}
\end{figure*}

The non-coplanarity of planet and disk planes might be a result of interaction of \object{HD\,114082 b} with the other planets residing inside the outer planetesimal belt. If such planets exist and move in the disk plane, they do not transit the star and can be currently detected only with RV measurements. We conclude thus that further photometric and RV monitoring of \object{HD\,114082} is needed to confirm the parameters of \object{HD\,114082 b} and its orbit and investigate the presence of other planets in this system.

\subsection{Companion candidates in the SPHERE image} \label{s_cc}
\begin{figure}
    \centering 
    \includegraphics[width=0.49\textwidth]{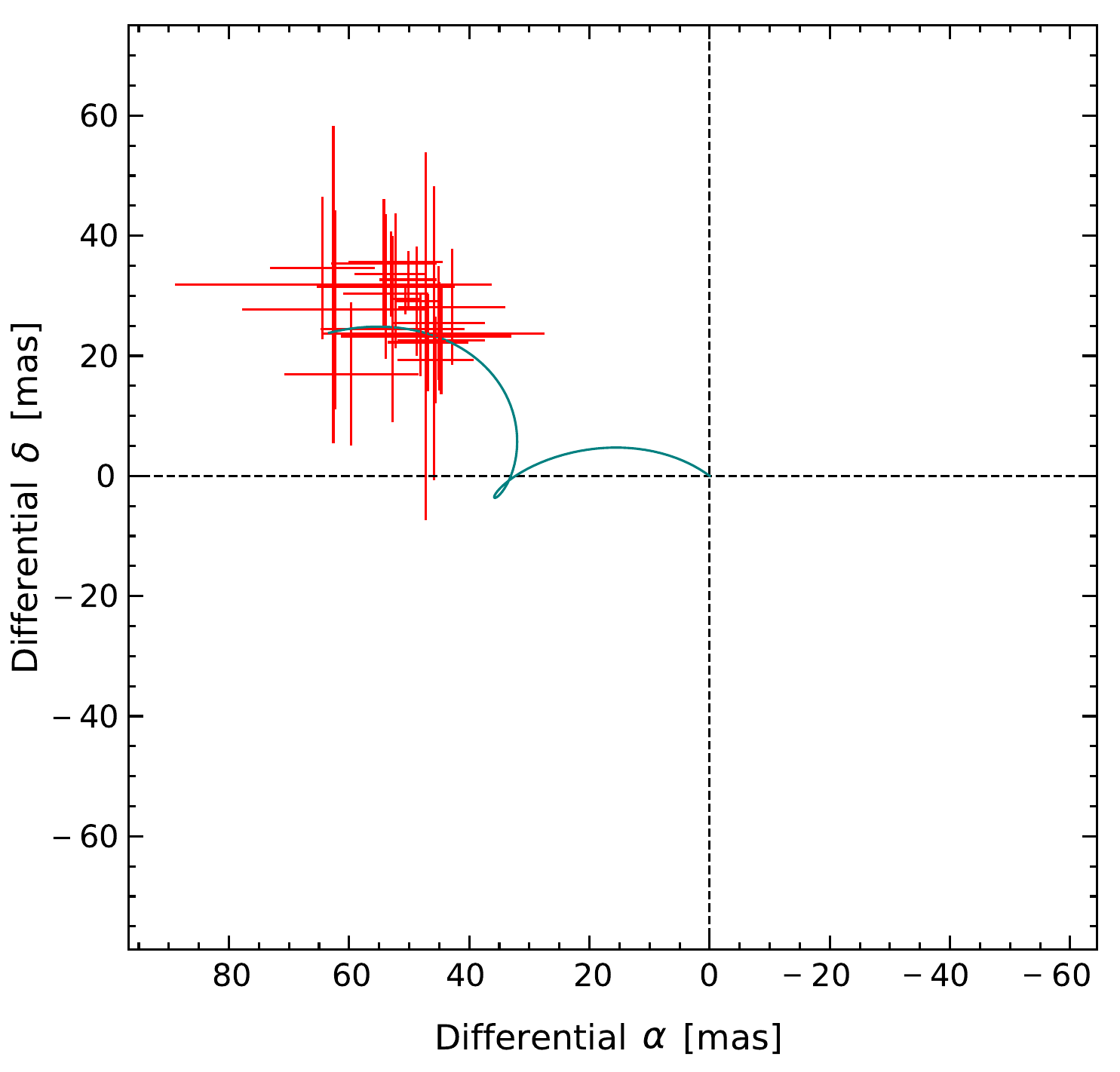}
    \caption{Relative motion of the candidate companions with respect to the central star between the SHARDDS and SHINE observations, on 2016-02-14 and 2017-05-17 respectively. The blue curve represents the track expected for stationary background objects. From this analysis we conclude that all candidates detected in the SHINE data are background stars.}
    \label{fig:cc}
\end{figure}
%\begin{figure}
%    \centering 
%    \includegraphics[width=0.5\textwidth]{./Figures/HIP_64184_mass_bex-cond-hot.pdf}
%    \caption{5$\sigma$ mass sensitivity limit of the SHINE observations in the K1 filter, based on %the contrast curve converted into mass using the BEX-COND-hot evolutionary tracks %\citep{Marleau2019}.}
%    \label{fig:mass_limit}
%\end{figure}

The SHINE observations of \object{HD\,114082} revealed 23 companion candidates in the IRDIS FOV, with SNR ranging from $\sim$3 up to $\sim$70 and projected semi-major axes from $\sim$60\,au up to more than 500\,au (Fig.~\ref{f_candidates}). To assess their status, the SHARDDS data from February 14, 2016, were reprocessed using the open source \texttt{vlt-sphere}\footnote{\url{https://github.com/avigan/SPHERE}} python pipeline \citep{Vigan2020pipeline}. A PCA analysis with 5 subtracted modes was applied to the reprocessed data cubes and all the SHINE candidates are clearly redetected in the SHARDDS data. Due to the time baseline of more than one year between the SHARDDS and SHINE observations, and to the relatively high proper motion of \object{HD\,114082} (44.4\,mas/yr), the motion of the candidates between the two epochs is obvious. We simply use a Gaussian fit of their PSF in the SHARDDS data to obtain their astrometry with an accuracy estimated to approximately half an IRDIS pixel (i.e. 6.1\,mas).

A plot of the relative motion of all the candidates with respect to the central star between the two epochs is presented in Fig.~\ref{fig:cc}. The detailed astrometry is also reported in Table~\ref{tab:cc}. The relative motion of all the candidates clearly follow the track expected for stationary background star, so we conclude that all candidates are background stars. Such a relatively large number of background stars in the IRDIS FOV is explained by the proximity of \object{HD\,114082} to the galactic plane: its galactic latitude is 2.5$^\circ$ ($epoch =J2000$).

%We also plot in Fig.~\ref{fig:mass_limit} the 5$\sigma$ mass sensitivity limit of the SHINE observations. We use the contrast curve obtained in the K1 filter, which takes into account the coronagraph throughput and the small sample statistics, and we convert it using the BEX-COND-hot evolutionary tracks \citep{Marleau2019}. The IRDIS sensitivity in $K$-band is very good and enables reaching masses down to $\sim$5~$M_{\rm Jup}$ at 50\,au, and $\sim$11~$M_{\rm Jup}$ at 20\,au. The sensitivity curve at small separation is probably affected by the structures of the disk, which means that the sensitivity to point sources is in fact slightly better than pictured here.

\subsection{Companion mass sensitivity limits combining direct imaging, radial velocity and Gaia} \label{s_di_rv_gaia}

To bring constraints on the presence of additional companions in the system, we combined the constraints from direct imaging with those from radial velocity (HARPS and FEROS) and from the GAIA astrometry. We used the Multi-epochs multi-purposes Exoplanet
Simulation System 3 (MESS3) tool (Kiefer et al, 2023, in prep.), an extension of the MESS2 tool \citep{Lannier2017} that combines only direct imaging and radial velocity data. 

In direct imaging, we used as input the sensitivity maps deduced from the IFS image of the night of 2017-05-16 reduced in spectral and angular differential imaging, and the IRDIS broad-band H image of the night of 2016-02-13 reduced with PCA. Two mass sensitivity maps were therefore created assuming the AMES-COND evolutionary model \citep{Baraffe2003}, and used as input for MESS3. We also retrieved the archival HARPS and FEROS radial velocity data of the target as described in \cite{Zakhozhay2022b}. We complemented that with the excess noise and proper motion anomaly from GAIA DR3 \citep{GaiaCollaboration2021}, with the technique described in Kiefer et al. (2023, in prep). For the generation of orbits of planetary companions, we assumed inclinations distributed between $0^\circ$ and $90^\circ$ and eccentricities between 0 and 0.9. 

\begin{figure}
    \centering 
    \includegraphics[width=0.5\textwidth]{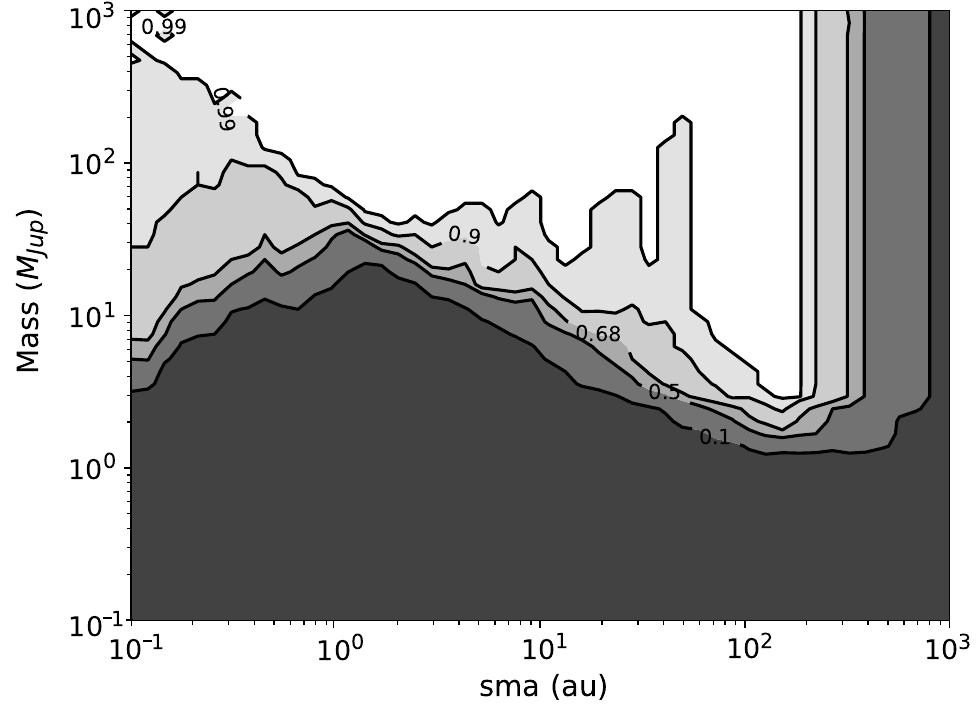}
    \caption{Sensitivity limits on bound companions, obtained by combining the constraints from IRDIS/IFS direct imaging, HARPS/FEROS radial velocity and Gaia astrometry. The darkest area show the parameter space where the probability to detect a planet is less than 10\% while the white area corresponds to a probability higher than 99\%.}
    \label{fig:mass_limit}
\end{figure}

The detection limits obtained when combining those three techniques are displayed in Fig.~\ref{fig:mass_limit} where we plot the isocontours of probability of detection, in a mass versus semi-major axis diagram. At a semi-major axis of 0.7\,au (the most likely semi-major axis of the transiting planet based on our analysis), the $1-\sigma$ sensitivity (68\% probability of detection) corresponds to companions more massive than $30\,M_{\rm Jup}$. We can exclude bound companions more massive than $40\,M_{\rm Jup}$ on any orbits with the same confidence interval.
Exterior to the disk, at 50\,au we reach a sensitivity down to $\sim$$5\,M_{\rm Jup}$, and interior to the disk at 30\,au we are sensitive to planets more massive than $\sim$$10\,M_{\rm Jup}$ .

\section{Summary} \label{s_Summary}
In this work, we present the new optical and near-IR images of scattered light (total and polarized intensity) from the debris disk surrounding a young F star \object{HD~114082}, a member of the Sco-Cen Association. The images were obtained using the SPHERE instruments IRDIS, IFS and ZIMPOL. They show a very bright axisymmetric planetesimal belt with a radius of $\sim$0.37$''$ (35\,au) and inclination of $\sim$83$^\circ$. The disk has a PA of $106^{\circ} \pm 2^{\circ}$ (EoN).

Our modeling results show that the distribution of disk brightness in the image of total intensity can be well reproduced with a SPF combining two HG functions with asymmetry parameters $g_1=0.83$ (weight 0.73) describing the forward-scattering peak and $g_2=0.06$ (weight 0.27) describing a flat middle region of the SPF. When compared to the diffraction part of theoretical SPFs calculated for the microasteroids (Fig.~\ref{f_all_SPFs}), the measured SPF of the \object{HD~114082} disk might indicate that the effective size of dust particles traced in the SPHERE observation at near-IR wavelengths is slightly larger or close to 5 $\upmu$m. 

We used polarimetric data of \object{HD~114082} to compare the disk morphologies seen in total and polarized intensity images (Figs.~\ref{f_imaging} and \ref{f_QphiUphi_IRDIS}) and to estimate the degree of linear polarization of scattered light. Comparing IRDIS DPI observations in the H band to the IFS data covering the same wavelength range, we obtained the maximum PF of $17 \pm 4$\% reached between scattering angles of $70^\circ$ and $115^\circ$ and a minimum PF less than $2$\% between $140^\circ$ and $155^\circ$. The decrease in PF at scattering angles larger than $115^\circ$ is similar to the behavior of cometary dust in the solar system which shows zero PF at $\theta \approx 160^\circ$. The overall shape of the disk PF resembles the bell-shaped PF measured for the cometary dust analogs.

A comparison of the near-IR spectra of two debris disks \object{HD\,114082} and HD 117214 reveals a higher reflectivity for HD\,114082 dust particles, with an albedo of $0.65 \pm 0.15$. Although both host stars are young F-type stars in the LCC subgroup, their debris disks most likely contain particles with different optical properties in the near-IR. This can be explained by different composition or structure of debris material that can change with the stellar age. The nearly twice higher luminosity of HD 117214 compared to the luminosity of \object{HD\,114082} suggests that HD 117214 is either a close binary or much younger than the adopted median age of 17 Myr for the LCC members. In the latter case HD 117214 would be at an earlier evolutionary stage than HD\,114082 and could harbour a more gas- and ice-rich disk.

The reflectance spectra of both disks, measured along the disk major axes at radial position of planetesimal belts, show a rather flat flux density contrast within the wavelength range between 0.96 and 1.66 $\upmu$m. It is lower than the flux density contrast derived for the IRDIS K1 and K2 filters ($\lambda_{\rm K1}=2.110\,\upmu$m, $\lambda_{\rm K2}=2.251\,\upmu$m) where the reflectance spectra show a red color. Apart from that, the \object{HD\,114082} spectrum appears to be featureless, whereas the HD 117214 spectrum may indicate absorption features. They could be interpreted as the absorption bands of CO$_2$ ice. However, to conclude about the actual content of CO$_2$ and other ices in the dust particles of HD 117214 or other debris disks, space-based observations with high spatial and angular resolutions in the near-IR are needed. For some debris disks, detection of the ice absorption bands will be possible with the instruments on-board the JWST.

The TESS photometric data show a dip in the light curve of \object{HD\,114082} indicating that a Jupiter-sized planet revolves around the star within the cavity of outer planetesimal belt in an orbit with a semi-major axis of $0.7 \pm 0.4$~au. Thus far, the transit of the planet was observed only in 2021, and its parameters remains to be confirmed via further photometric and radial velocity monitoring. 

Within the IRDIS FOV, we detected 23 companion candidates in the HD\,114082 data. Based on comparison with available SHARDDS images, we conclude that all these candidates are background stars. 

By combining the SPHERE data with archival radial velocity data and astrometry, we reach a high sensitivity down to $\sim$5~$M_{\rm Jup}$ at 50\,au, and $\sim$10~$M_{\rm Jup}$ at 30\,au when using the AMES-COND evolutionary models to convert the derived contrast limits into companion mass.   

\begin{acknowledgements} 
We thank the anonymous referee for many thoughtful comments which helped to improve this paper.\\
This work has made use of the SPHERE Data Centre, jointly operated by OSUG/IPAG (Grenoble), PYTHEAS/LAM/CeSAM (Marseille), OCA/Lagrange (Nice), Observatoire de Paris/LESIA (Paris), and Observatoire de Lyon/CRAL, and supported by a grant from Labex OSUG@2020 (Investissements d’avenir -- ANR10 LABX56). This research has made use of the Direct Imaging Virtual Archive (DIVA), operated at CeSAM/LAM, Marseille, France. \\
AV acknowledges funding from the European Research Council (ERC) under the European Union's Horizon 2020 research and innovation program (grant agreement No.~757561). JPM acknowledges research support by the Ministry of Science and Technology of Taiwan under grants MOST107-2119-M-001-031-MY3 and
MOST109-2112-M-001-036-MY3, and Academia Sinica under grant AS-IA-106-M03. CdB acknowledges support by Mexican CONACYT research grant FOP16-2021-01-320608. AZ acknowledges support from the FONDECYT Iniciaci\'on en investigaci\'on project number 11190837 and ANID -- Millennium Science Initiative Program -- Center Code NCN2021\_080. \\
SPHERE is an instrument designed and built by a consortium consisting
of IPAG (Grenoble, France), MPIA (Heidelberg, Germany), LAM (Marseille,
France), LESIA (Paris, France), Laboratoire Lagrange (Nice, France), 
INAF – Observatorio di Padova (Italy), Observatoire de Gen`eve 
(Switzerland), ETH Zurich (Switzerland), NOVA (Netherlands), ONERA (France) 
and ASTRON (Netherlands), in collaboration with ESO. 
SPHERE was funded by ESO, with additional contributions from CNRS (France), 
MPIA (Germany), INAF (Italy), FINES (Switzerland) and NOVA (Netherlands). 
SPHERE also received funding from the European Commission Sixth and 
Seventh Framework Programmes as part of the Optical Infrared 
Coordination Network for Astronomy (OPTICON)
under grant number RII3-Ct-2004-001566 for FP6 (2004–2008), grant number
226604 for FP7 (2009–2012) and grant number 312430 for FP7 (2013–2016).\\
This work has made use of data from the European Space Agency (ESA) mission
{\it Gaia} (\url{https://www.cosmos.esa.int/gaia}), processed by the {\it Gaia}
Data Processing and Analysis Consortium (DPAC,
\url{https://www.cosmos.esa.int/web/gaia/dpac/consortium}). Funding for the DPAC
has been provided by national institutions, in particular the institutions
participating in the {\it Gaia} Multilateral Agreement.

\end{acknowledgements}

\bibliographystyle{aa} 
\bibliography{HD114082_reference.bib} 

\newpage 
\appendix
\section{$U_\varphi$ images} \label{s_Uphi_app} 

\begin{figure}
\centering
\includegraphics[width=7.5cm]{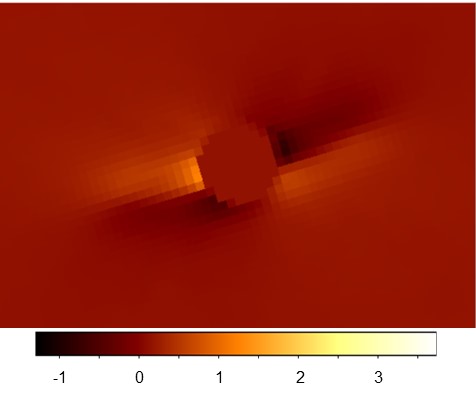} 
\caption{IRDIS $U_\varphi$ model image.  \label{f_Uphi_model}}
\end{figure} 
\begin{figure}
\centering
\includegraphics[width=7.5cm]{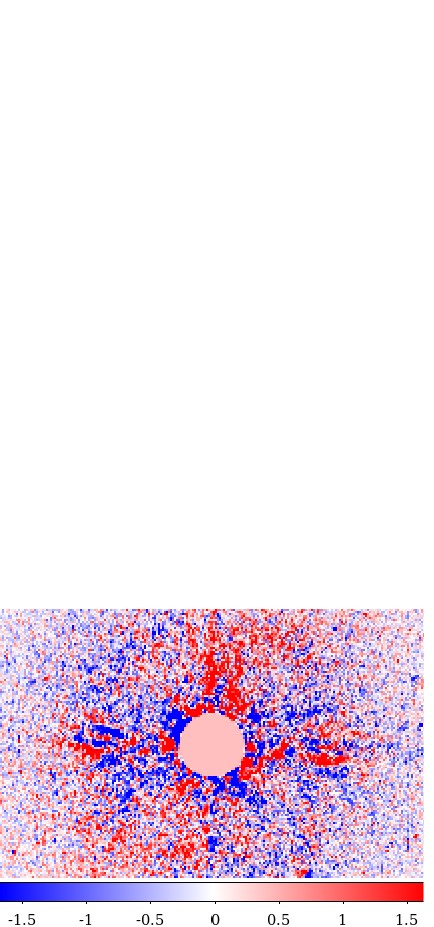} 
   \caption{ZIMPOL $U_\varphi$ image.  \label{f_Uphi_colored}}
\end{figure} 

Figure~\ref{f_Uphi_model} shows the modeled image of the IRDIS $U_\varphi$ signal after convolution of the modeled Stokes parameters $Q$ and $U$ with the IRDIS PSF from the night of observation. Model 2 was used to create the images.

The convolution of the measured Stokes parameters $Q$ and $U$ with the instrument PSF always generates an artificial $U_\varphi$ signal and results in reduction of the polarized flux measured in the $Q_\varphi$ image. This effect is stronger for the disks with smaller angular size such as \object{HD\,114082} disk \citep{Engler2018}. 

Figure~\ref{f_Uphi_colored} shows the ZIMPOL $U_\varphi$ image, the same as in Fig.~\ref{f_QphiUphi_IRDIS} but here with different scale range and colors. This image does not contain a signal from the disk. 

\section{Modeling the scattered (polarized) light from a debris disk} 
\label{s_modeling_app}
\begin{figure}
\centering
\includegraphics[width=7.5cm]{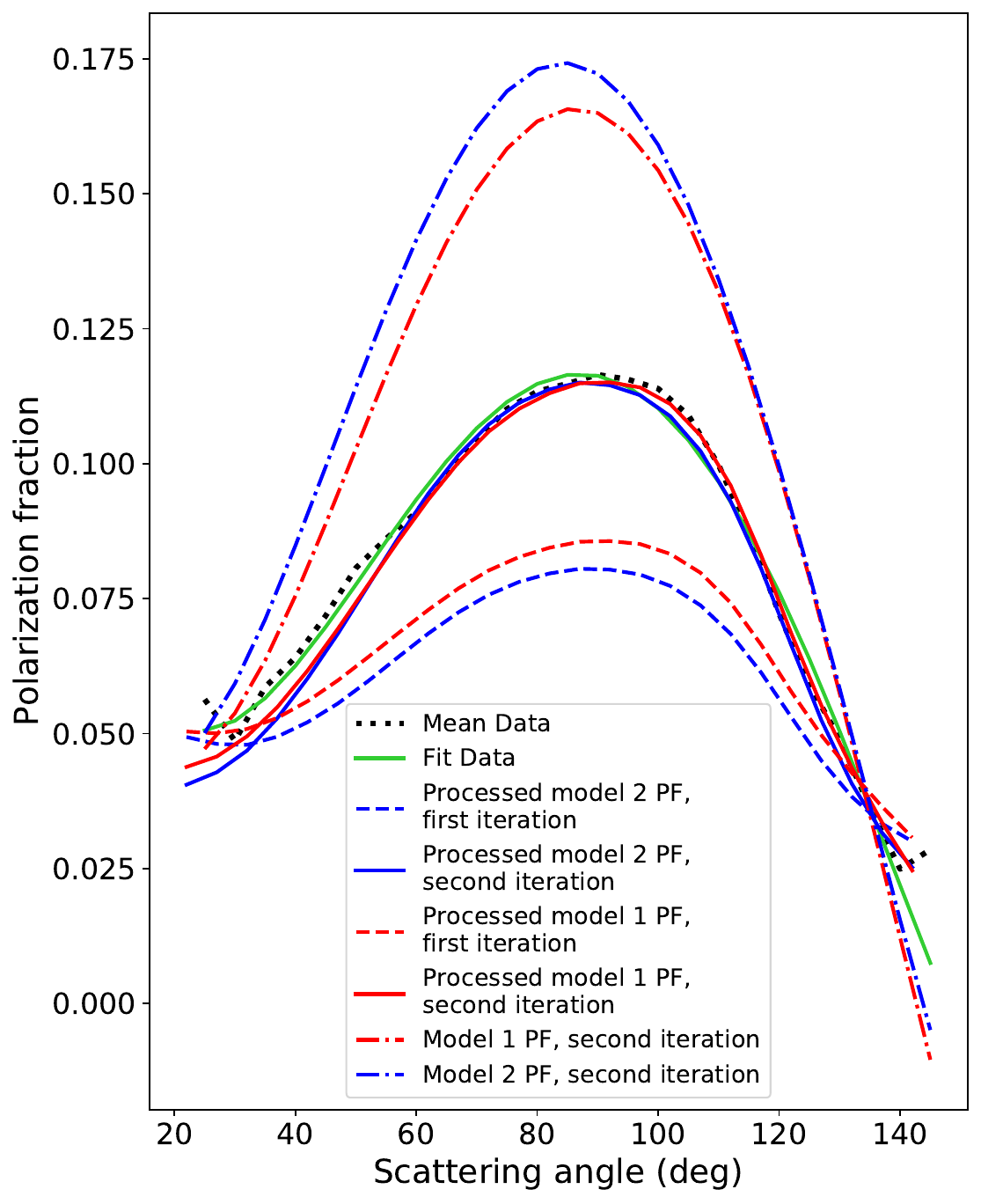} 
   \caption{Comparison of the polarization fraction derived with models 1 and 2. \label{f_pol_frac_2models}}
\end{figure}  

\subsection*{Model for the image of scattered (polarized) light}
In order to create a synthetic image of light scattered by small dust particles, we use a 3D rotationally symmetric model for the spatial distribution of particle number density in a debris disk. We describe such a distribution by the product of the radial power law $R(r)$ \citep[Eq.~\ref{eq_R},][]{Augereau1999} and the Lorentzian profile $Z(r,h)$ for the vertical distribution of particle density (Eq.~\ref{eq_Z}). Compared to the exponential profile adopted by \cite{Augereau1999} for the study of \object{HR 4796\,A} disk, the Lorentzian profile offers a model with a smaller number of parameters. This profile seems to fit the vertical cuts of debris disks well, and was used for the disk modeling in previous studies \citep[e.g.,][]{Krist2005, Engler2018}.  
Our model is based on the theory of the parent body belt that consists of massive planetesimals moving on a circular orbit with a radius $R_{\rm belt}$. Mutual collisions between the planetesimals produce a large amount of micron-sized dust particles that are radially redistributed with time. The maximum density of particles occurs at the site of their production, i.e. at radial distance $R_{\rm belt}$. Therefore the particle number density in the disk midplane can be described by the radial power laws with the exponents $\alpha_{\rm in}> 0$ inside the belt and $\alpha_{\rm out}< 0$ outside the belt:

\begin{equation} \label{eq_R}
R(r) \propto {\left( \left(\frac{r}{r_0}\right)^{-2\alpha_{in}}+\\\ \left(\frac{r}{r_0}\right)^{-2\alpha_{out}} \right)}^{-1/2},
\end{equation} 
where $r_0$ is the reference radius that is related to the belt radius $R_{\rm belt}$  as: 
\begin{equation} \label{eq_r_belt}
R_{\rm belt} = r_0 \, \left(-\frac{\alpha_{in}}{\alpha_{out}}\right)^{1/(2\,\alpha_{in}-2\, \alpha_{out})} 
\end{equation} 
and thus $R_{\rm belt} = r_0 $ if $\alpha_{\rm in} = - \alpha_{\rm out}$. 

We define the vertical profile for the particle density distribution in disk cross-sections perpendicular to the disk midplane according to:
\begin{equation}  \label{eq_Z}
Z(r,h) \propto {\left[ 1 + \left(\frac{h}{H(r)}\right)^2\right]}^{-1},
\end{equation}
where $h$ is the height above the disk midplane. 
%and $a_L$ is the peak number density of particles in the disk midplane. 
The scale height of the disk $H(r)$ is defined as a half-width 
at half-maximum of the vertical profile at radial distance 
$r$ and scales like $H(r)=H_0\,(r/r_0)\,^\beta$, where $H_0= H(r_0)$ and $\beta$ is the disk flare index. 

How much stellar light is scattered into LOS of observer depends on the scattering angle $\theta$. This dependence is given by the SPF which is often approximated by the HG function \citep{Henyey1941}:
\begin{equation} \label{e_HG}
f(\theta, g)=\frac{1-g^2}{4 \pi (1+g^2-2 g cos (\theta))^{3/2}},
\end{equation}
where $g$ is the HG scattering asymmetry parameter ($-1~\leqslant~g~\leqslant~1$), and $\theta$ is the scattering angle measured between direction of the incoming stellar radiation and direction of the scattered radiation. The latter is the LOS of the observer. 

We assume that dust particles scatter more radiation in the forward direction (at smaller scattering angles). This means that the asymmetry parameter $g$ has a positive value and the brighter side of the disk is closer to the observer. We also consider that dust particles have the same properties everywhere in the disk and  scattered-light images preferentially trace dust particles with a size comparable to the wavelength of observation. 

As mentioned in Sect.~\ref{s_Modeling}, in this work we tested two models with different SPFs. The phase function of model 1 is the HG function (Eq.~\ref{e_HG}). The phase function of model~2 is a linear combination of two HG functions \citep[][]{Engler2017}:
\begin{equation} \label{eq:phase func}
f(\theta,g_1,g_2)= w \cdot f(\theta,g_1) + (1-w) \cdot f(\theta,g_2)\, ,
\end{equation} 
where the first parameter $g_1$ describes a forward-scattering peak of the SPF, the second parameter $g_2$ represents the more isotropic part of the SPF, and $w$ is the scaling parameter, $0~\leqslant~w\leqslant~1$.

For the modeling of polarized scattered flux the pSPF is used instead of SPF. The pSPF is given by a product of SPF and PF (Eq.~\ref{eq_p}). The latter can be approximated by an analytic expression for the small dust particles \citep[e.g.,][]{Graham2007, Engler2017} or measured from the data, if the data quality is good. 

Using this model we create a 2D image of scattered or polarized light from a debris disk by summing up all contributions to the scattered (polarized) flux $F_i$ along the LOS of the observer at each pixel (resolution element) $i$ of the image: 
\begin{equation} \label{eq_F}
F_i = A \int\limits_{LOS}\frac{f(\theta, g)\, R(r)\ Z(r, h)}{r^2}, 
\end{equation}
where $A$ is a scaling factor, and the SPF $f(\theta, g)$ shall be replaced by the pSPF for the polarized flux image.

The obtained model image can be compared to the science image using the reduced $\chi_\nu^2$ metric (Eq.~\ref{eq_chi2}). %A detailed description of this modeling approach is given in \cite{Engler2017}.} 

To determine the best-fit parameter set for the total intensity image in the K1 band (Fig.~\ref{f_imaging}, left panel), we used a custom MCMC code employing the Python package \textit{emcee} \citep{Foreman-Mackey2013}. We ran the MCMC sampler with 1000 walkers for each model family separately using the uniform priors as specified in Col.~2 of Table~\ref{t_results}.

At each walker step, a synthetic image of scattered light for a model with a given parameter set $\vec{p}=(p_{1}, p_{2}, ..., p_{N_{\rm par}})$ is created. This image is then inserted in a data cube at different PAs to reproduce the rotation of the sky field during the observation and perform the cADI forward-modeling. The forward model is used to estimate the likelyhood $\mathcal{L}$ of parameter set \vec{p} by calculating the $\chi^2$-metric according to Eq.~\ref{eq_chi2} and the $\ln{\mathcal{L}}$ according to \cite{Hogg2010}:
\begin{equation} \label{eq_lnL}
\ln{\mathcal{L}} = -\dfrac{1}{2}\, \Big(  \chi_\nu^2 \nu + \sum_{i=1}^{N_{\rm data}} \ln{(2\pi \, \sigma_{i,\, {\rm data}}^2}) \Big).
\end{equation}

%where $N_{\rm data}$ is a number of image pixels used to perform the $\chi^2_{\nu}$ minimization, $F_{i,\, {\rm data}}$ is the flux measured in pixel $i$ with an uncertainty $\sigma_{i,\, {\rm data}}$ and $F_{i,\, {\rm model}}(\vec{p})$ is the modeled flux of the $i$ pixel. The degree of freedom of the fit is denoted by $\nu = N_{\rm data}-N_{\rm par}$, where $N_{\rm par}$ represents the number of variable parameters. In a total, the model 1 has nine parameters, and the model~2 eleven (Col.~1 of Table~\ref{t_results}). The uncertainties $\sigma_{i,\, {\rm data}}$ are taken from the noise map computed as the standard deviation of the flux distribution in concentric annuli of the science image, excluding regions with the disk flux. 

The MCMC sampler converges to the parameter values with the highest likelyhood after 2000 steps providing the posterior distributions of the fitted parameters, which we obtain after removing the burn-in phase consisting of first 400 walker steps. The derived posterior distributions are shown in Figs.~\ref{f_mcmc_1g} and \ref{f_mcmc_2g_1}. Their median values are given in Cols.~3 and 4 of Table~\ref{t_results}. The lower and
upper uncertainties on these values are defined as the 16th and 84th percentiles of the posterior distributions.

\subsection*{Polarization fraction phase functions of models 1 and 2}
%The measured \textcolor{red}{PF phase function (H band)} is shown by black dotted line in Fig.~\ref{f_pol_frac_2models}. We fitted a polynomial curve to this line (green solid line) to get a numeric expression which could be applied in the model. 
Fig.~\ref{f_pol_frac_2models} shows a comparison between forward-modeled PF for model 1 and model 2 which are plotted as red and blue lines, respectively: dashed lines show the PF used in the first iteration, and dashdotted lines show the PFs used in the second iteration as discussed in Sect.~\ref{s_pol_frac_Hband}. After two iterations a very good match between the measured PF from the data (black dotted line) and measured PF from both models (red and blue solid lines) was obtained.  

\begin{figure}
\centering
\includegraphics[width=8.5cm]{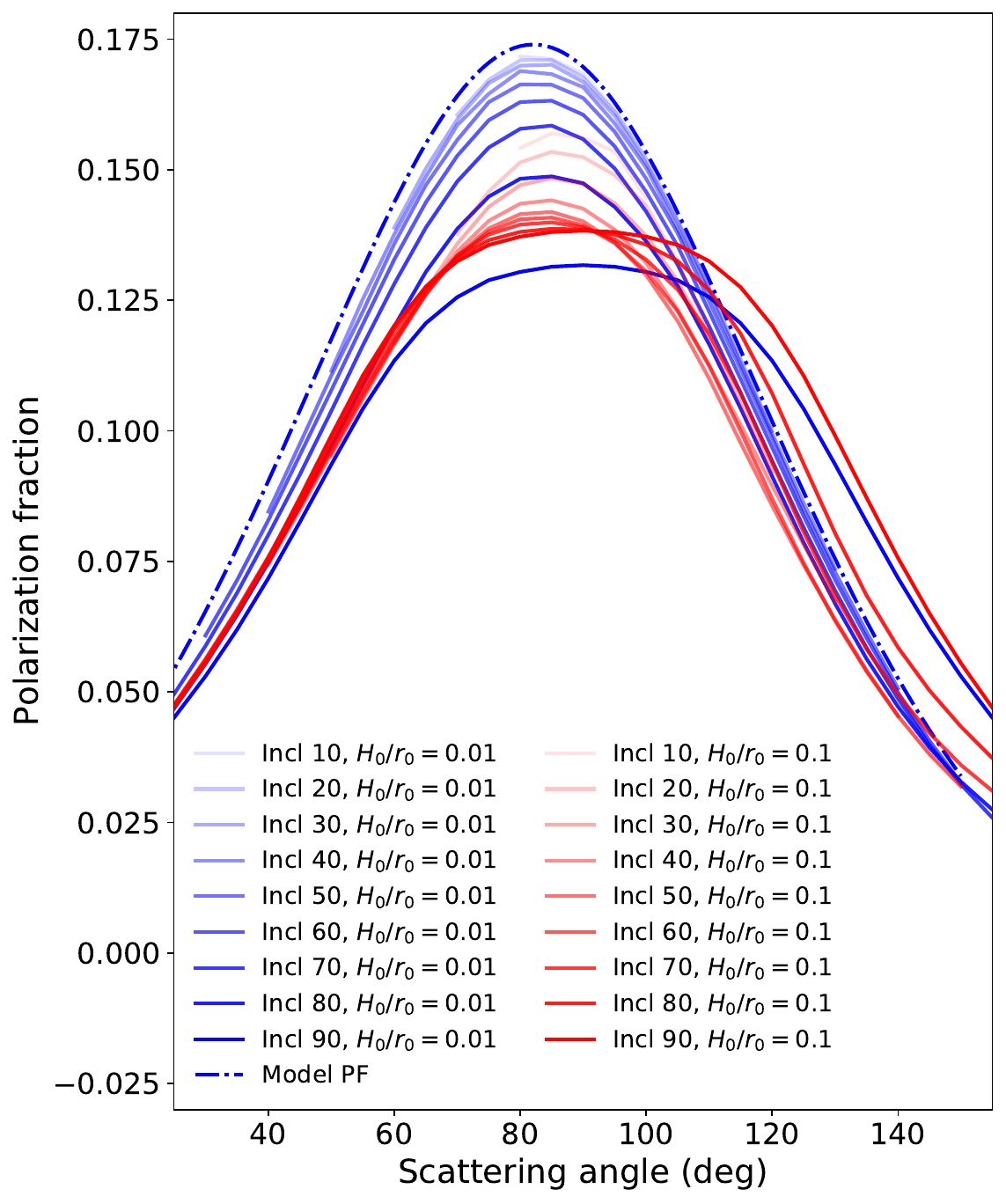} 
\caption{Polarization fraction of scattered light derived for model 2 with different inclinations and scale heights of the disk. } \label{f_LOS}
\end{figure}  
\subsection*{Impact of disk inclination and scale height on measured polarization fraction of scattered light}
The outcome of PF measurement from the images of debris disks in scattered and polarized light is significantly affected by the disk morphology, in particular by its radial and vertical extent. We referred to this effect as the LOS effect in Sect.~\ref{s_LOS_effect}.
Since this is a geometrical effect, we expect that inclination and scale height are the two most important disk parameters which influence the measured PF value. To investigate this, we used model 2 to obtain the measured PFs for disks with different inclinations and scale heights. The inclinations were varied from 10$^\circ$ (nearly pole-on disk) to 90$^\circ$ (edge-on disk). The scale heights were set to 0.004$''$ ($H_0/r_0=0.01$) and 0.04$''$ ($H_0/r_0=0.1$). 

The PFs shown in Fig.~\ref{f_LOS} were derived applying aperture photometry to the non-convolved model images of the total and polarized intensities. They demonstrate that the measured PF decreases with the growing disk inclination and scale height and that the position of maximum value is shifted to the larger scattering angles for the disks with inclination larger than $80^\circ$.

\section{Why is HD 117214 more luminous?} \label{s_HD117214}
In Sect.~\ref{s_SED} we discussed different reflectivity of dust particles in debris disks \object{HD\,114082} and HD 117214. The difference in optical properties between these two disks could arise from different properties of their host stars. It is assumed that HD 117214 and \object{HD\,114082} are approximately 17 Myr old \citep{Pecaut2012}. Since they are both F-type stars with a similar surface temperature, the much higher brightness of HD 117214 (1.5-2 times brighter than \object{HD\,114082}) raises the question of why this star is much more luminous than expected for its estimated age.

There are two possible answers to this question: (1) the star is a close (spectroscopic) binary with two stars of nearly equal masses, or (2) the HD 117214 is significantly younger than 17 Myr old, the age which is commonly adopted for the LCC members.
\begin{figure}
  %  \centering 
\includegraphics[width=8.5cm]{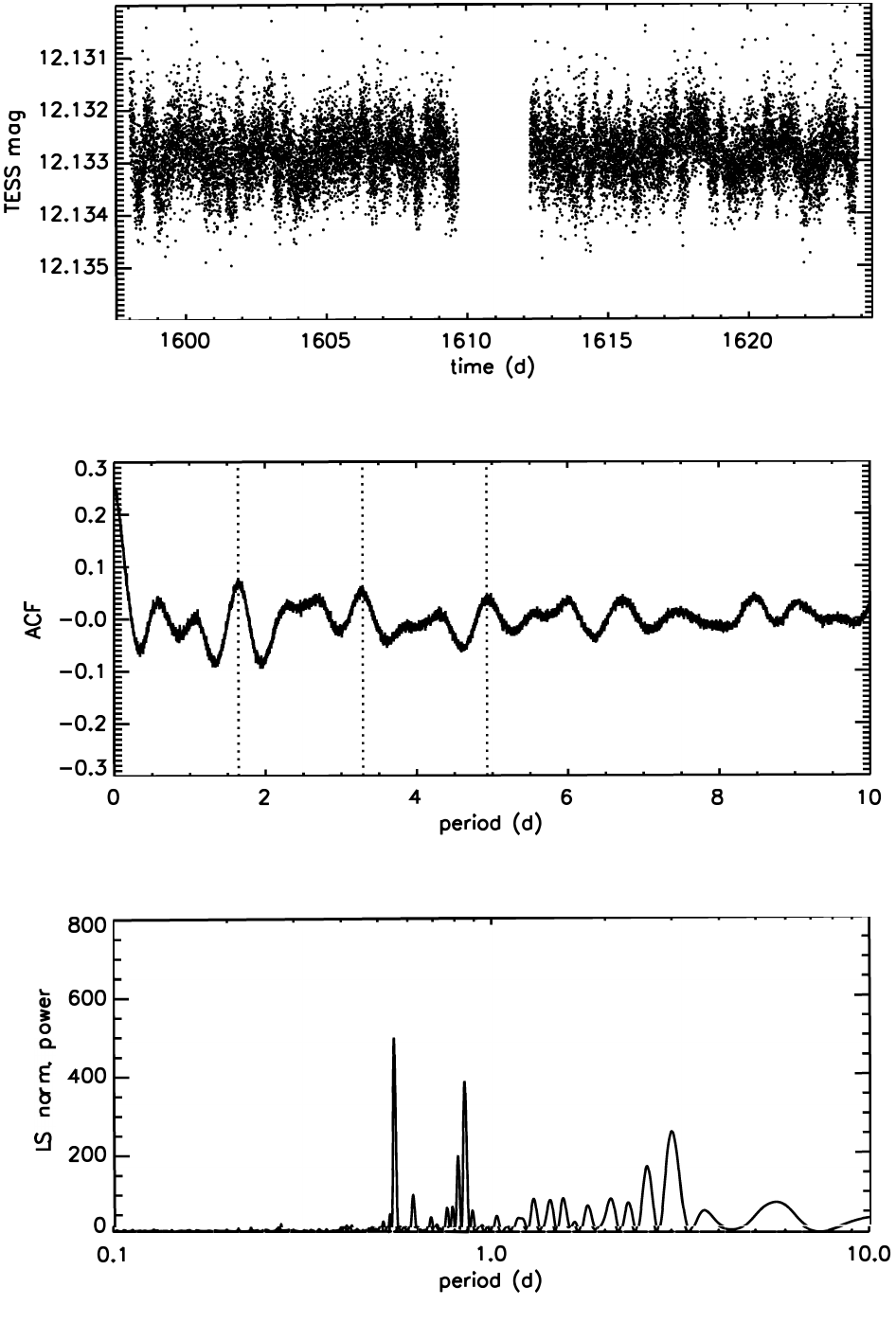}
\caption{TESS photometric timeseries ({\textit{top panel}}), the ACF ({\textit{middle panel}}) and  periodogram of HD 117214 ({\textit{bottom panel}}). }  \label{f_TESS}
\end{figure}

Indeed, HD 117214 is not a single star. According to the Gaia eDRH3 catalog \citep{GaiaCollaboration2021}, it is accompanied by at least one M dwarf (2MASS J13300763-5829026), which has a G magnitude of 13.4$^{m}$ and a contrast to the central star of ${\rm dG} = 5.44^{m}$. This is certainly a bound companion to the HD 117214 since both stars have the same parallax of about 9.3 mas within the errorbars and a similar proper motion of 38-39 mas/year. However, this late-type companion has a projected separation of 10.72$''$ or 1150 au \citep{GaiaCollaboration2021} and therefore cannot be responsible for the over-luminosity of HD 117214. If this companion is on a circular orbit with the same inclination of $\sim$70$^\circ$ as the debris disk \citep{Engler2020}, then its actual separation from the primary star is $\sim$3500 au. 

%%%%%%%%%%%%%%%%%%%%%%%%%%%%%%%%%%%%%%%%%%%%%%%%%%%%%%%%%%%%%%%%%
However, HD\,117214 might be itself an unresolved spectroscopic binary. If so, the two components of star should be separated by at most a few au. This follows from our analysis of the SHINE survey data \citep{Engler2020} and proper motion analysis of this system based on \textit{Hipparcos} and Gaia DR2 data by \cite{Kervella2019}. In such a case, the binarity of HD~117214 can be tested by means of an analysis of the photometric variability or RV data. 

% from the high-contrast imaging \citep{Janson2013} no HD 117214

\begin{figure}
 %\centering 
\includegraphics[width=7.5cm]{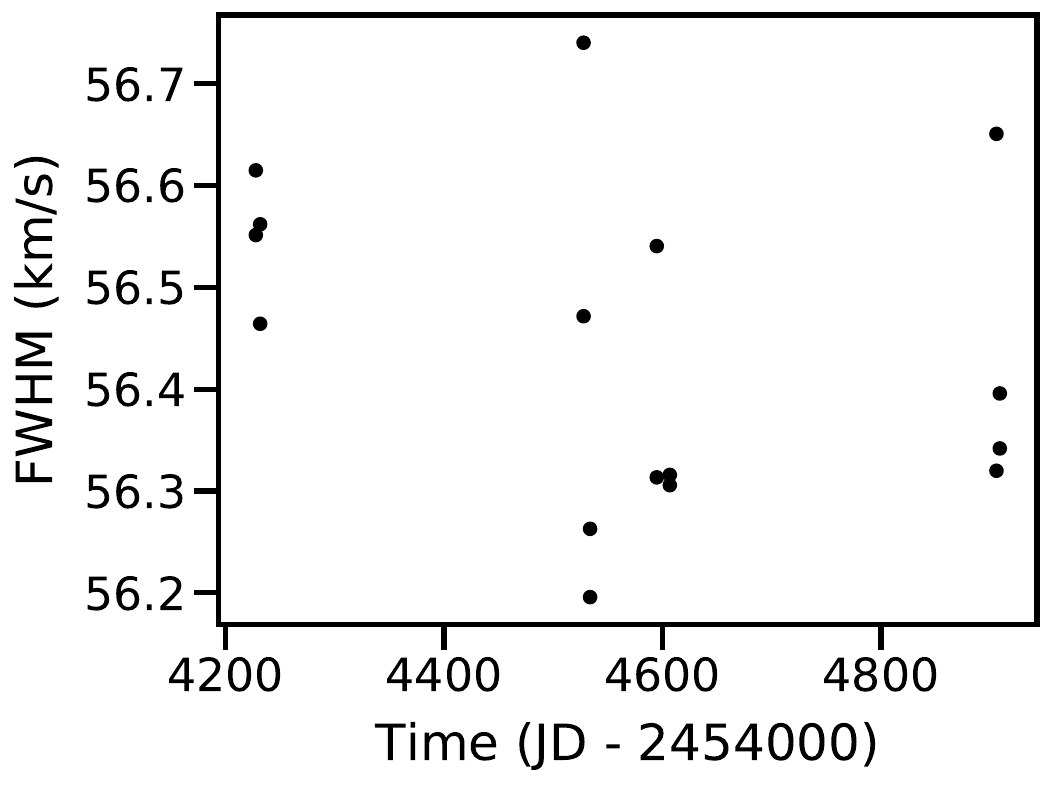}
\caption{FWHM time variations for HD 117214.     \label{f_rv_HD117214} }
\end{figure}

\begin{figure}
\centering 
\includegraphics[width=7.5cm]{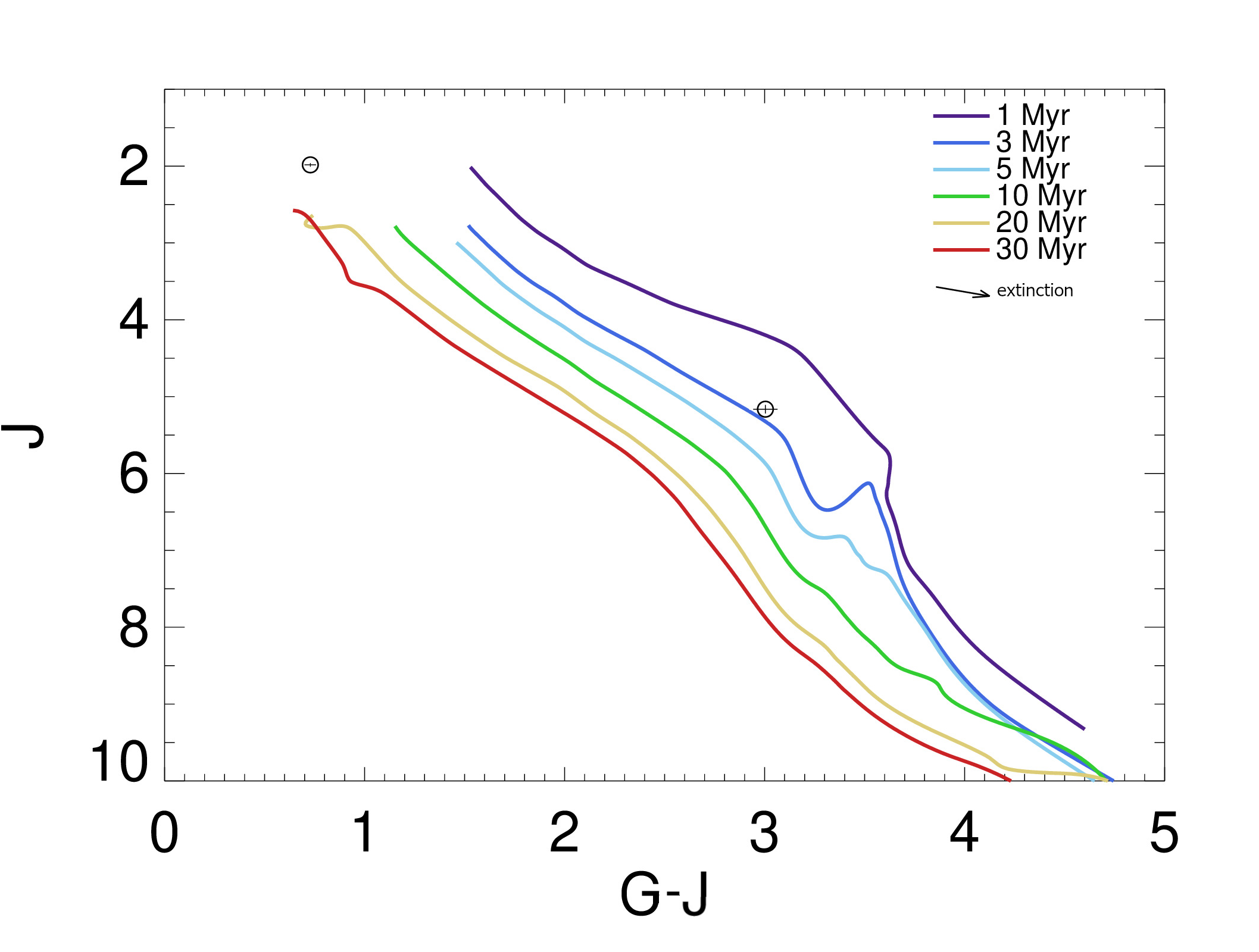}
\caption{CMD for the HD 117214 primary and secondary stars. The extinction vector just indicates the direction of reddening. \label{f_CMD_HD117214} }
\end{figure}

In fact, we know that unresolved close binaries consisting of components of similar brightness, as assumed in our hypothesis, and of late spectral type (equal to or later than the F type), generally host activity phenomena, like starspots, which can induce quasi-periodic variation of their flux.
A periodogram analysis of photometric timeseries can reveal the rotation period of each component, thus unveiling the binary nature of the system \citep[see, e.g.,][]{Rebull2018, Messina2019}.
Indeed, our target was observed by TESS, and we analysed its photometric timeseries (top panel of Fig\,\ref{f_TESS}) using both the auto correlation function \citep[ACF,][]{McQuillan2013} and the Lomb-Scargle \citep[][]{Horne1986} periodogram analysis.

The ACF (second panel of Fig\,\ref{f_TESS}) displays the typical pattern with one or two lower-amplitude local maxima followed by the highest amplitude maximum. This  pattern is generally produced when the spots responsible of the photometric variability are unevenly distributed across the stellar surface. The highest local maximum at P = 1.644\,d is assumed to represent the stellar rotation period.

This interpretation is supported by the results of the LS periodogram (bottom panel of  Fig\,\ref{f_TESS}) in which we observe two major peaks at P = 0.551$\pm$ 0.006\,d and P = 0.844 $\pm$ 0.012\,d. The latter period is half the rotation period inferred from ACF, confirming the presence of two spot groups on opposite hemispheres. The P = 0.551\,d is a beat period of P = 0.844\,d as well as the other lower-power peaks at P = 2.94$\pm$0.13\,d and P = 2.56$\pm$0.10\,d are all aliases. Therefore, no evidence of binarity is inferred from our periodogram analysis. % and the possibility that the over-luminosity of HD\,117214 may arise from binarity seems to be ruled out.

The RV measurements of this star do not show significant variations on the time baseline of several years. \cite{Chen2011} reported a RV of $9.7 \pm 0.3$ km/s whereas Gaia's measurements resulted in $10.2 \pm 0.2$ km/s. The IR spectra of HD 117214 obtained with the HARPS have been analysed by Grandjean et al. (in prep.), who found no indication of RV variation over a period of 682 days (measured amplitude of RV variations is 96 m/s). This makes it quite difficult but not completely impossible for the star to be an equal mass binary. In fact, if two stellar components have very similar brightness, with a RV difference not larger than a half of $v \sin i$ \citep[40 km/s,][]{Chen2011}, lines of two components in spectrum are blended. In this case the barycenter of the observed line does not change along the orbit but the width of the line should change. However, the FWHM is stable around 56.4 km/s over time in HARPS data (Fig.~\ref{f_rv_HD117214}), and so the variation of the line width is not observable.

Since available RV measurements do not suggest HD 117214 being a close binary, 
the analysis of photometric variability and RV data seems to rule out the possibility for HD 117214 to be a close binary. We suggest thus that the star may be much younger than the median age of $17 \pm0-8$ Myr found for the F-type members of the LCC \citep{Pecaut2012}. 

\begin{table*}  
      \caption[]{\object{HD 117214} photometry.  \label{t_photometry} }
          \centering
             \begin{tabular}{lccccc}
            \hline
            \hline
            \noalign{\smallskip}
                   &  &  & Primary star & Secondary star & \\
            Filter & $\lambda$ & Bandwidth & magnitude  & magnitude & Ref. \\
                   & ($\mu$m) & ($\mu$m) & (mag) &  (mag) &  \\
            \hline
            \noalign{\smallskip}
            Gaia G  & 0.673 & 0.440 & 7.97 & 13.41 &  1\\
            ZIMPOL VBB  & 0.735 & 0.290 & 7.72 & (...) &2  \\  
            2MASS J & 1.235 & 0.162 & 7.17 & 10.35 &3  \\
            2MASS H & 1.662 & 0.251 & 6.97 & 9.69 &3  \\
            2MASS K & 2.159 & 0.262  & 6.90 & 9.49 & 3 \\            
  \noalign{\smallskip}
            \hline
            \noalign{\smallskip}
         \end{tabular}
\tablebib{ (1)~\citet{GaiaCollaboration2021}; (2) \citet{Engler2020}; (3) \citet{Cutri2003}. }    
\end{table*}  

%J=10.345, H=9.693, K=9.488   \citet{ESA1997}  2MASS J13300897-5829043   2MASS J13300763-5829026
%                                              5870290798365685248        5870290793999026560

Both the central star and the late-type companion are much redder than expected for the age of 17 Myr (see the compilation of available photometry for the primary and secondary stars in Table~\ref{t_photometry}). 
If corrected for the distance, the absolute G magnitude of companion is $8.258^m$ and its bp-rp color is 2.864. This is inconsistent with the color-magnitude diagram (CMD) isochrones from the BT-Settl CIFIST evolutionary models by \cite{Baraffe2015}. Such inconsistency can be explained either by much younger age of the system or by large interstellar reddening toward the system which makes stars much redder compared to the stars on the main sequence. If the last case were true, the stars would be intrinsically much brighter and massive, but this would be inconsistent with their spectral classification. 

From comparison of the M-type companion position with the isochrones in the CMD (Fig.~\ref{f_CMD_HD117214}), the age of this star would be between 2 and 3 Myr. However extensive comparisons for stars in Sco-Cen show that isochrones underestimate the ages for M-stars \citep{Feiden2016}, likely because they do not properly take into account the impact of stellar activity. Once corrected for this effect, a more appropriate age
for the secondary star should be about 5 Myr, if it is not an equal mass binary and does not possess a circumstellar disk itself. Such a young age is not an exception for the LCC members: \cite{Murphy2015} identified two accreting M-dwarfs as young as 5 - 10 Myr in this subgroup.

The primary star is too massive for comparison with the BT-Settle CIFIST isochrones. Using the PARSEC models \citep{Bressan2012}, we find a good match of stellar photometry for stellar ages between 7 and 9 Myr, and masses between 1.63 and 1.53 $M_\odot$, respectively. Based on the analysis of the primary and secondary stars, we conclude that the age of the HD 117214 might be $7 \pm 2$ Myr. Our conclusion is supported by the results of the RV Survey for Planets around Young stars \citep[RVSPY,][]{Zakhozhay2022} which classifies this target as a pre-main sequence star. Given the large intrinsic age spread for the LCC members, the HD 117214 might be one of the last formed stellar systems across the region. 

%%S. Messina %%%%%%%%%%%%%%%%%%%%%%%%%%%%%%%%%%%%%%%%%%%%%%%%%%%%%%%%%%%%%%%%%%%%%%%%%%%%%%%%%%%
%%The other possibility is that HD\,117214 is over-luminous because it is younger than LCC. Again, we can %take advantage of the inferred rotation period to give some constrain on the age.
%In Fig.\,\ref{fig_age}, we compare the rotation period with the evolution sequence of rotation periods of %F-type stars in the age range from 8 (Upper Scorpius) to 125 Myr (Pleiades). Rotation periods of F stars %to track their time evolution are taken from \citet{Rebull2018}(Upper Scorpius); \citet{Moraux2013} (h %Persei); \citet{Messina2018} ($\beta$ Pic); \citet{Messina2011} (Argus+IC2391+IC2602); \citet{Rebull2016} %(Pleiades). Although a comparison is meaningful when data from an ensemble of coeval F stars is %available, nonetheless we see that the rotation period of HD117241 is expected for ages younger than 10 %Myr, that is  younger than the LCC and closer to the younger age we inferred from the isochrone fitting. %Therefore, the younger age to explain the over luminosity of HD117214 seems to be most likely explanation.
%%%%%%%%%%%%%%%%%%%%%%%%%%%%%%%%%%%%%%%%%%%%%%%%%%%%%%%%%%%%%%%%%%%%%%%%%%%%%%%%%%%

If the star is younger than 10 Myr, the upper limits to the masses of companions possibly responsible for the disk gap presented in \cite{Engler2020} (their Fig.~7) should be slightly reduced. In this case the configuration of planetary system with several small planets rather than a single one is even more compelling.

\section{Companion candidates astrometry}
\label{sec:cc_astrometry}

Table~\ref{tab:cc} presents the astrometry of all detected companion candidates around \object{HD\,114082}. The 2016-02-14 points correspond to the SHARDDS data, while the 2017-05-17 points correspond to the SHINE data. All candidates are classified as background.
\begin{table*}
  \caption[]{Companion candidates astrometry}
  \label{tab:cc}
  \centering
  \begin{tabular}{lllrrrrrrrrc}
    \hline\hline
    ID   & Date   & MJD    &  $\delta$RA  & $\delta$RA err. & $\delta$DEC & $\delta$DEC err. & Sep. & Sep. err & PA & PA err. & Status \\
       &    &   &  (mas)  & (mas) & (mas) & (mas) & (mas) & (mas) & ($^\circ$) & ($^\circ$) &  \\    
    \hline
    0    & 2016-02-13 & 57431 &  1757.1 &   6.1 & -4276.9 &   6.1 &  4623.7 &   6.1 &  157.67 &  0.08 & B\tablefootmark{a} \\
         & 2017-05-17 & 57890 &  1809.8 &  12.0 & -4252.5 &  15.5 &  4621.6 &  11.0 &  156.95 &  0.20 & B \\
    \hline
    1    & 2016-02-14 & 57431 &  5126.1 &   6.1 & -1829.2 &   6.1 &  5442.7 &   6.1 &  109.64 &  0.06 & B \\
         & 2017-05-17 & 57890 &  5172.0 &  18.4 & -1805.5 &  24.5 &  5478.1 &  25.2 &  109.24 &  0.18 & B \\
    \hline
    2    & 2016-02-14 & 57431 & -4497.3 &   6.1 & -1413.1 &   6.1 &  4714.0 &   6.1 &  252.56 &  0.07 & B \\
         & 2017-05-17 & 57890 & -4452.3 &   7.2 & -1389.8 &   9.0 &  4664.2 &   9.2 &  252.66 &  0.09 & B \\
    \hline
    3    & 2016-02-14 & 57431 &  4362.3 &   6.1 &   639.6 &   6.1 &  4409.0 &   6.1 &   81.66 &  0.08 & B \\
         & 2017-05-17 & 57890 &  4416.2 &   7.1 &   669.9 &   9.7 &  4466.7 &   9.8 &   81.37 &  0.09 & B \\
    \hline
    4    & 2016-02-14 & 57431 &  3854.6 &   6.1 & -4086.9 &   6.1 &  5618.0 &   6.1 &  136.68 &  0.06 & B \\
         & 2017-05-17 & 57890 &  3917.3 &  26.4 & -4055.1 &  26.4 &  5638.1 &  26.4 &  135.99 &  0.27 & B \\
    \hline
    5    & 2016-02-14 & 57431 &  2964.7 &   6.1 & -4902.8 &   6.1 &  5729.5 &   6.1 &  148.84 &  0.06 & B \\
         & 2017-05-17 & 57890 &  3027.0 &  15.5 & -4875.1 &  16.6 &  5738.4 &  14.9 &  148.16 &  0.17 & B \\
    \hline
    6    & 2016-02-14 & 57431 &  1461.5 &   6.1 & -2089.7 &   6.1 &  2550.1 &   6.1 &  145.03 &  0.14 & B \\
         & 2017-05-17 & 57890 &  1515.4 &  11.5 & -2058.2 &  12.0 &  2555.9 &  10.7 &  143.64 &  0.28 & B \\
    \hline
    7    & 2016-02-14 & 57431 &  3618.8 &   6.1 &  2988.2 &   6.1 &  4693.1 &   6.1 &   50.45 &  0.07 & B \\
         & 2017-05-17 & 57890 &  3671.0 &   7.9 &  3023.8 &   8.1 &  4756.1 &   8.5 &   50.52 &  0.09 & B \\
    \hline
    8    & 2016-02-14 & 57431 &  4916.8 &   6.1 & -1392.7 &   6.1 &  5110.2 &   6.1 &  105.82 &  0.07 & B \\
         & 2017-05-17 & 57890 &  4969.8 &   6.1 & -1359.1 &   7.1 &  5152.3 &   7.1 &  105.29 &  0.07 & B \\
    \hline
    9    & 2016-02-14 & 57431 &  2984.0 &   6.1 & -5726.2 &   6.1 &  6457.1 &   6.1 &  152.48 &  0.05 & B \\
         & 2017-05-17 & 57890 &  3043.6 &  11.2 & -5709.2 &  11.9 &  6469.9 &  10.8 &  151.94 &  0.11 & B \\
    \hline
    10   & 2016-02-14 & 57431 &  -344.8 &   6.1 &  2084.5 &   6.1 &  2112.8 &   6.1 &  350.61 &  0.17 & B \\
         & 2017-05-17 & 57890 &  -296.1 &   3.8 &  2113.6 &   9.1 &  2134.2 &   3.6 &  352.03 &  0.25 & B \\
    \hline
    11   & 2016-02-14 & 57431 &  -709.7 &   6.1 &  -380.9 &   6.1 &   805.4 &   6.1 &  241.78 &  0.44 & B \\
         & 2017-05-17 & 57890 &  -657.4 &   8.9 &  -349.2 &  10.5 &   744.4 &  11.0 &  242.02 &  0.63 & B \\
    \hline
    12   & 2016-02-14 & 57431 &  3037.6 &   6.1 &  1253.2 &   6.1 &  3286.0 &   6.1 &   67.58 &  0.11 & B \\
         & 2017-05-17 & 57890 &  3091.8 &   8.8 &  1288.5 &  10.7 &  3349.6 &  11.1 &   67.38 &  0.14 & B \\
    \hline
    13   & 2016-02-14 & 57431 &  1258.0 &   6.1 & -6900.6 &   6.1 &  7014.4 &   6.1 &  169.67 &  0.05 & B \\
         & 2017-05-17 & 57890 &  1305.2 &  14.2 & -6877.4 &  30.6 &  7000.1 &  13.2 &  169.25 &  0.25 & B \\
    \hline
    14   & 2016-02-14 & 57431 &  -596.1 &   6.1 &   910.7 &   6.1 &  1088.5 &   6.1 &  326.79 &  0.32 & B \\
         & 2017-05-17 & 57890 &  -550.5 &   6.3 &   930.0 &   7.2 &  1080.8 &   5.9 &  329.38 &  0.40 & B \\
    \hline
    15   & 2016-02-14 & 57431 & -4918.8 &   6.1 &  3007.4 &   6.1 &  5765.3 &   6.1 &  301.44 &  0.06 & B \\
         & 2017-05-17 & 57890 & -4875.9 &   8.9 &  3035.5 &   9.7 &  5743.6 &  10.1 &  301.90 &  0.08 & B \\
    \hline
    16   & 2016-02-14 & 57431 &  1735.2 &   6.1 &  2271.5 &   6.1 &  2858.4 &   6.1 &   37.38 &  0.12 & B \\
         & 2017-05-17 & 57890 &  1785.3 &   4.8 &  2304.2 &   4.7 &  2914.9 &   4.8 &   37.77 &  0.09 & B \\
    \hline
    17   & 2016-02-14 & 57431 & -6227.9 &   6.1 &  1281.0 &   6.1 &  6358.3 &   6.1 &  281.62 &  0.06 & B \\
         & 2017-05-17 & 57890 & -6182.8 &   7.8 &  1306.5 &   9.4 &  6319.3 &   9.5 &  281.93 &  0.07 & B \\
    \hline
    18   & 2016-02-14 & 57431 & -4690.6 &   6.1 & -1816.7 &   6.1 &  5030.1 &   6.1 &  248.83 &  0.07 & B \\
         & 2017-05-17 & 57890 & -4642.5 &   6.0 & -1793.2 &   6.8 &  4976.8 &   7.0 &  248.88 &  0.07 & B \\
    \hline
    19   & 2016-02-14 & 57431 & -5897.5 &   6.1 &  -984.9 &   6.1 &  5979.2 &   6.1 &  260.52 &  0.06 & B \\
         & 2017-05-17 & 57890 & -5852.9 &   7.3 &  -962.4 &   8.9 &  5931.4 &   9.0 &  260.66 &  0.07 & B \\
    \hline
    20   & 2016-02-14 & 57431 &  4756.5 &   6.1 & -1395.3 &   6.1 &  4957.0 &   6.1 &  106.35 &  0.07 & B \\
         & 2017-05-17 & 57890 &  4821.0 &   8.8 & -1360.7 &  11.9 &  5009.3 &  12.1 &  105.76 &  0.10 & B \\
    \hline
    21   & 2016-02-14 & 57431 & -5343.3 &   6.1 & -1258.5 &   6.1 &  5489.5 &   6.1 &  256.75 &  0.06 & B \\
         & 2017-05-17 & 57890 & -5296.4 &   6.7 & -1236.3 &   8.1 &  5438.8 &   8.2 &  256.86 &  0.07 & B \\
    \hline
    22   & 2016-02-14 & 57431 &   728.2 &   6.1 & -1264.3 &   6.1 &  1459.0 &   6.1 &  150.06 &  0.24 & B \\
         & 2017-05-17 & 57890 &   778.9 &   2.5 & -1234.8 &   2.6 &  1459.9 &   2.5 &  147.76 &  0.10 & B \\
    \hline
  \end{tabular} 
  \tablefoot{\tablefoottext{a}{``B'' stands for background.}}
\end{table*}

\section{Posterior distributions of the fitted parameters}
\begin{figure*}
\centering
\includegraphics[width=16cm]{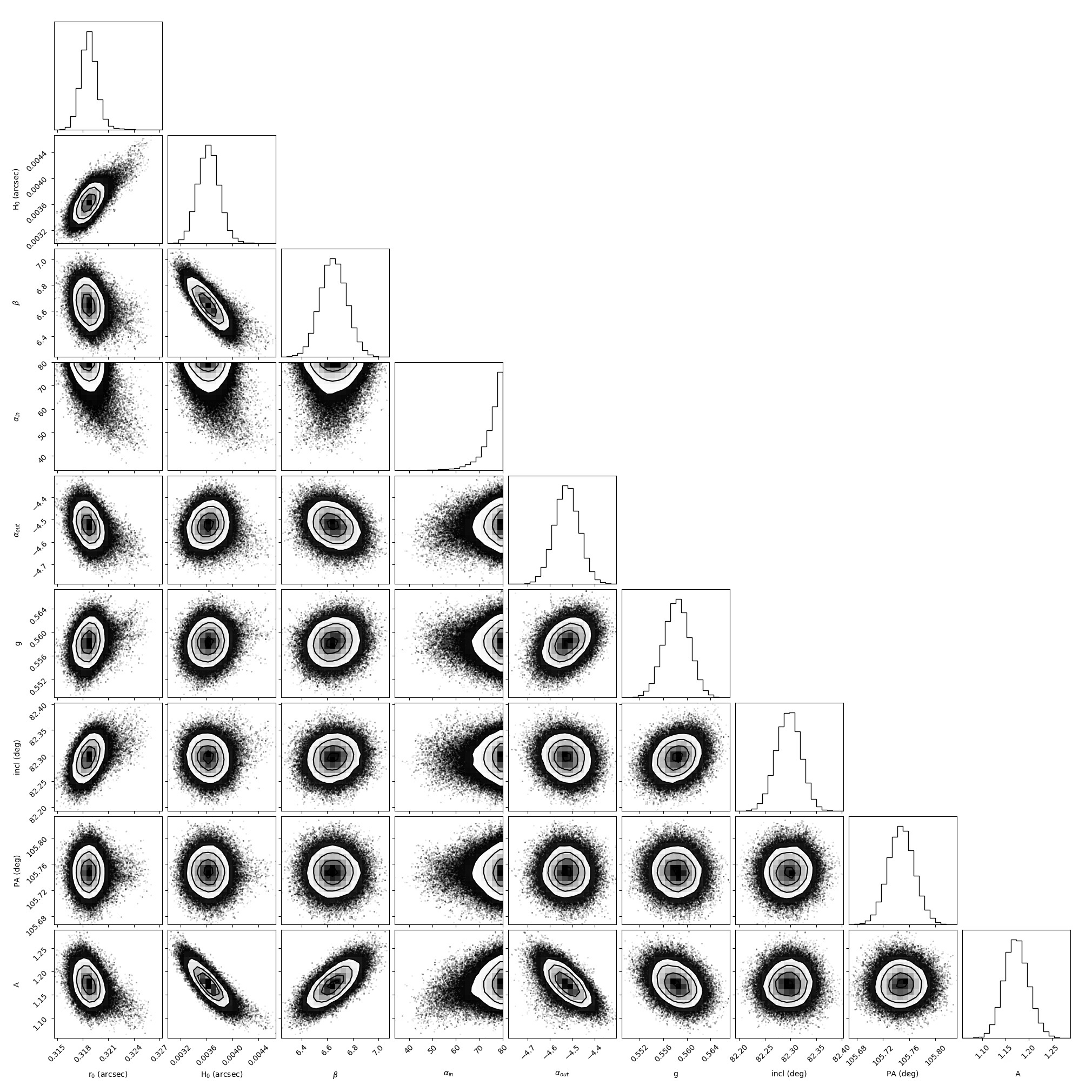} 
\caption{Posterior distributions of the fitted parameters for a model with one HG parameter (model 1). \label{f_mcmc_1g}}
\end{figure*}

\begin{figure*}
\centering
    \includegraphics[width=11.5cm]{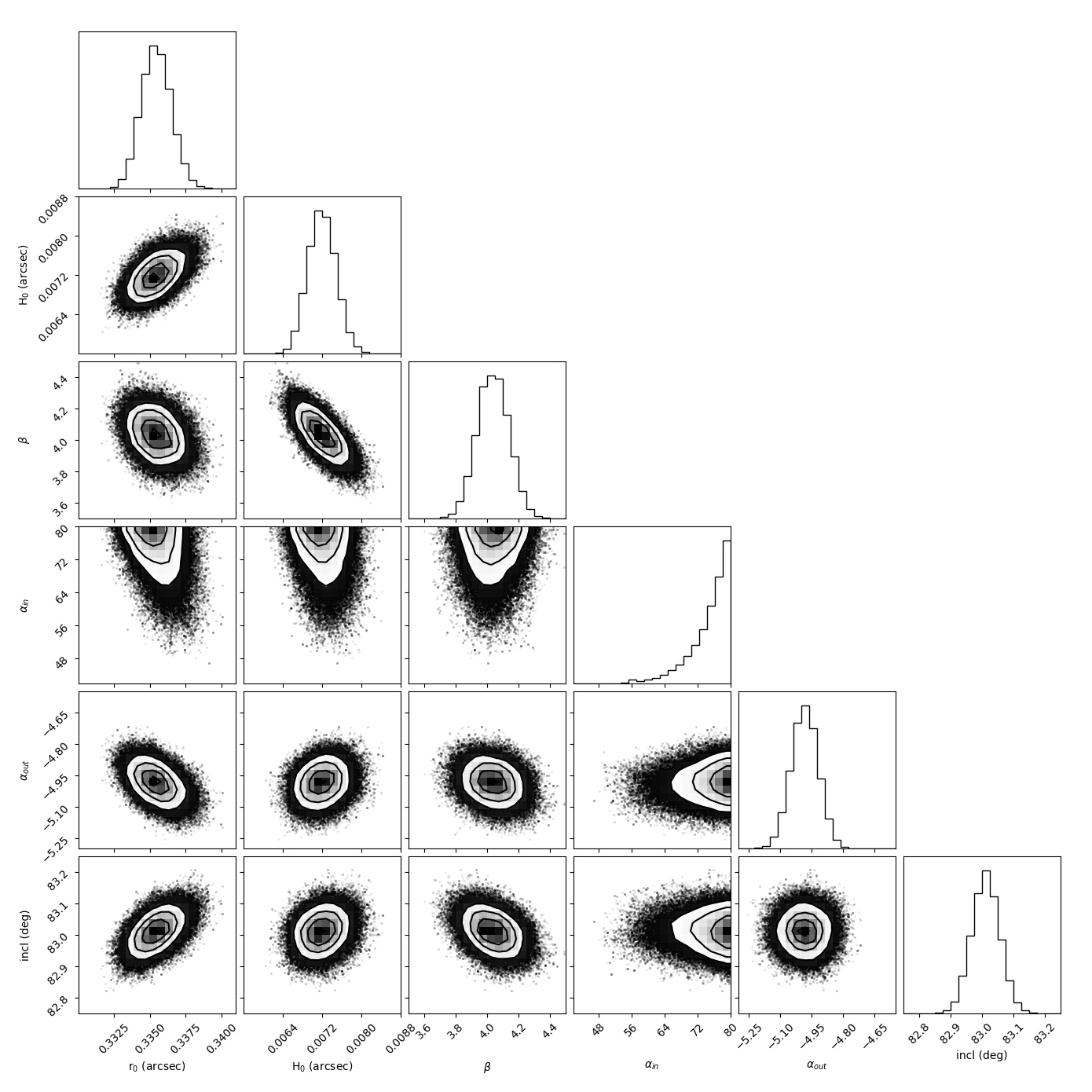} 
    \caption{Posterior distributions of the fitted parameters for a model with two HG parameters (model 2). \label{f_mcmc_2g_1}}
\end{figure*}

Figures \ref{f_mcmc_1g} and \ref{f_mcmc_2g_1} show the posterior distributions of the fitted parameters for models 1 and 2.

\renewcommand{\thefigure}{E.\arabic{figure} (Cont.)}
\addtocounter{figure}{-1}
\begin{figure*}
\centering
    \includegraphics[width=11.5cm]{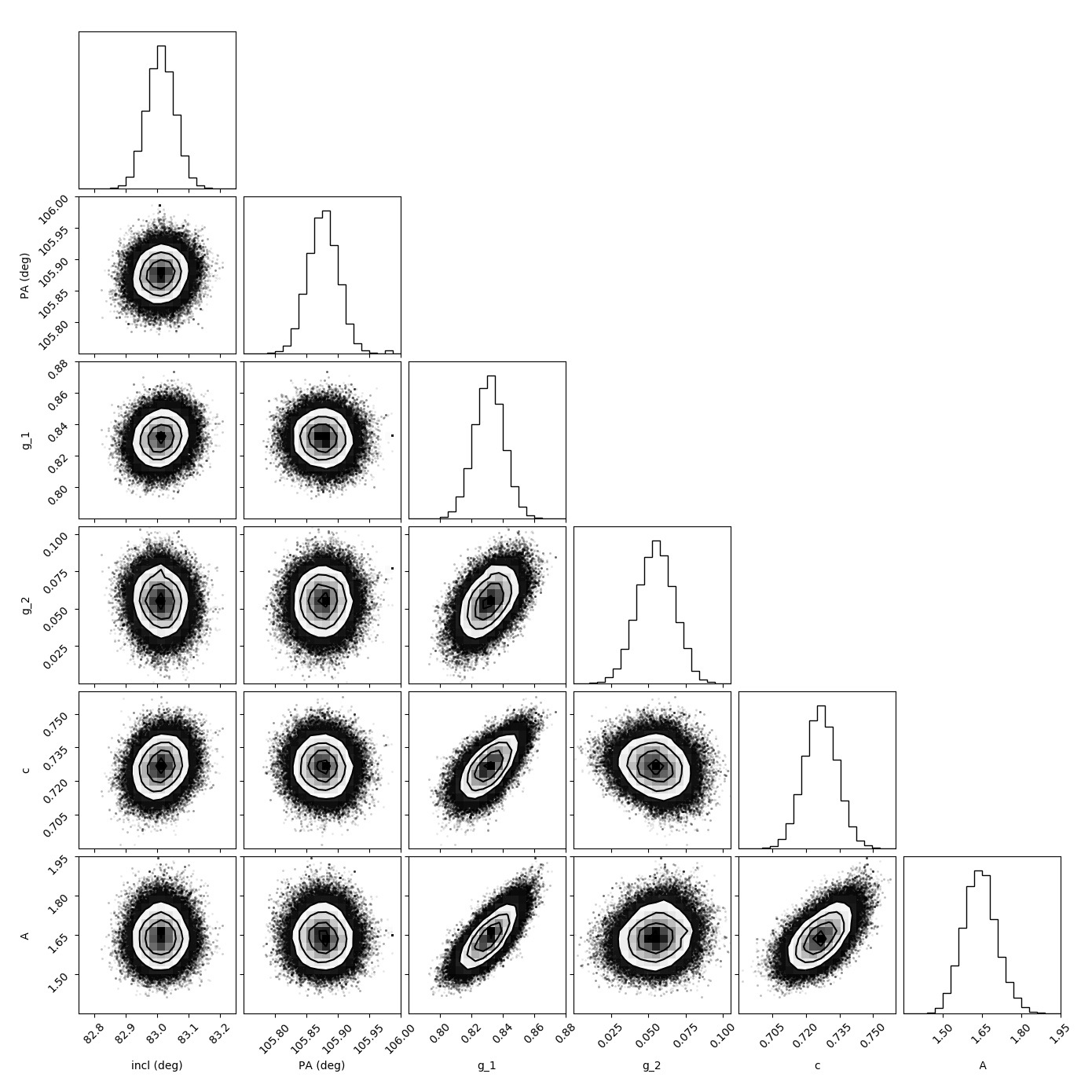} 
    \caption{Posterior distributions of the fitted parameters for a model with two HG parameters (model 2). \label{f_mcmc_2g_2}}
\end{figure*} 

\end{document}